\documentclass[iop]{emulateapj}

\newcommand{\gps}{\ensuremath{g_{\rm P1}}}
\newcommand{\rps}{\ensuremath{r_{\rm P1}}}
\newcommand{\ips}{\ensuremath{i_{\rm P1}}}
\newcommand{\zps}{\ensuremath{z_{\rm P1}}}
\newcommand{\yps}{\ensuremath{y_{\rm P1}}}

\newcommand{\spitzer}{{\it Spitzer~}}
\def\ra#1#2#3{#1$^{\rm h}$#2$^{\rm m}$#3$^{\rm s}$}
\def\dec#1#2#3{$#1^\circ#2'#3''$}

\usepackage{epsfig,graphicx}
\usepackage{amsmath}
\usepackage{natbib}
\usepackage{comment}

\begin{document}

\title{Hydrogen-Poor Superluminous Supernovae and Long-Duration Gamma-Ray Bursts Have Similar Host Galaxies}
\submitted{ApJ in press}
\def\cfa{1}
\def\stsci{2}
\def\ila{3}
\def\ilp{4}
\def\ifa{5}
\def\ucsc{6}
\def\qub{7}
\def\jhu{8}
\def\pri{9}

\author{R.~Lunnan\altaffilmark{\cfa},
 R.~Chornock\altaffilmark{\cfa},
 E.~Berger\altaffilmark{\cfa},
 T.~Laskar\altaffilmark{\cfa},
 W.~Fong\altaffilmark{\cfa},
 A.~Rest\altaffilmark{\stsci},
 N.~E.~Sanders\altaffilmark{\cfa},
 P.~M.~Challis\altaffilmark{\cfa},
 M.~R.~Drout\altaffilmark{\cfa},
 R.~J.~Foley\altaffilmark{\cfa,\ila,\ilp},
 M.~E.~Huber\altaffilmark{\ifa},
 R.~P.~Kirshner\altaffilmark{\cfa},
 C.~Leibler\altaffilmark{\cfa,\ucsc},
 G.~H.~Marion\altaffilmark{\cfa},
 M.~McCrum\altaffilmark{\qub},
 D.~Milisavljevic\altaffilmark{\cfa},
 G.~Narayan\altaffilmark{\cfa},
 D.~Scolnic\altaffilmark{\jhu},
 S.~J.~Smartt\altaffilmark{\qub},
 K.~W.~Smith\altaffilmark{\qub},
 A.~M.~Soderberg\altaffilmark{\cfa},
  J.~L.~Tonry\altaffilmark{\ifa}, 
 W.~S.~Burgett\altaffilmark{\ifa}, 
 K.~C.~Chambers\altaffilmark{\ifa},
 H.~Flewelling\altaffilmark{\ifa},
 K.~W.~Hodapp\altaffilmark{\ifa},
 N.~Kaiser\altaffilmark{\ifa},
 E.~A.~Magnier\altaffilmark{\ifa},
 P.~A.~Price\altaffilmark{\pri}, and 
 R.~J.~Wainscoat\altaffilmark{\ifa}
}

\altaffiltext{\cfa}{Harvard-Smithsonian Center for Astrophysics, 60 Garden St., Cambridge, MA 02138, USA}
\altaffiltext{\stsci}{Space Telescope Science Institute, 3700 San Martin Dr., Baltimore, MD 21218, USA} 
\altaffiltext{\ila}{Astronomy Department, University of Illinois at Urbana-Champaign, 1002 W. Green Street, Urbana, IL 61801, USA}
\altaffiltext{\ilp}{Department of Physics, University of Illinois at Urbana-Champaign, 1010 W. Green Street, Urbana, IL 61801, USA}
\altaffiltext{\ifa}{Institute for Astronomy, University of Hawaii, 2680 Woodlawn Drive, Honolulu, HI 96822, USA}
\altaffiltext{\ucsc}{Department of Astronomy and Astrophysics, UCSC, 1156 High Street, Santa Cruz, CA 95064, USA}
\altaffiltext{\qub}{Astrophysics Research Centre, School of Mathematics and Physics, Queen's University Belfast, Belfast BT7 1NN, UK}
\altaffiltext{\jhu}{Department of Physics and Astronomy, Johns Hopkins University, 3400 North Charles Street, Baltimore, MD 21218, USA}
\altaffiltext{\pri}{Department of Astrophysical Sciences, Princeton University, Princeton, NJ 08544, USA} 

\email{rlunnan@cfa.harvard.edu}

\begin{abstract}
We present optical spectroscopy and optical/near-IR photometry of 31 host galaxies of hydrogen-poor superluminous supernovae (SLSNe), including 15 events from the Pan-STARRS1 Medium Deep Survey. Our sample spans the redshift range $0.1 \lesssim z \lesssim 1.6$ and is the first comprehensive host galaxy study of this specific subclass of cosmic explosions. Combining the multi-band photometry and emission-line measurements, we determine the luminosities, stellar masses, star formation rates and metallicities. We find that as a whole, the hosts of SLSNe are a low-luminosity ($\langle M_B \rangle  \approx -17.3~{\rm mag}$), low stellar mass ($\langle M_* \rangle \approx 2 \times 10^8~{\rm M}_{\odot}$) population, with a high median specific star formation rate $(\langle {\rm sSFR} \rangle \approx 2~{\rm Gyr}^{-1})$. The median metallicity of our spectroscopic sample is low, $12 + \log({\rm O/H}) \approx 8.35 \approx 0.45 Z_{\odot}$, although at least one host galaxy has solar metallicity.  The host galaxies of H-poor SLSNe are statistically distinct from the hosts of GOODS core-collapse SNe (which cover a similar redshift range), but resemble the host galaxies of long-duration gamma-ray bursts (LGRBs) in terms of stellar mass, SFR, sSFR and metallicity. This  result indicates that the environmental causes leading to massive stars forming either SLSNe or LGRBs are similar, and in particular that SLSNe are more effectively formed in low metallicity environments. We speculate that the key ingredient is large core angular momentum, leading to a rapidly-spinning magnetar in SLSNe and an accreting black hole in LGRBs.

\end{abstract}


\section{Introduction}
\label{sec:intro}

The advent of wide-field time-domain surveys like the Panoramic Survey Telescope and Rapid Response System (Pan-STARRS; PS1), the Palomar Transient Factory (PTF) and the Catalina Real-Time Transient Survey (CRTS) has led to the discovery of a growing number of ``superluminous'' supernovae (SLSNe), characterized by luminosities $\sim 10-100$ times larger than ordinary Type Ia and core-collapse SNe. Their spectra are diverse, though distinct subclasses are emerging (e.g., \citealt{gal12}). For example, members of the subclass of the SLSNe that show hydrogen in their spectra can be classified as Type IIn SNe, with the origin of the extreme luminosity being interaction with a dense circumstellar medium \citep[e.g.][]{ock+07,slf+07, scs+10, rfg+11, mbt+13}.

For the SLSNe without hydrogen, however, the energy source(s) remains a matter of debate. Many of these objects form a spectroscopic subclass characterized by a blue continuum with a few broad rest-frame UV absorption features from intermediate-mass elements; in some cases the spectra develop Ic features at late times \citep{psb+10,qkk+11,ccs+11}. An interaction scenario similar to the H-rich SLSNe has been proposed for these objects as well \citep{ci11, gb12, mm12}, but this requires extreme mass loss episodes ( $ > 1~{\rm M}_{\odot}/{\rm yr}$) shortly before the explosion and may be at odds with the lack of intermediate-width lines seen in the spectra \citep{ci11,ccs+11,qkk+11}. Another proposed mechanism is energy injection by a newborn, rapidly-spinning magnetar \citep{kb10, woo10, dhw+12}, which can explain a wide range of luminosities and timescales \citep{ccs+11, lcb+13, isj+13}. Other H-poor SLSNe, most notably SN\,2007bi, have been proposed to be examples of pair-instability SNe \citep{gmo+09,ysv+10}, but this interpretation is controversial and the events can also be explained in an interaction or magnetar scenario (\citealt{dhw+12,msm+13,nsj+13,msk+14}). The superluminous SN PS1-10afx did not resemble any previously seen SLSNe, and may represent a new class of transients \citep{cbr+13}; its unusual properties lead \citet{qwo+13} to conclude that it may have been a lensed Type Ia SN rather than a SLSN.

An important clue to the origin of the H-poor SLSNe may come from their host galaxy properties. An early study by \citet{nsg+11} utilized GALEX near-UV and Sloan Digital Sky Survey (SDSS) $r'$-band photometry of SLSN hosts, and used this to argue for a preference for low-luminosity (and by extension, possibly low-metallicity) galaxies. However, this study was limited in several ways. First, all SLSNe were analyzed as a group regardless of spectral properties. Second, it was based on limited data: of the seven H-poor SLSN hosts considered, only three were actually detected in either of the two photometric bands they considered. Third, it relied on the luminosities in only two bands to draw inferences about underlying properties of interest (e.g. metallicity), which were not measured directly. A possible trend of low-metallicity galaxies was also pointed out by \citet{sps+11}, who determined the metallicities for two SLSN host galaxies, and found them to be low and comparable to the host galaxies of long-duration gamma-ray burst (LGRBs)\citep[e.g.][]{sgl09,lbk+10,lkb+10}. Recently, detailed studies of the host galaxies of two individual H-poor SLSNe \citep{csb+13,lcb+13} revealed low metallicities and high specific star formation rates, similar to LGRB host galaxies.

Despite these initial results it is clear that to fully examine the physical properties of SLSN host environments, make a meaningful comparison to other classes of transients and draw conclusions about the progenitors requires several key improvements on the existing data. First, high-quality spectroscopic data and optical/NIR photometric data are needed to accurately determine the host galaxy luminosities, stellar masses, metallicities, star formation rates (SFRs) and specific SFRs. Second, a comprehensive study, examining the SLSN host galaxies as a population rather than a few individual objects is essential. With the large number of SLSNe being discovered by Pan-STARRS and other surveys, this is now feasible.

Here, we present such observations and analysis of a sample of 31 SLSN hosts, spanning a redshift range of $0.1 \lesssim z \lesssim 1.6$. Our sample includes 15 objects from the Pan-STARRS1 Medium Deep Survey (PS1/MDS), and 16 targets from other surveys available in the literature. We only include hosts of H-poor SLSNe, as the energy source of Type IIn SLSNe is better understood and possibly distinct from the H-poor SLSNe. We do however include all types of H-poor SLSNe, so as not to make any initial assumptions about potentially different energy sources. This is the most comprehensive systematic study of SLSN hosts so far. 

This paper is organized as follows. We describe our sample of SLSN hosts and follow-up photometric and spectroscopic observations in Section~\ref{sec:obs}. We describe our comparison samples and statistical methods in Section~\ref{sec:samples}. We detail how the various galaxy properties are derived from the data, and discuss them in Section~\ref{sec:galprop}. Implications for the progenitors and caveats are discussed in Section~\ref{sec:disc}, and we summarize our conclusions in Section~\ref{sec:conc}. All calculations in this paper assume a $\Lambda$CDM cosmology with $H_0 = 70$~km~s$^{-1}$~Mpc$^{-1}$, $\Omega_{\rm M} = 0.27$ and $\Omega_{\Lambda} = 0.73$ \citep{ksd+11}.

\section{Observations}
\label{sec:obs}

\subsection{Targets}
Our sample consists of 15 H-poor SLSNe discovered in the PS1/MDS transient search. To supplement the PS1 sample, which covers the redshift range $0.5 \lesssim z \lesssim 1.6$, we also include events from the literature, extending the redshift coverage down to $z \approx 0.1$ and bringing the total number of targets up to 31. Table~\ref{tab:gallist} lists all targets, including references to the SN discoveries.

For the purposes of this paper, we define a H-poor SLSN as a SN with a peak absolute magnitude $ M \lesssim -20.5$, and without evidence of hydrogen in the spectrum. The majority of objects belong to the subclass of SLSNe with spectra resembling SN\,2005ap and SCP06F6 \citep{qkk+11}. However, we include other H-poor SLSNe such as SN\,2007bi \citep{gmo+09} and PS1-10afx \citep{cbr+13}, to explore their environments and relation to the other events.

\subsubsection{SLSNe from the PS1/MDS Transient Survey}
The PS1 telescope on Haleakala is a high-etendue wide-field survey
instrument with a 1.8-m diameter primary mirror and a $3.3^\circ$
diameter field of view imaged by an array of sixty $4800\times 4800$
pixel detectors, with a pixel scale of $0.258''$
\citep{PS1_system,PS1_GPCA}.  The observations are obtained through
five broadband filters ($grizy_{\rm P1}$); details of the filters and photometric system are described in \citet{tsl+12}. 

The PS1/MDS consists of 10 fields (each with a
single PS1 imager footprint) observed in \gps\rps\ips\zps with a typical cadence of 3~d in each filter, to a $5\sigma$ depth of $\sim 23.3$ mag; \yps is used near full moon with a typical depth of $\sim 21.7$ mag.
The standard reduction, astrometric solution, and stacking of the
nightly images is done by the Pan-STARRS1 Image Processing Pipeline (IPP) system \citep{PS1_IPP, PS1_astrometry} on a computer cluster at the Maui High Performance Computer Center. The nightly MDS stacks are transferred
to the Harvard FAS Research Computing cluster, where they are
processed through a frame subtraction analysis using the {\tt photpipe}
pipeline developed for the SuperMACHO and ESSENCE surveys \citep{rsb+05, gsc+07, mpr+07,rsf+13}.

A subset of targets is chosen for spectroscopic follow-up, using Blue Channel spectrograph on the 6.5m MMT telescope \citep{swf89}, the Gemini Multi-Object Spectrograph (GMOS; \citealt{hja+04}) on the 8-m Gemini telescopes, and the Low Dispersion Survey Spectrograph (LDSS3) and Inamori-Magellan Areal Camera and Spectrograph (IMACS; \citealt{dhb+06}) on the 6.5m Magellan telescopes. Since the beginning of the survey in 2010, we have discovered over 15 SLSNe in the PS1/MDS data (\citealt{ccs+11,bcl+12,cbr+13,lcb+13,msr+14,msk+14}; Lunnan et al., in prep.). The combination of a relatively small survey area and deep photometry provides sensitivity primarily to SLSNe at higher redshifts; the current sample spans $ 0.5 \lesssim z \lesssim 1.6$. Thus, the PS1 sample is a great complement to the SLSNe from other surveys,  which are generally found at $z \lesssim 0.5$ due to shallower photometry (Figure~\ref{fig:rmag_z}).

While PS1/MDS is an untargeted survey the spectroscopic follow-up is not complete. The SLSNe in our sample were targeted by some combination of light curve and host properties, in particular for having long observed rise times, or standing out as being several magnitudes brighter than any apparent host.  We discuss to what extent selection effects may affect our results in Section~\ref{sec:sel}.

\subsubsection{SLSNe from the Literature}
In addition to the PS1/MDS SLSNe, we also include in our sample H-poor SLSNe reported by other surveys, most notably the Palomar Transient Factory (PTF; \citealt{lkd+09}), the Catalina Real-Time Transient Survey (CRTS; \citealt{ddm+09}) and the Robotic Optical Transient Search Experiment (ROTSE-III; \citealt{akm+03}). Table~\ref{tab:gallist} lists these objects, with references. Since not all of the objects in this list have published spectra available, we include objects that are reported with a peak absolute magnitude $\lesssim -21~{\rm mag}$, and classified as Type Ic or described as having a spectrum similar to known H-poor SLSNe.

\subsection{Host Galaxy Photometry}

\subsubsection{Ground-Based Optical Photometry}
For targets from the PS1/MDS SLSN sample, we stack the pre-explosion images and obtain deep $grizy_{\rm P1}$ photometry of the host galaxies. The results are listed in Table~\ref{tab:ps1phot}. In addition, a number of the literature hosts are detected in the Sloan Digital Sky Survey (SDSS), and we use available DR9 model magnitudes for these objects \citep{sdssdr9}.

To complement the survey photometry, we obtained deep imaging observations of a number of targets that were either not covered by or undetected in either PS1/MDS or SDSS. This was mainly done with LDSS and IMACS on Magellan, as well as with MMTCam\footnote{http://www.cfa.harvard.edu/mmti/wfs.html}, an f/5 imager on the 6.5m MMT telescope.

We processed and stacked all images using standard routines in IRAF\footnote{IRAF is
  distributed by the National Optical Astronomy Observatory,
    which is operated by the Association of Universities for Research
    in Astronomy, Inc., under cooperative agreement with the National
    Science Foundation.}. We measured host galaxy magnitudes using aperture-matched photometry, with zeropoints determined either from observations of standard star fields taken at similar airmass on the same night, or from photometry of stars with listed SDSS and/or PS1/MDS magnitudes. In cases where the host galaxy was not detected, a $3\sigma$ upper limit was determined by measuring the mean magnitude of objects at the detection threshold (S/N of 3). Images of the hosts are shown in Figures~\ref{fig:galpix} and \ref{fig:hstpix} with the instrument and filter noted on each image. All non-PS1 and non-SDSS photometry is listed in Table~\ref{tab:morephot}, and the host galaxy apparent magnitude distribution is shown in Figure~\ref{fig:rmag_z} as a function of redshift.

\begin{figure*}
\begin{center}
\begin{tabular}{ccccc}
\includegraphics[width=3.0cm]{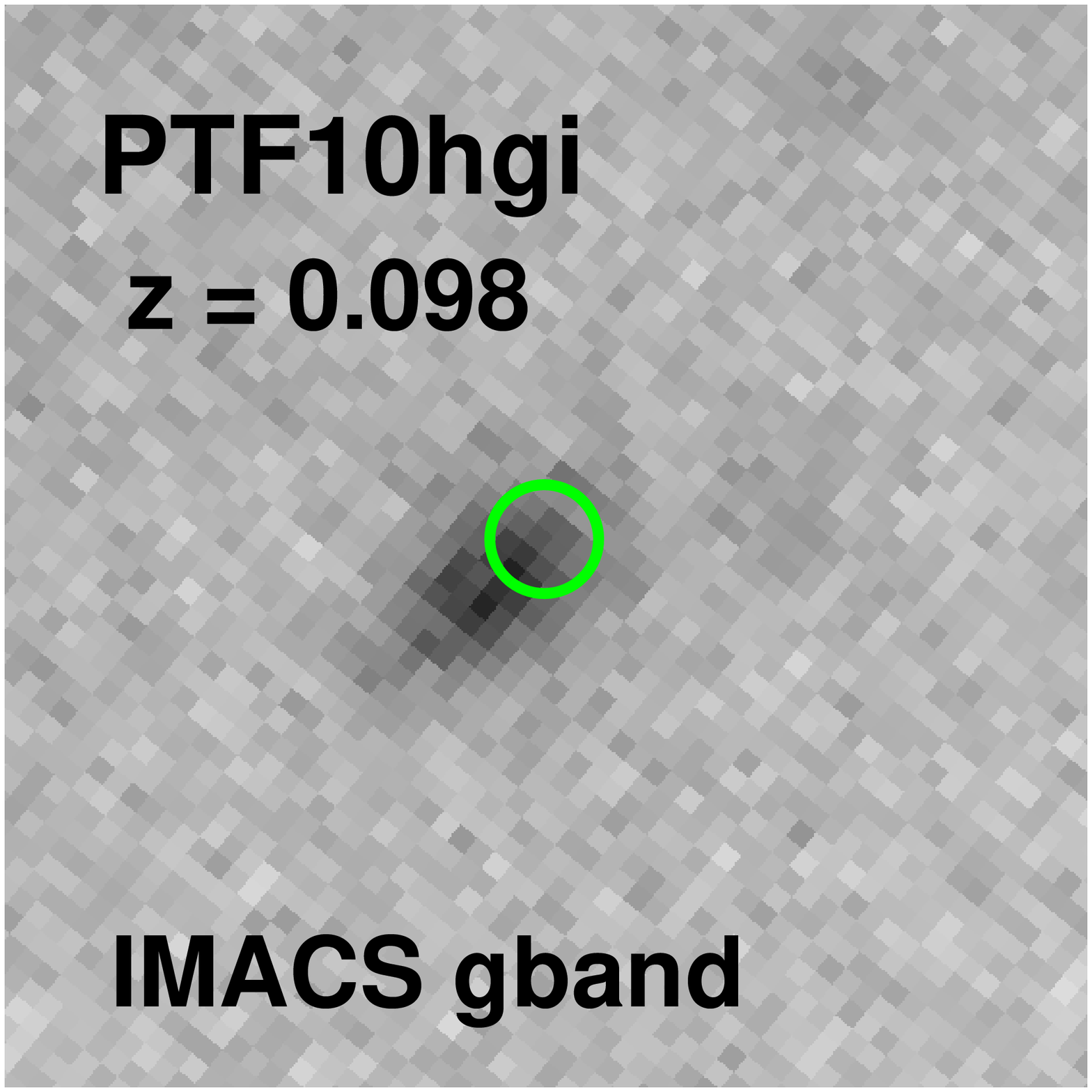} & \includegraphics[width=3.0cm]{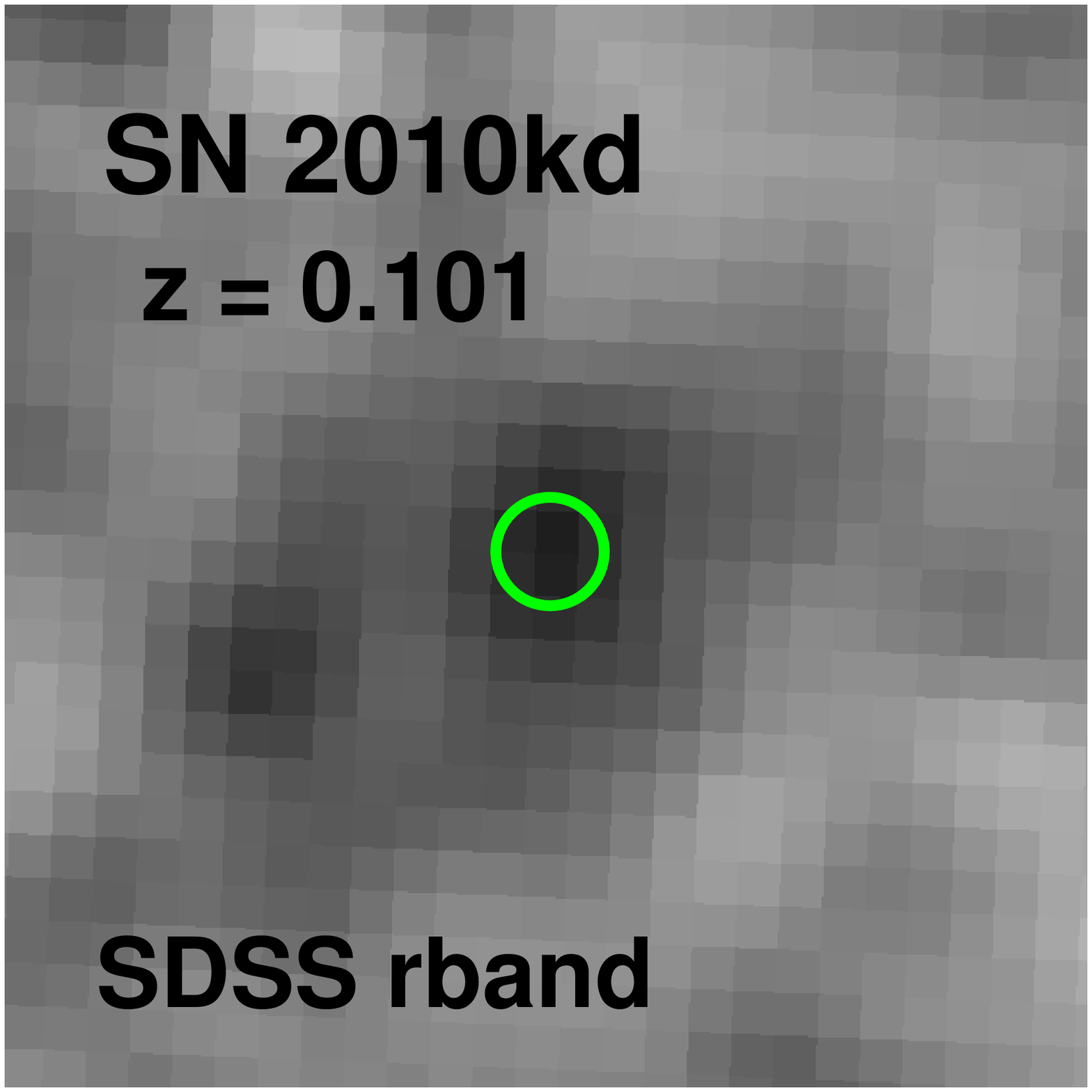} & \includegraphics[width=3.0cm]{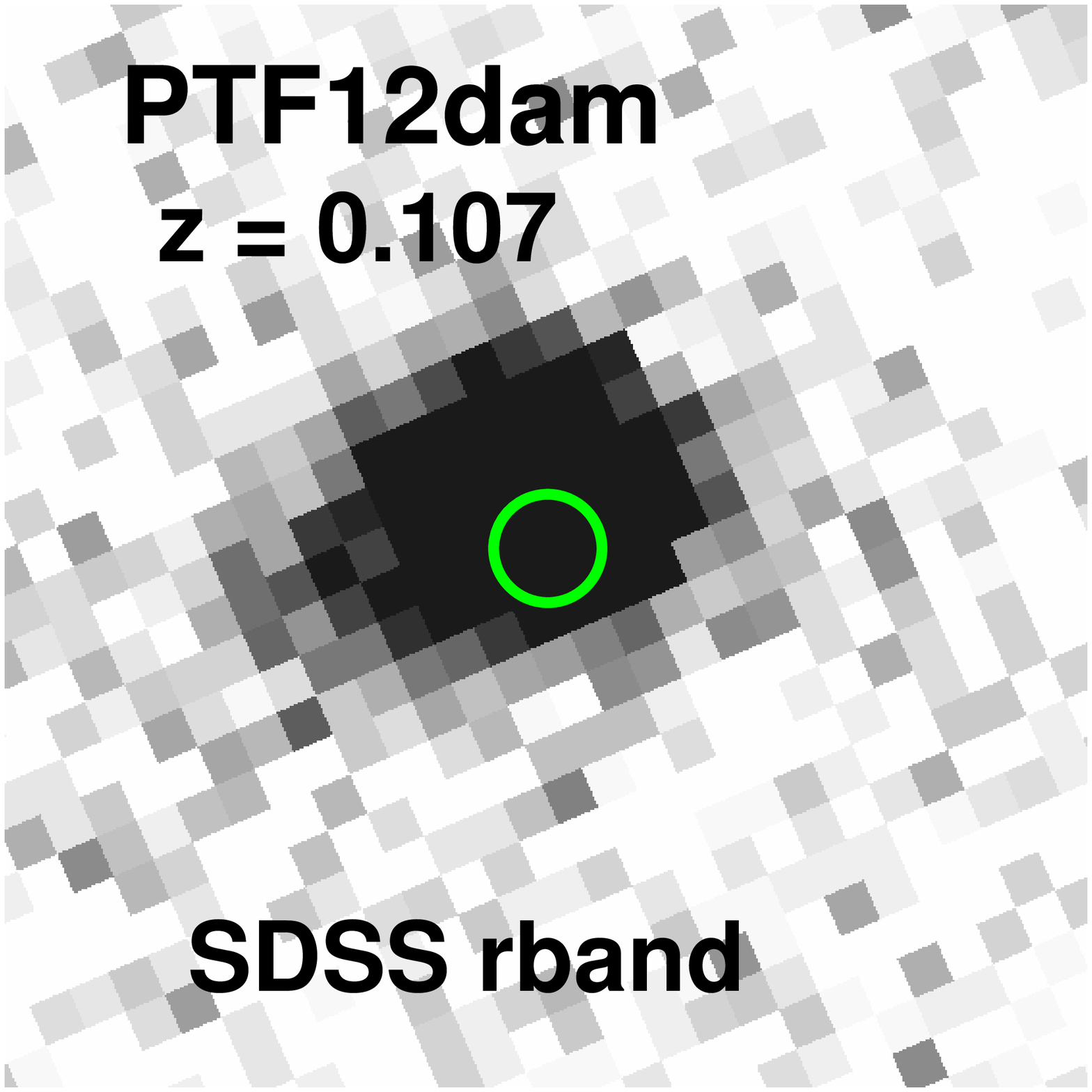} & \includegraphics[width=3.0cm]{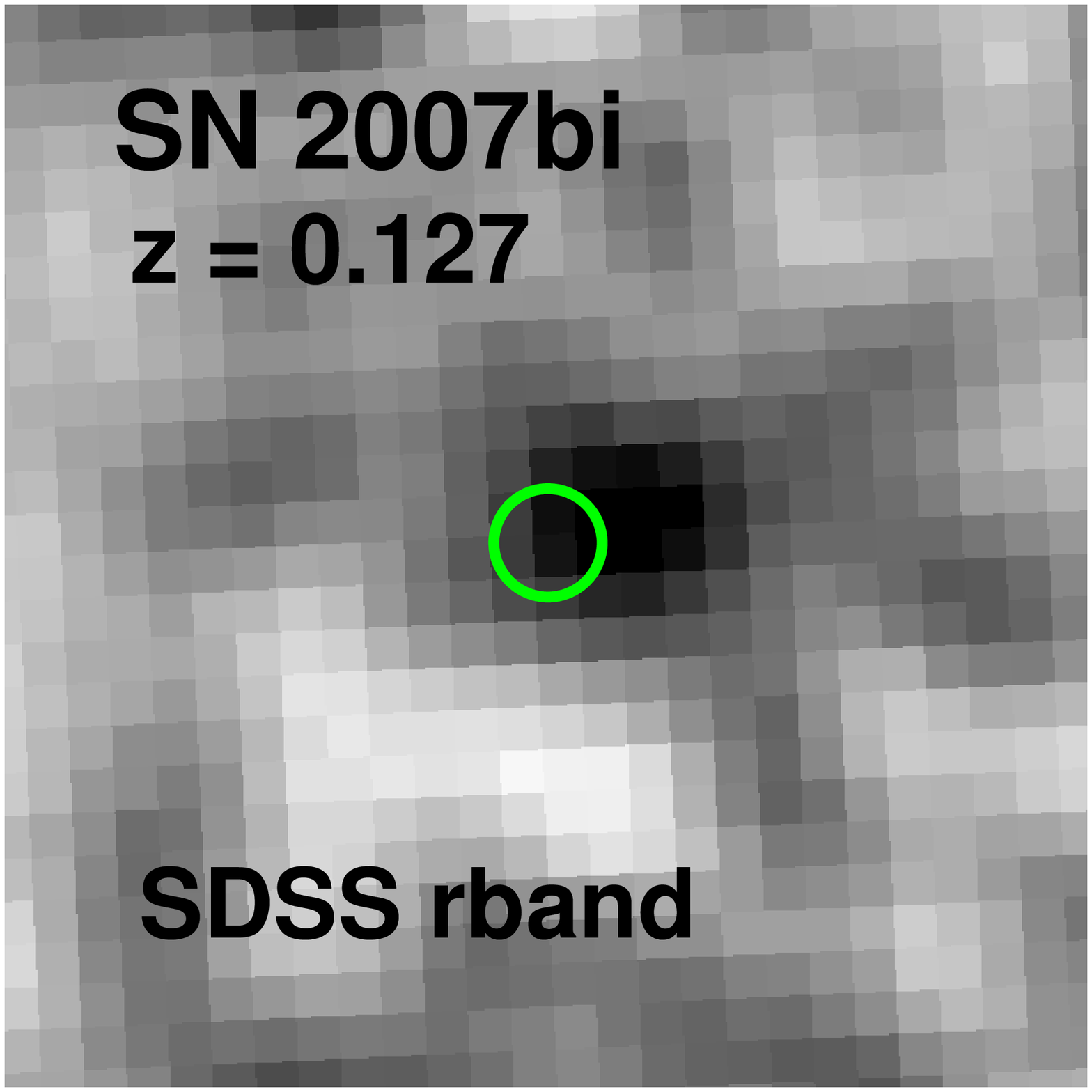} & \includegraphics[width=3.0cm]{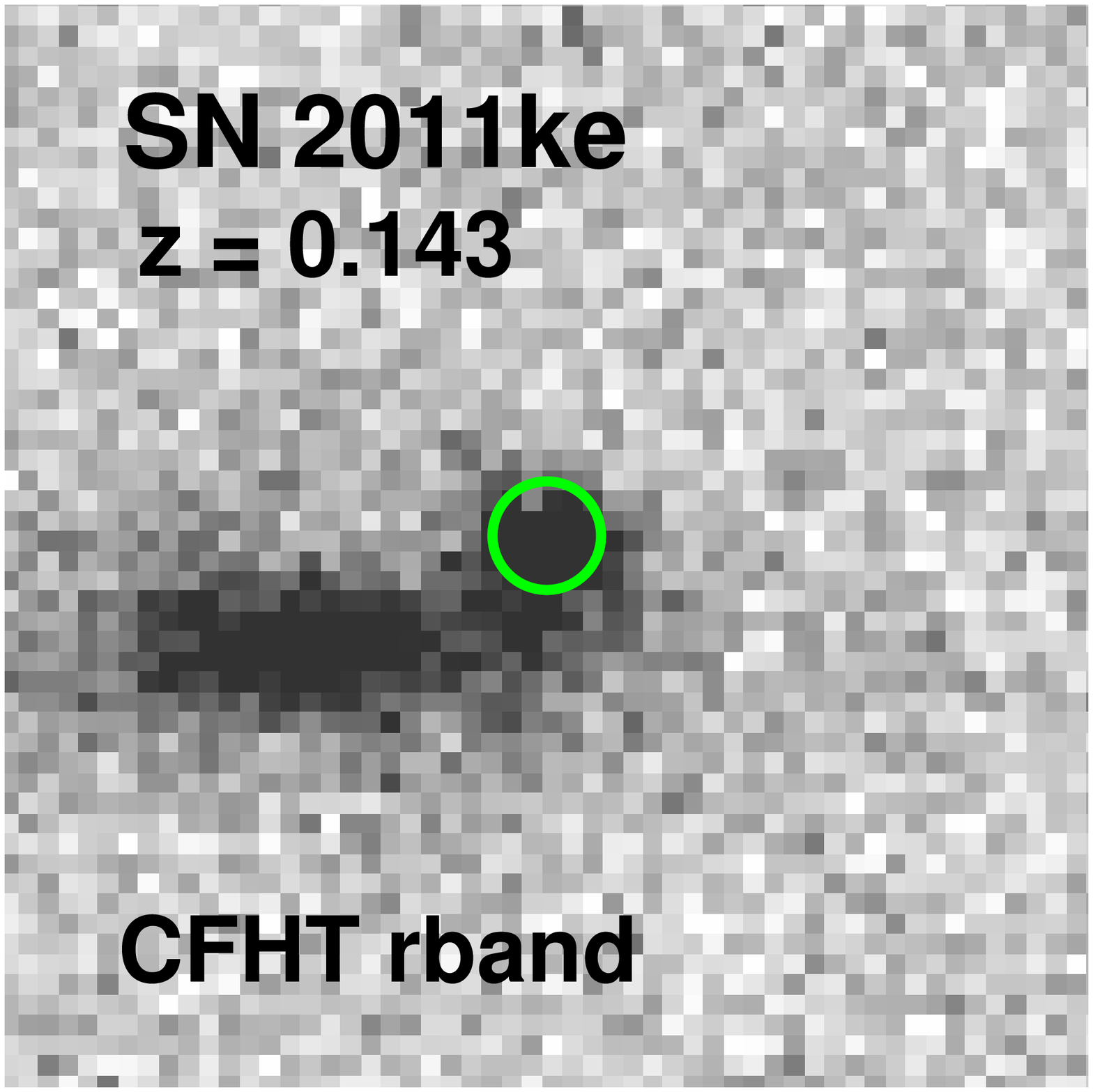} \\
 \includegraphics[width=3.0cm]{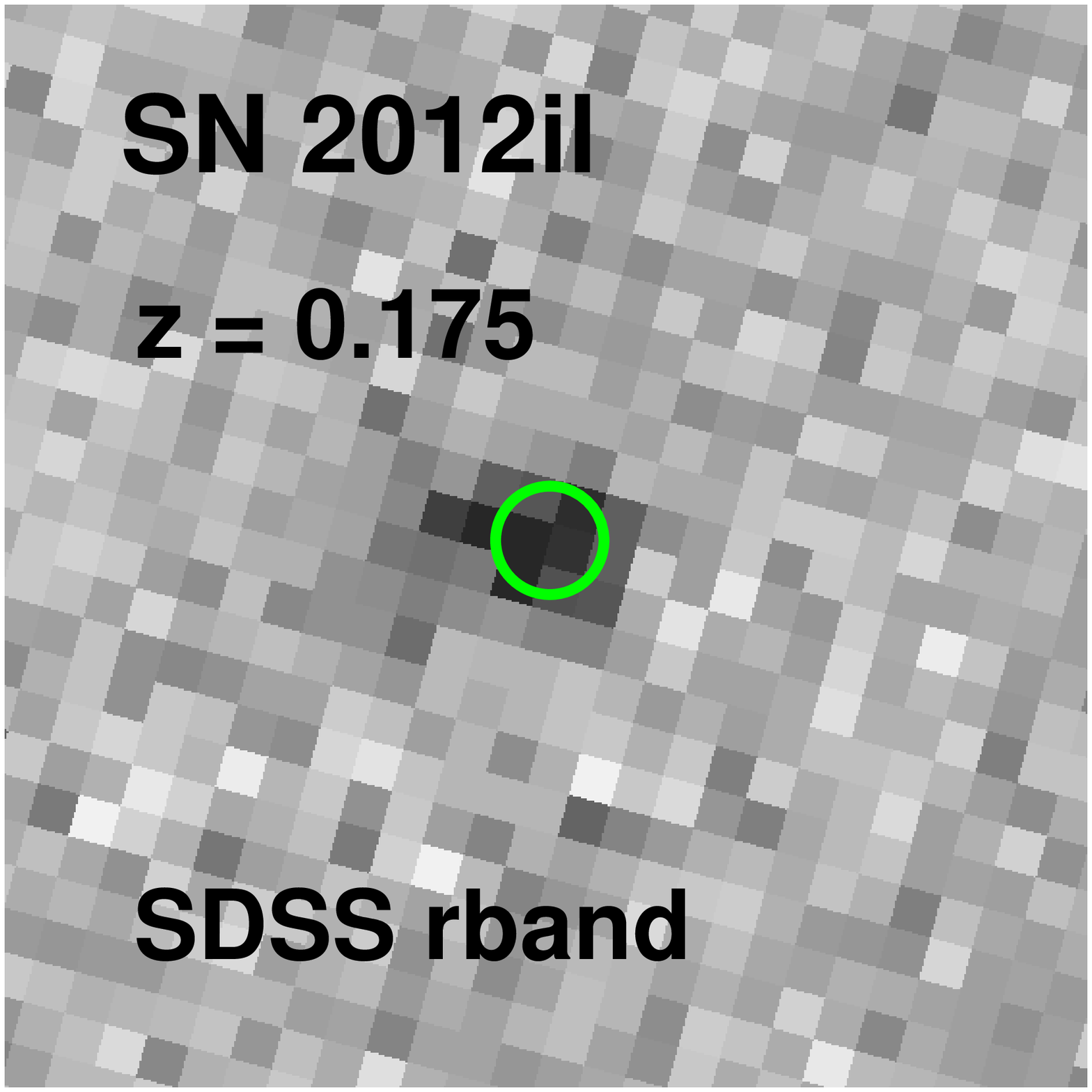} & \includegraphics[width=3.0cm]{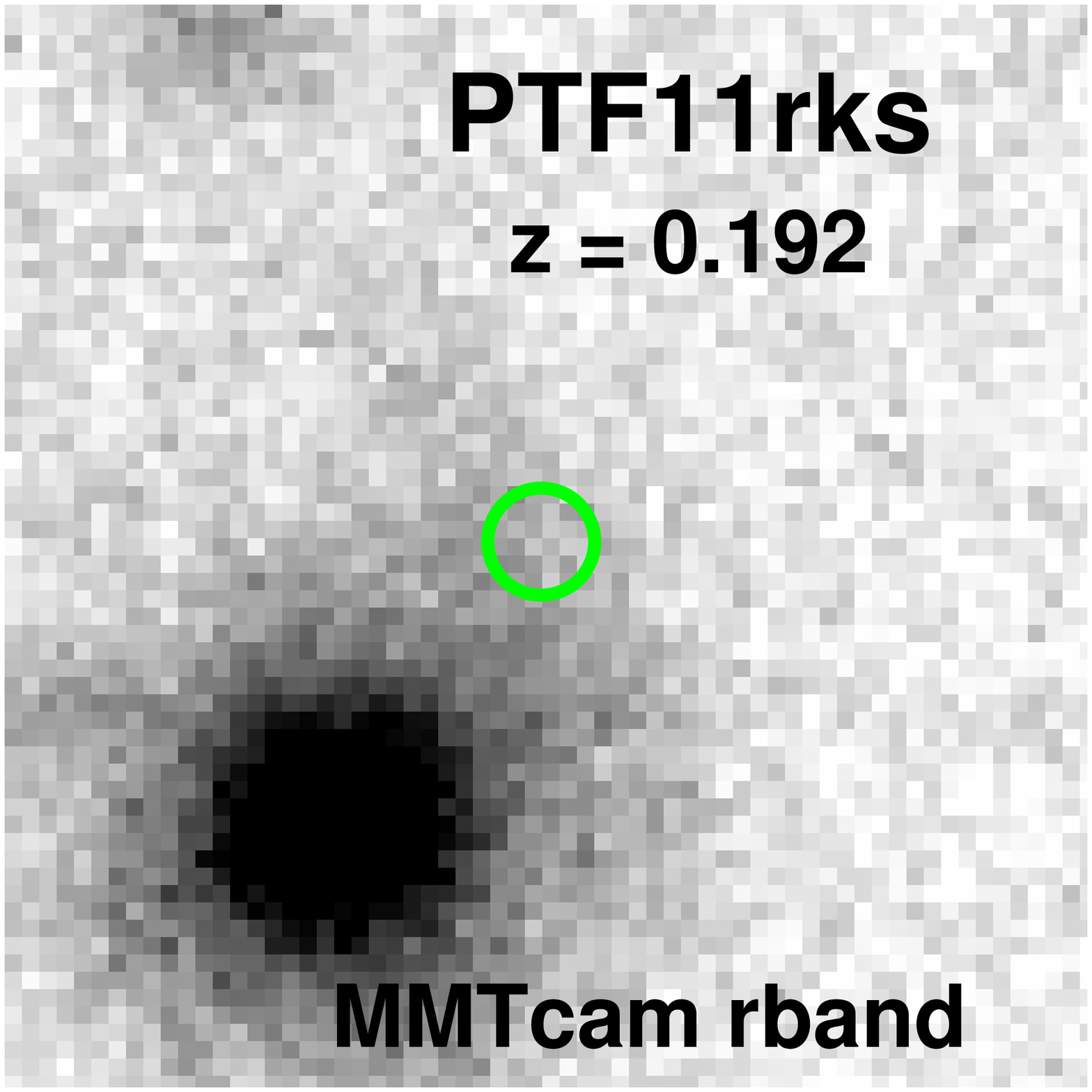} & \includegraphics[width=3.0cm]{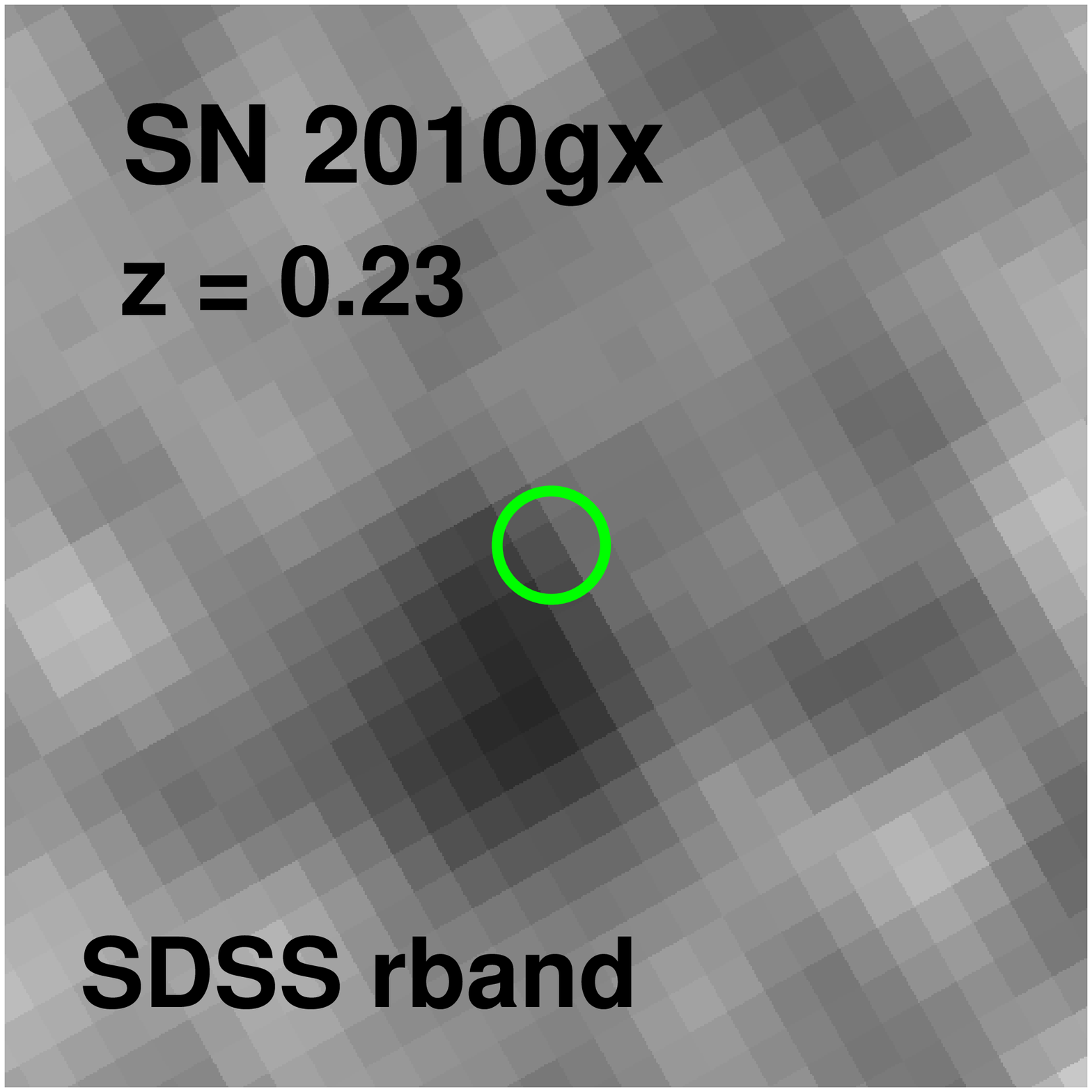} & \includegraphics[width=3.0cm]{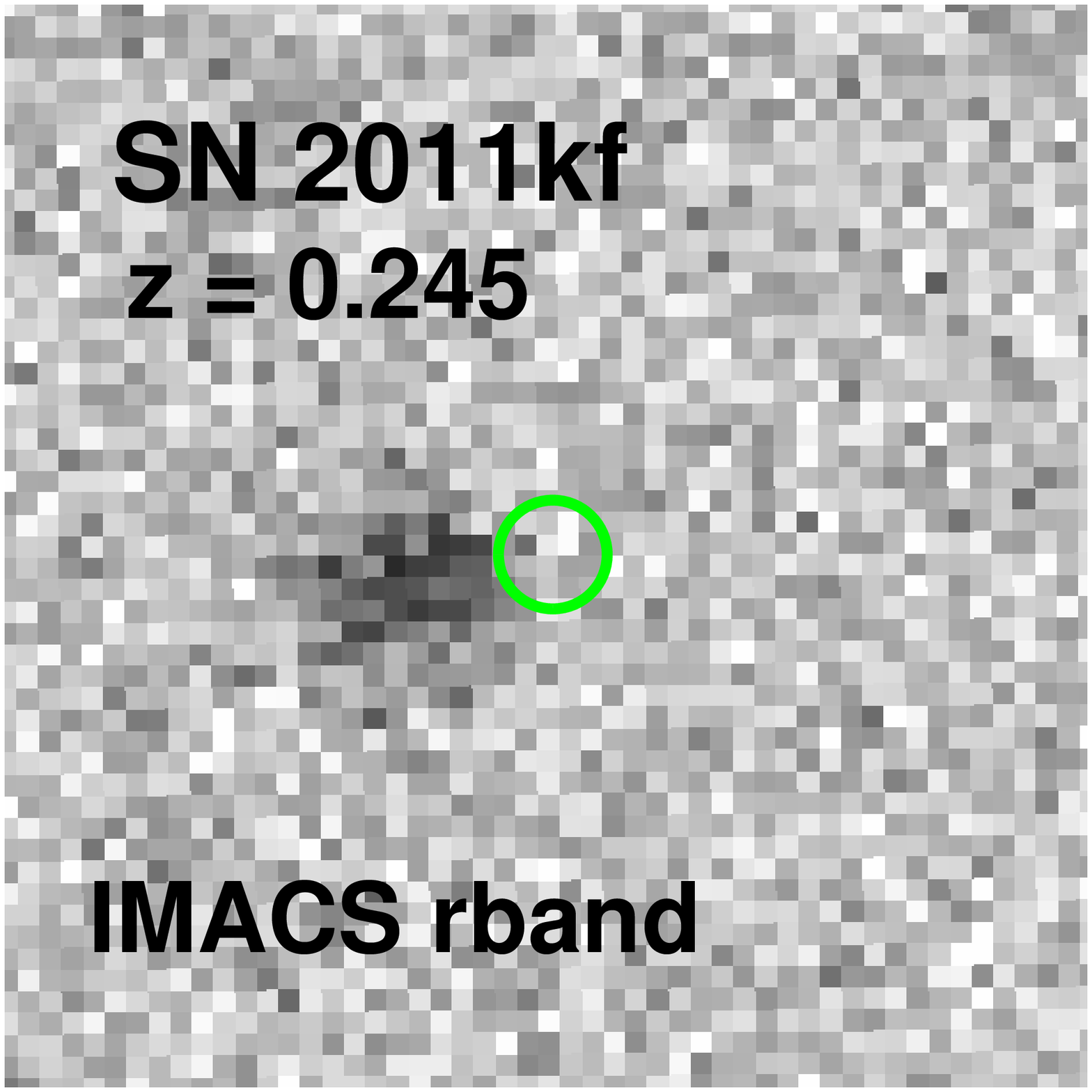} & \includegraphics[width=3.0cm]{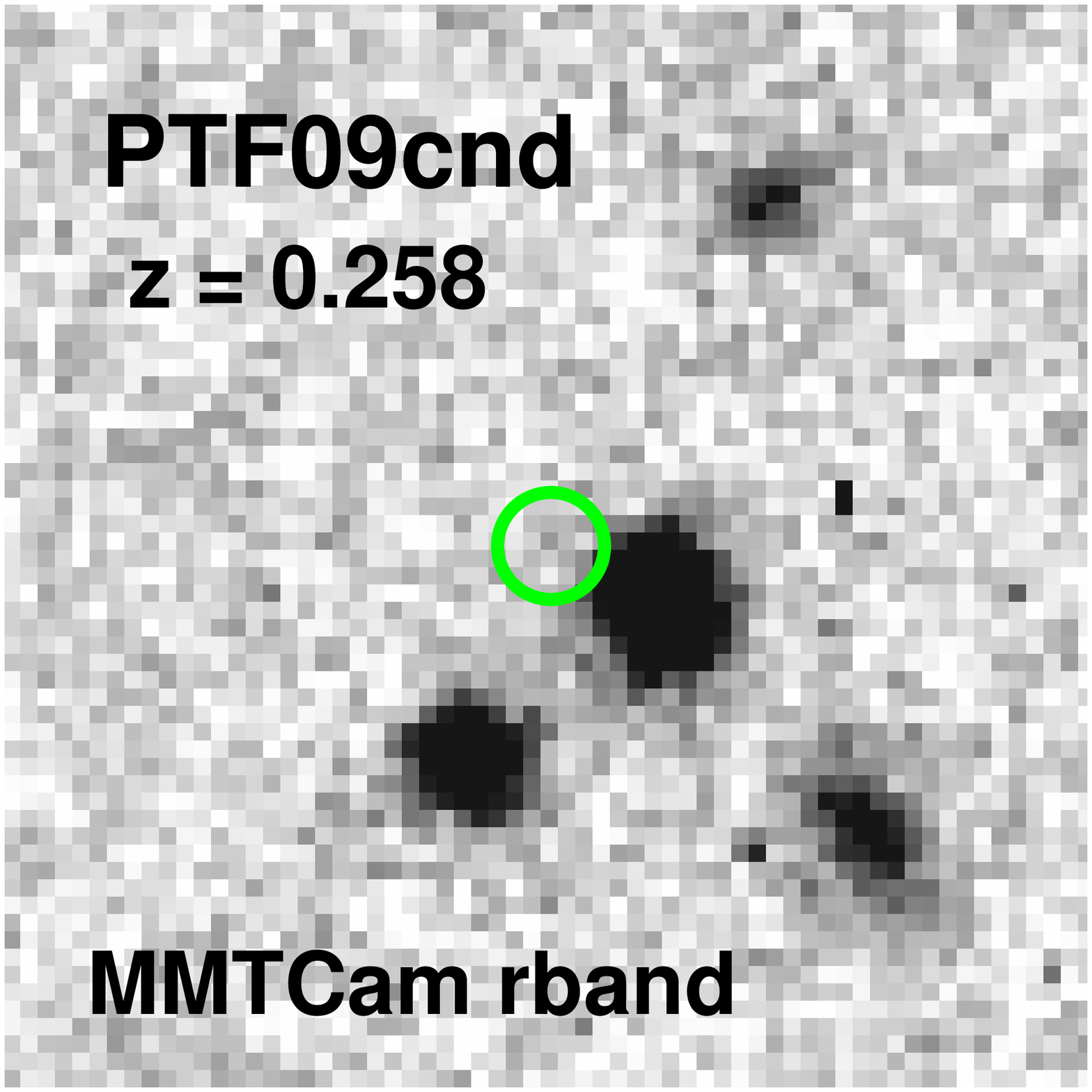} \\
  \includegraphics[width=3.0cm]{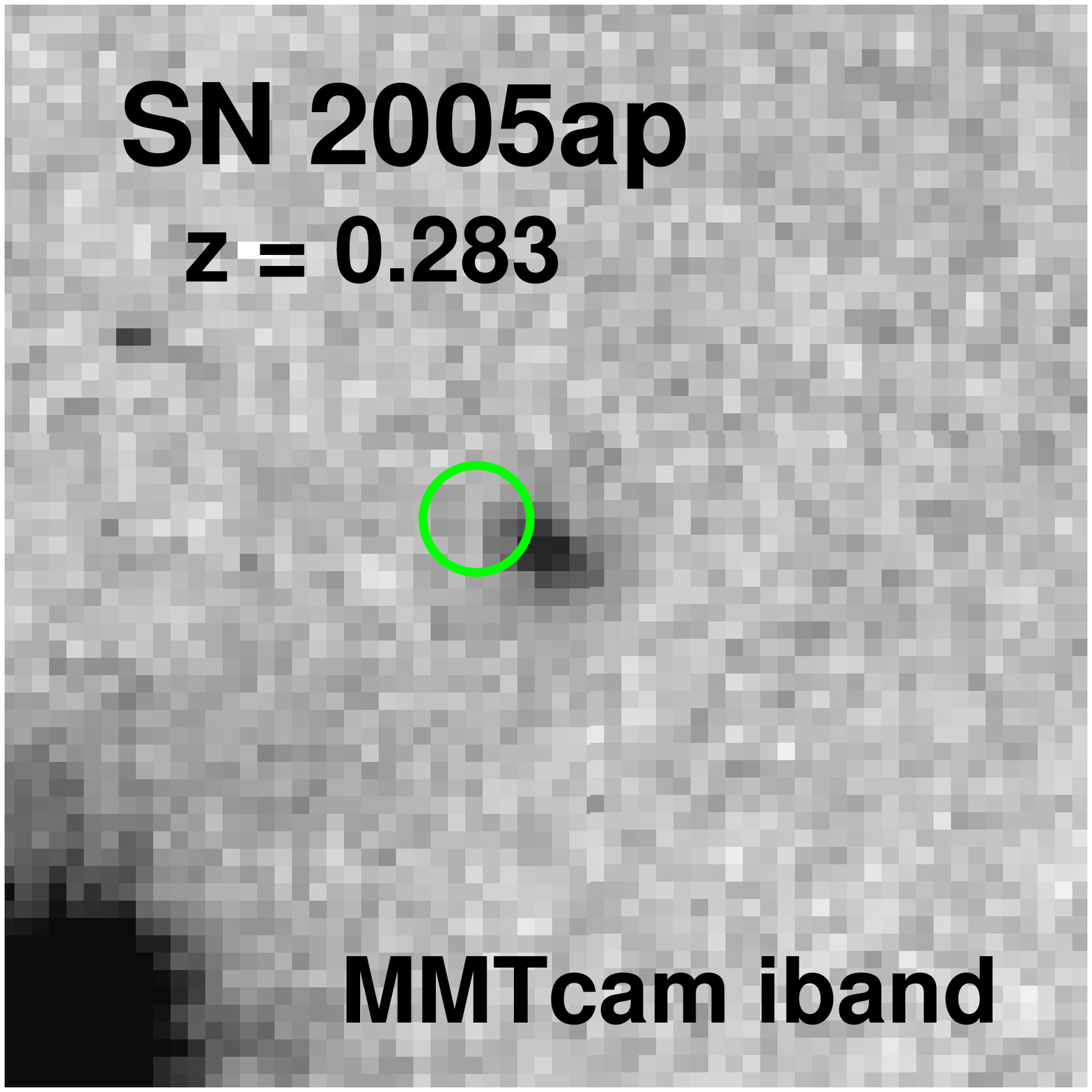} & \includegraphics[width=3.0cm]{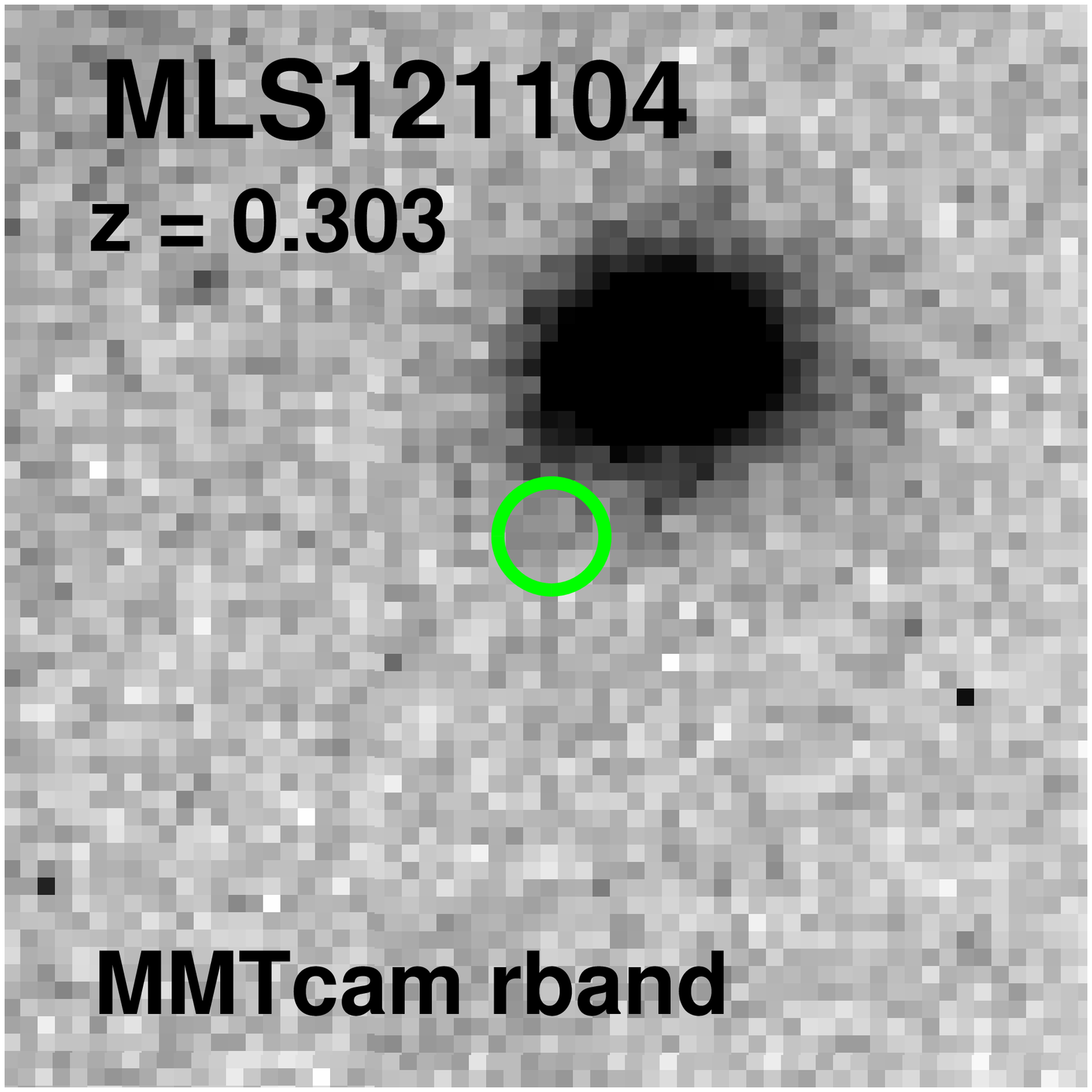} & \includegraphics[width=3.0cm]{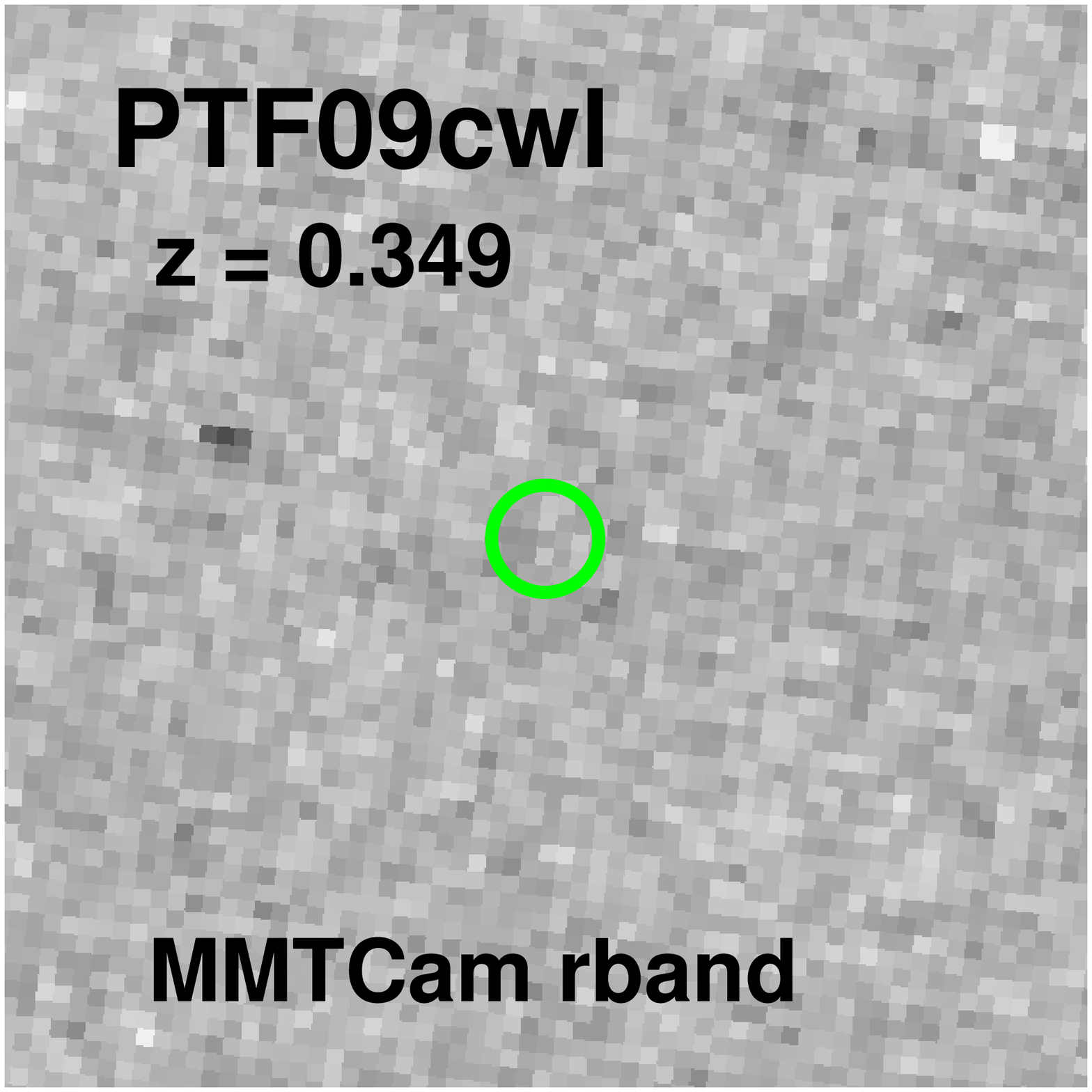} & \includegraphics[width=3.0cm]{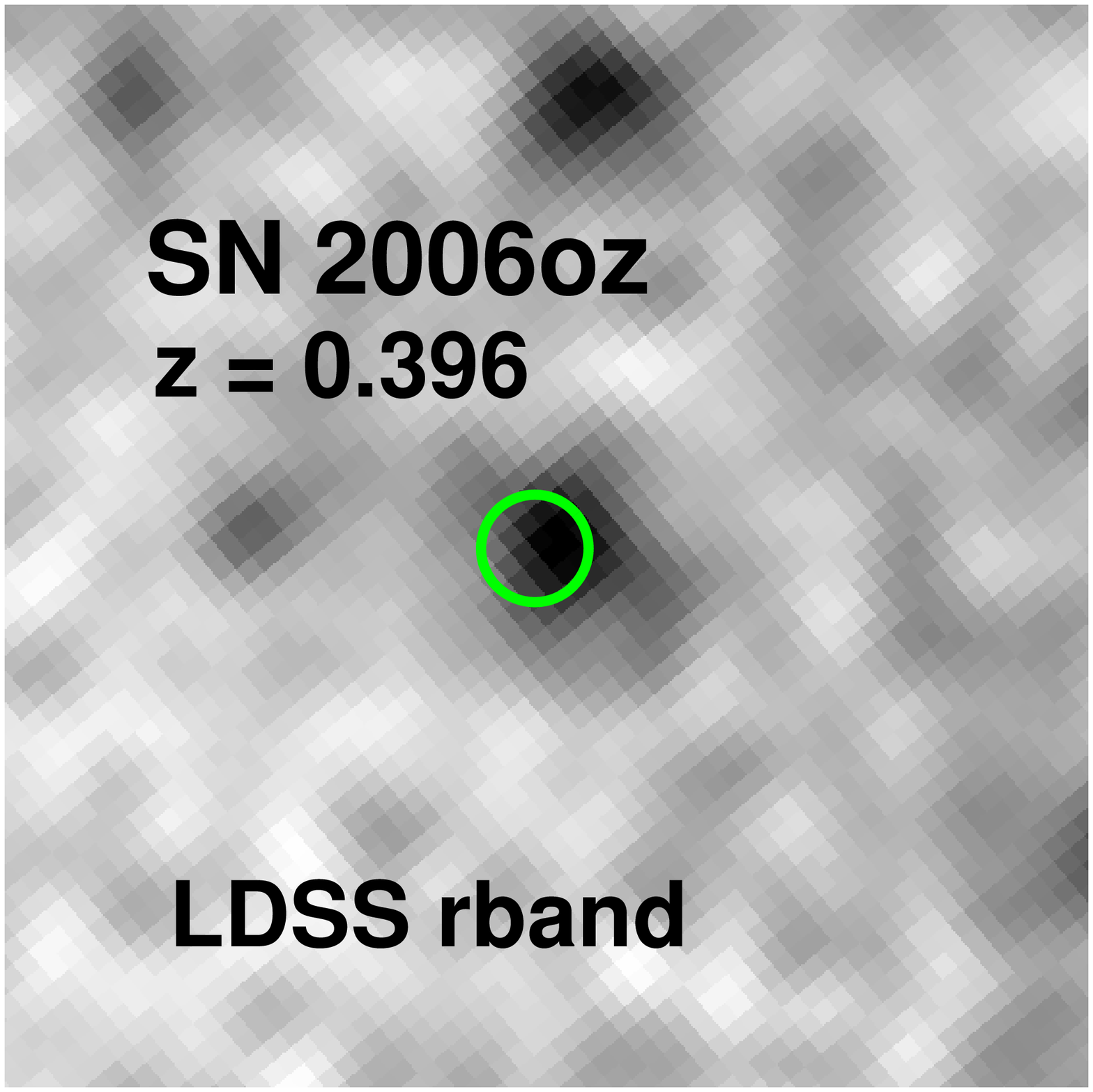} & \includegraphics[width=3.0cm]{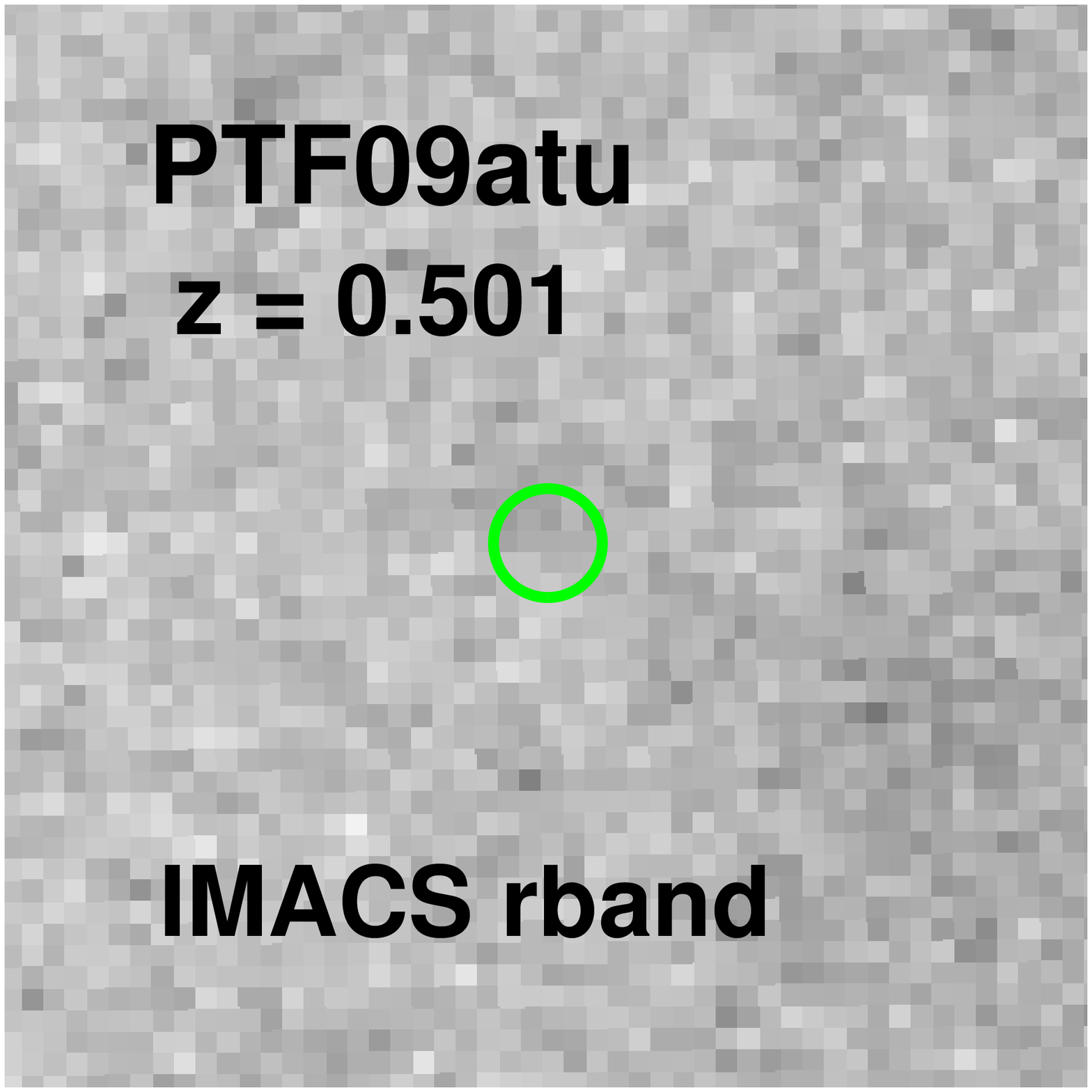} \\
   \includegraphics[width=3.0cm]{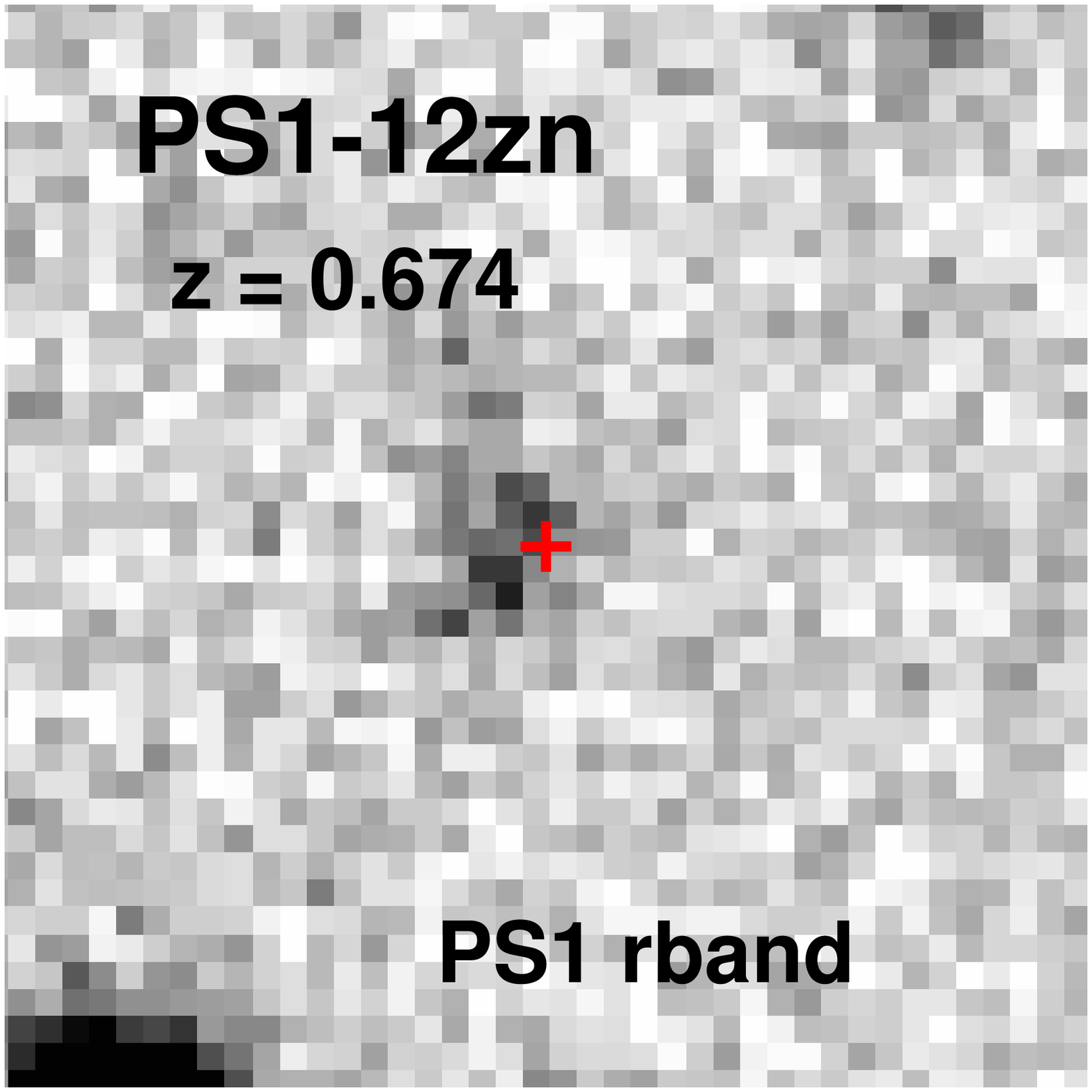} &  \includegraphics[width=3.0cm]{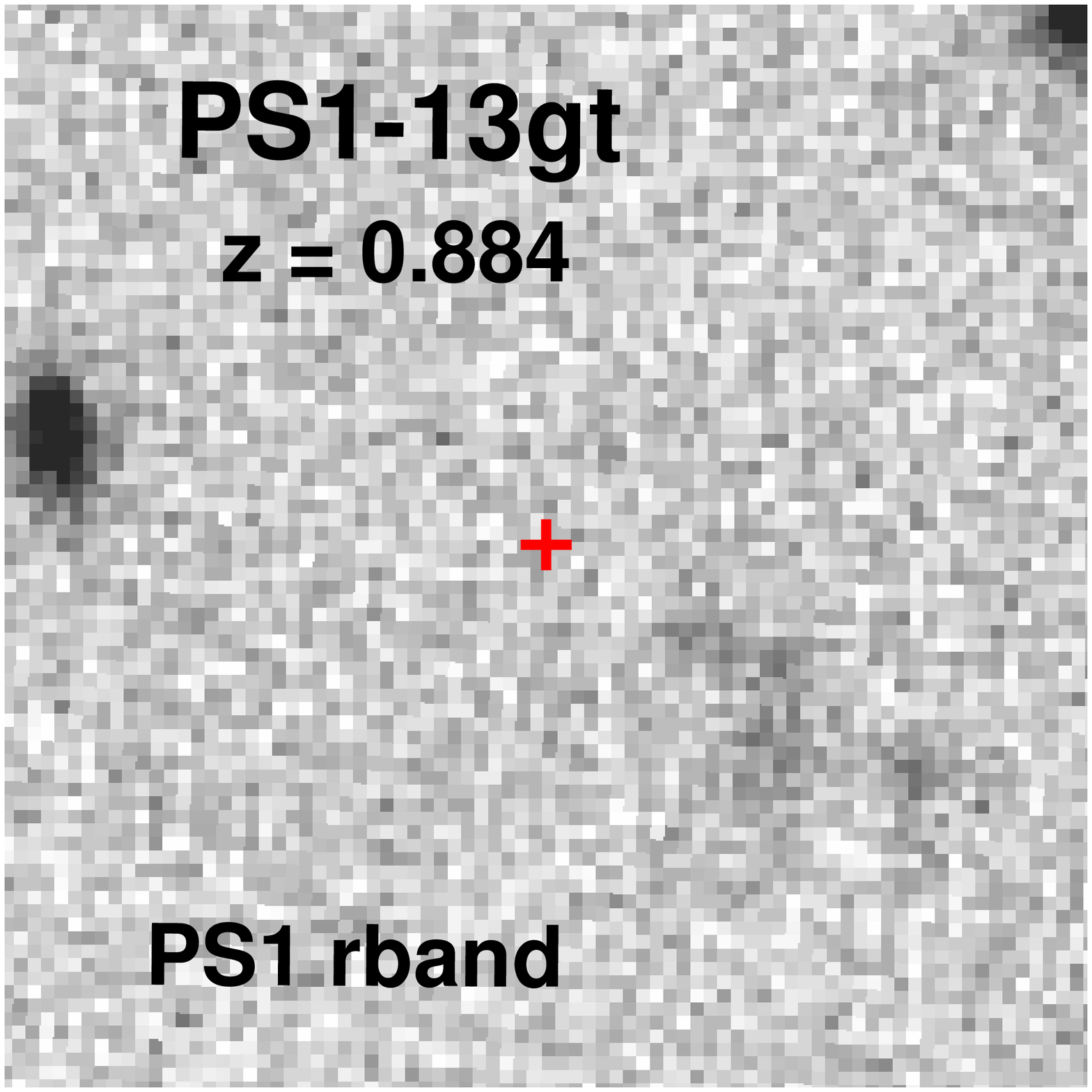}  &  \includegraphics[width=3.0cm]{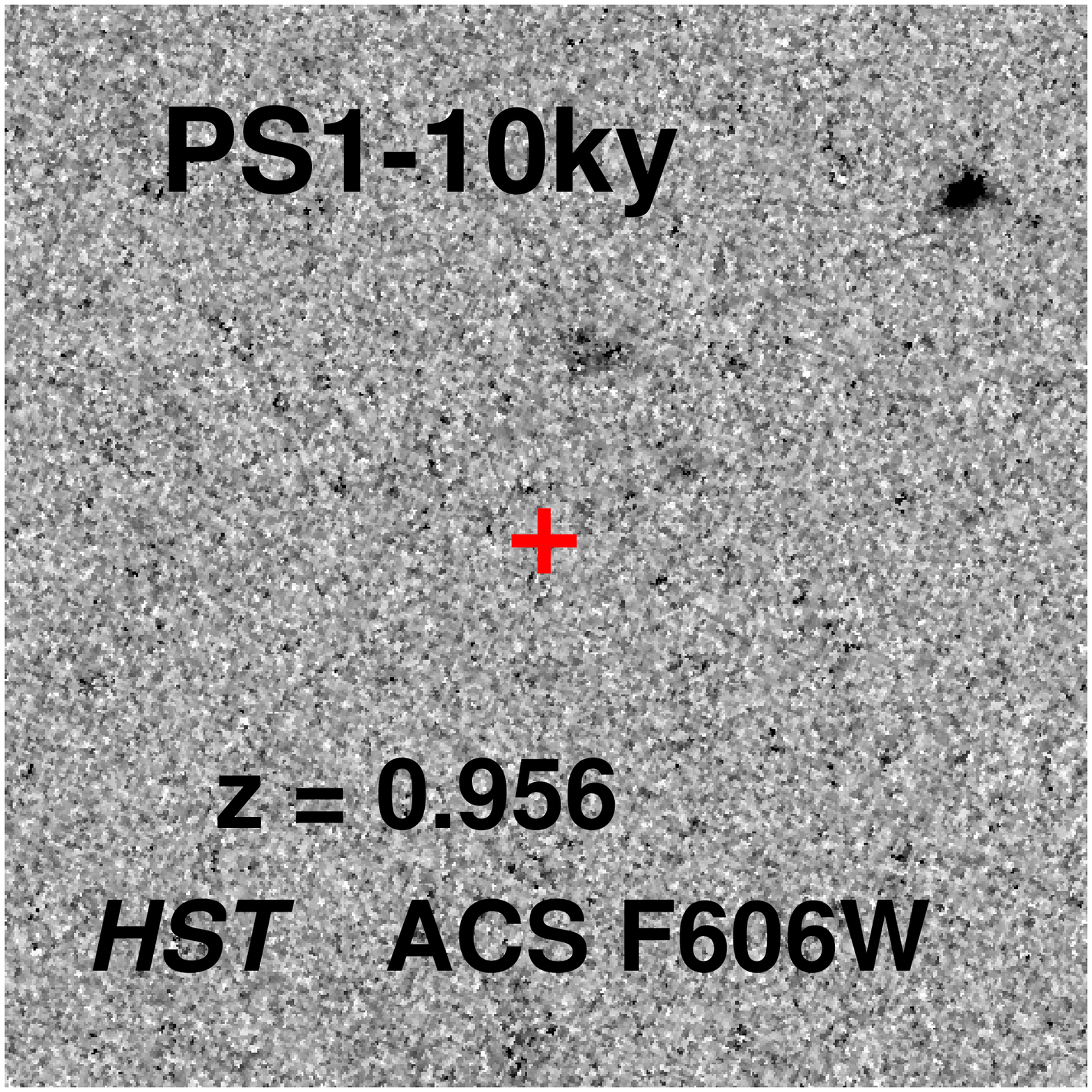} &
 \includegraphics[width=3.0cm]{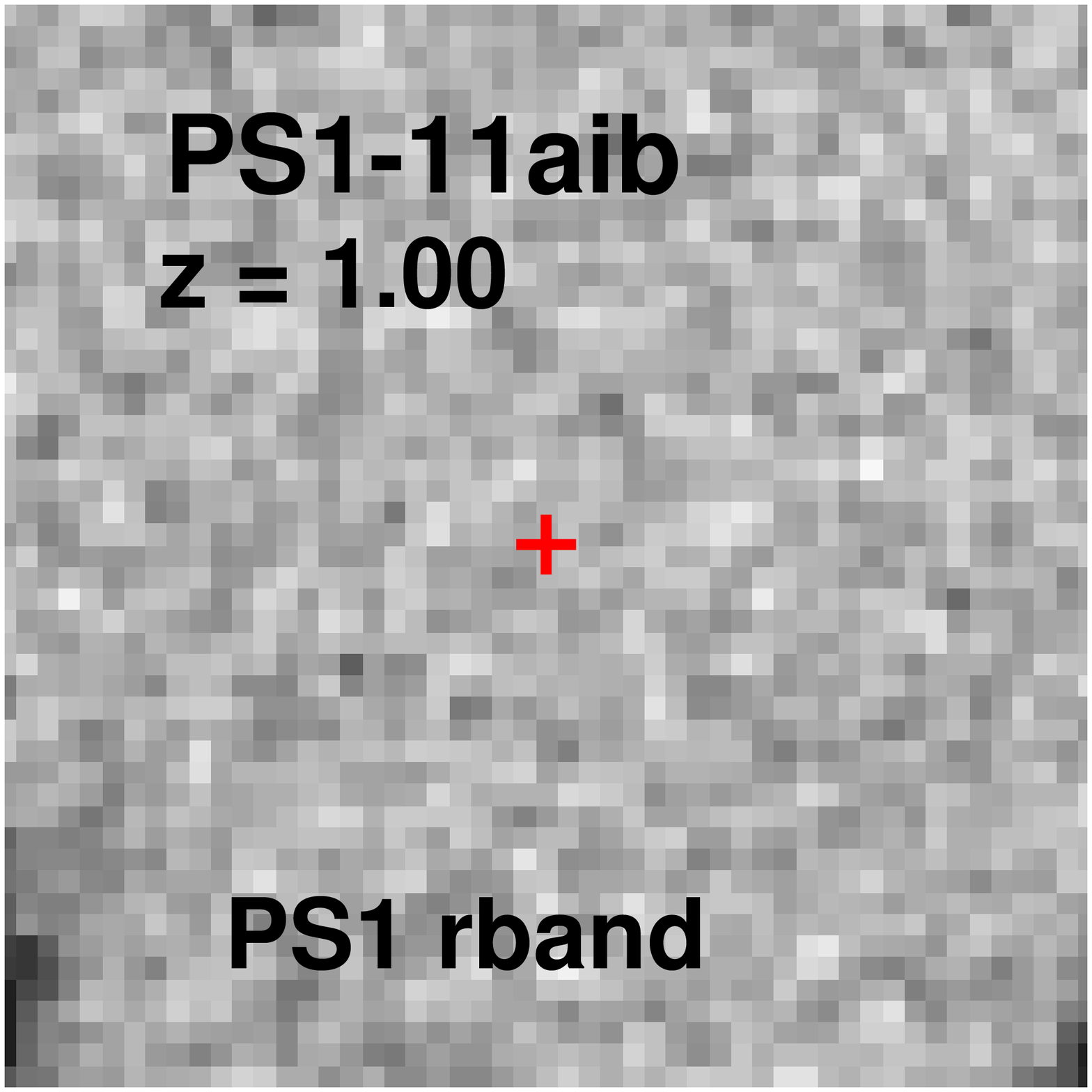} & \includegraphics[width=3.0cm]{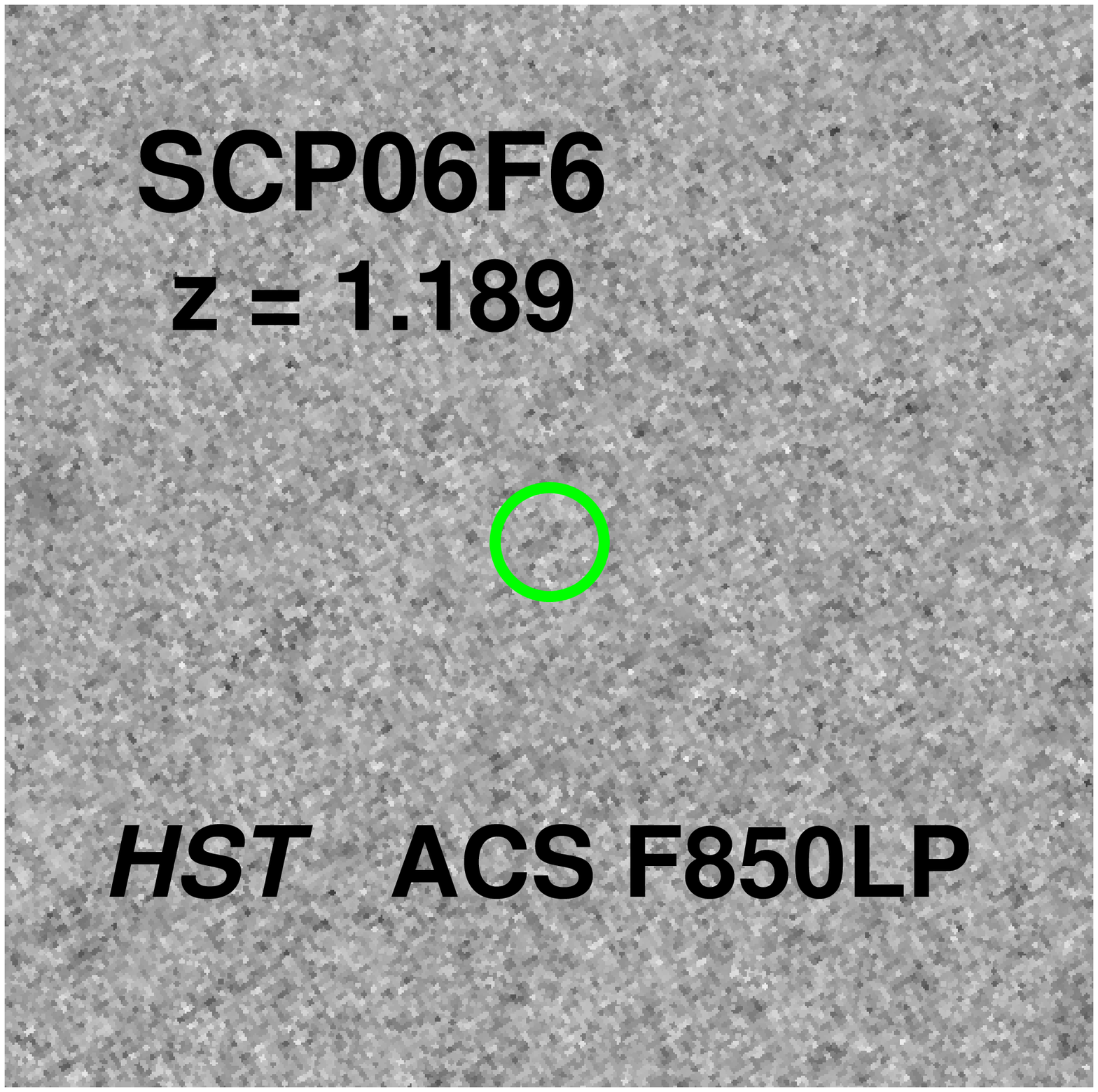} \\
 \includegraphics[width=3.0cm]{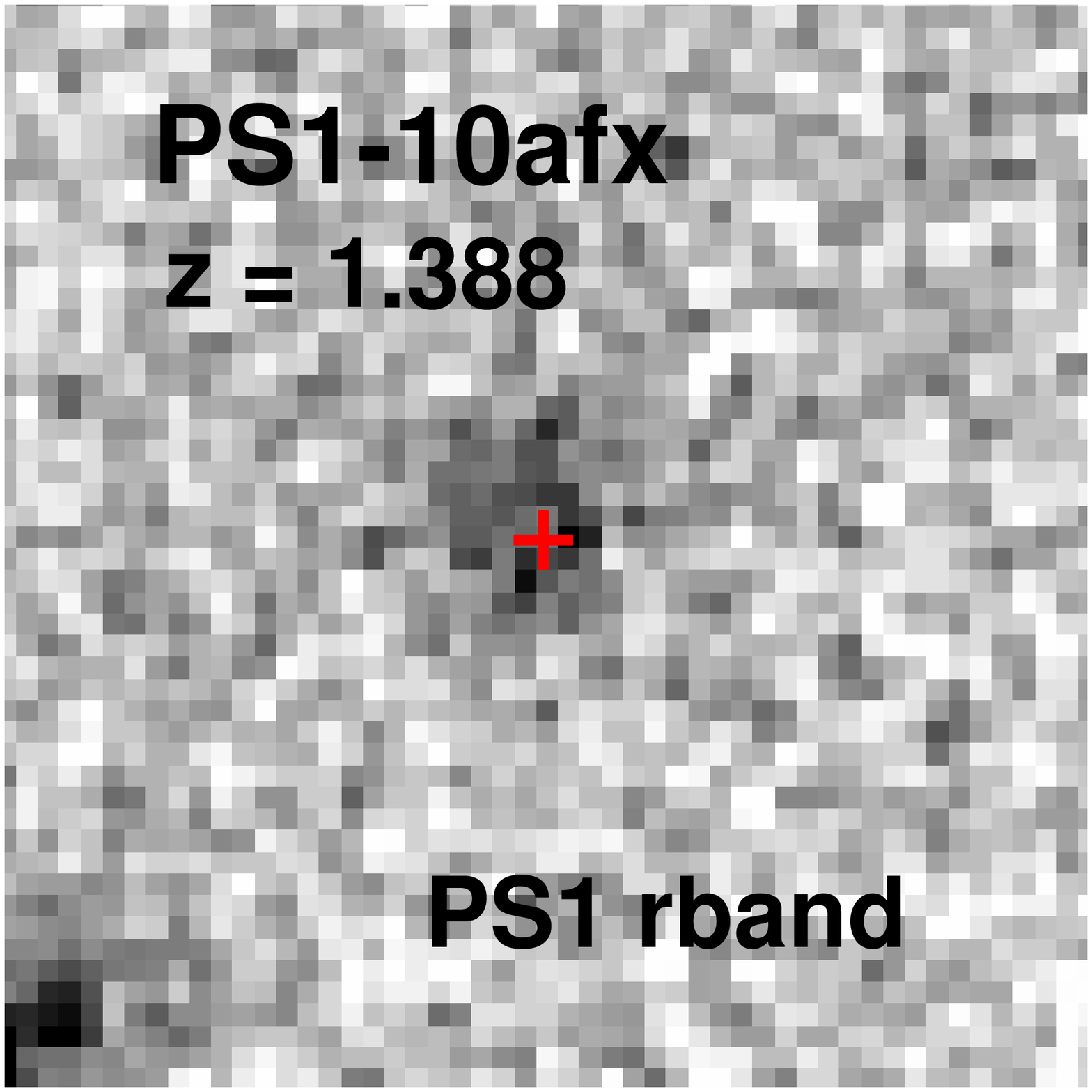} &  & & & \\
\end{tabular}
\caption{Images of 21 SLSN hosts considered in this paper (10\arcsec $\times$ 10\arcsec). All images are oriented with north corresponding to up and east to the left. The 10 remaining objects in our sample were detected in {\it HST} imaging, and are shown in Figure~\ref{fig:hstpix}. The PS1 objects have the SN position marked by red crosses, as determined by relative astrometry. For the non-PS1 objects, we mark the absolute position reported in the literature with a green circle (radius 0.5\arcsec). Six hosts remain undetected: PTF09cwl, PTF09atu, PS1-13gt, PS1-11aib, PS1-10ky and SCP06F6. The latter two remain undetected even in deep {\it HST} imaging.}
\label{fig:galpix}
\end{center}
\end{figure*}

\begin{figure*}
\begin{center}
\begin{tabular}{ccccc}
\includegraphics[width=3.0cm]{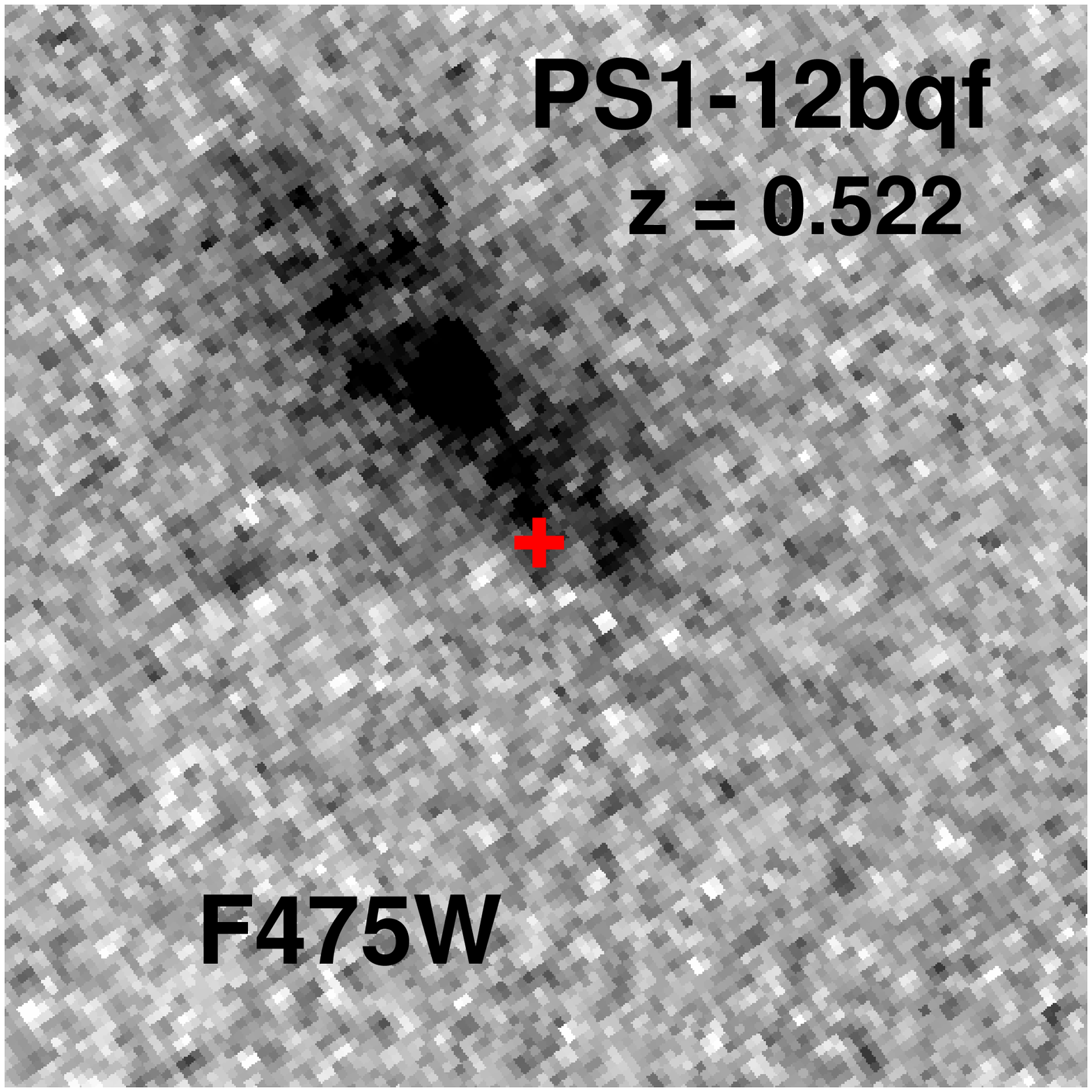} & \includegraphics[width=3.0cm]{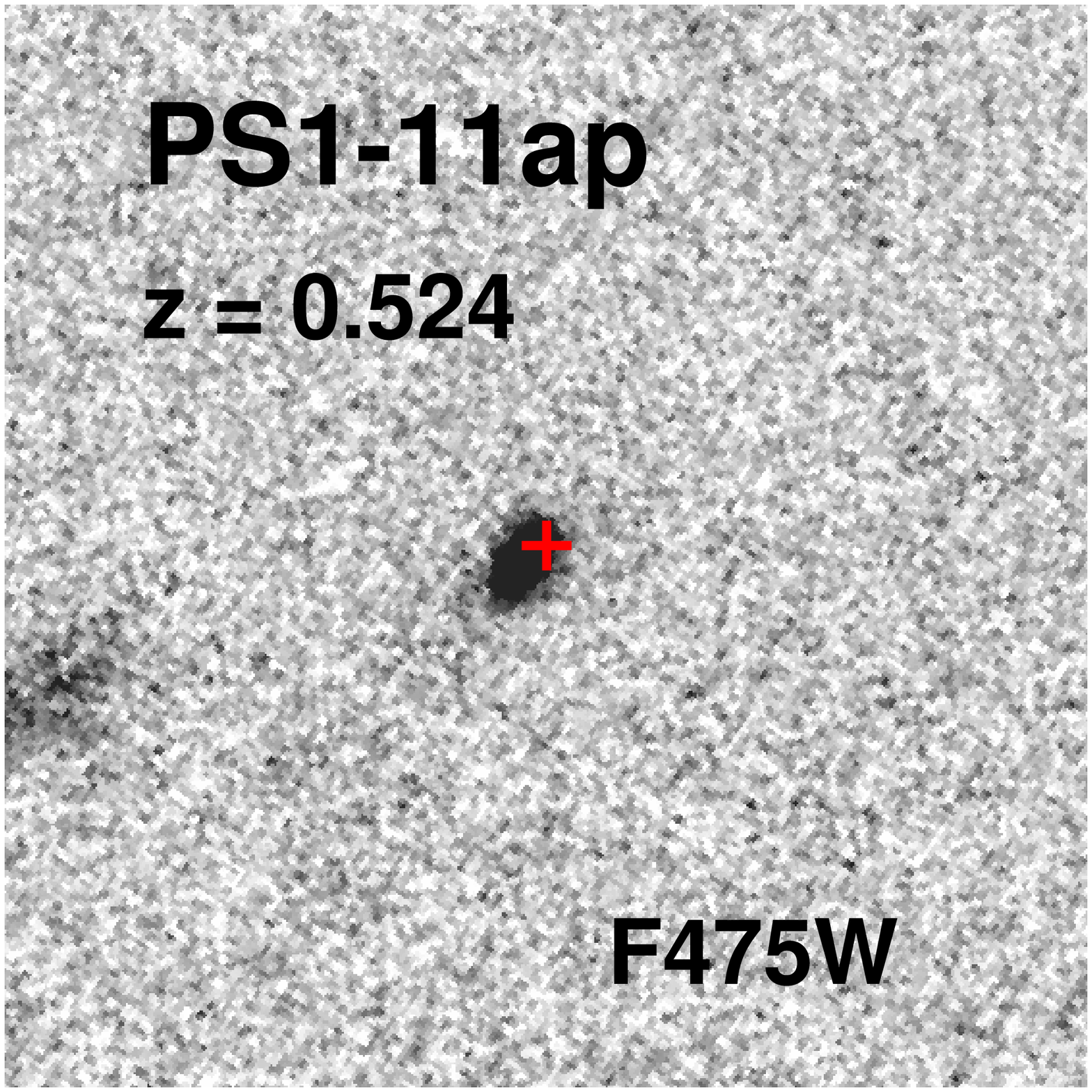} & \includegraphics[width=3.0cm]{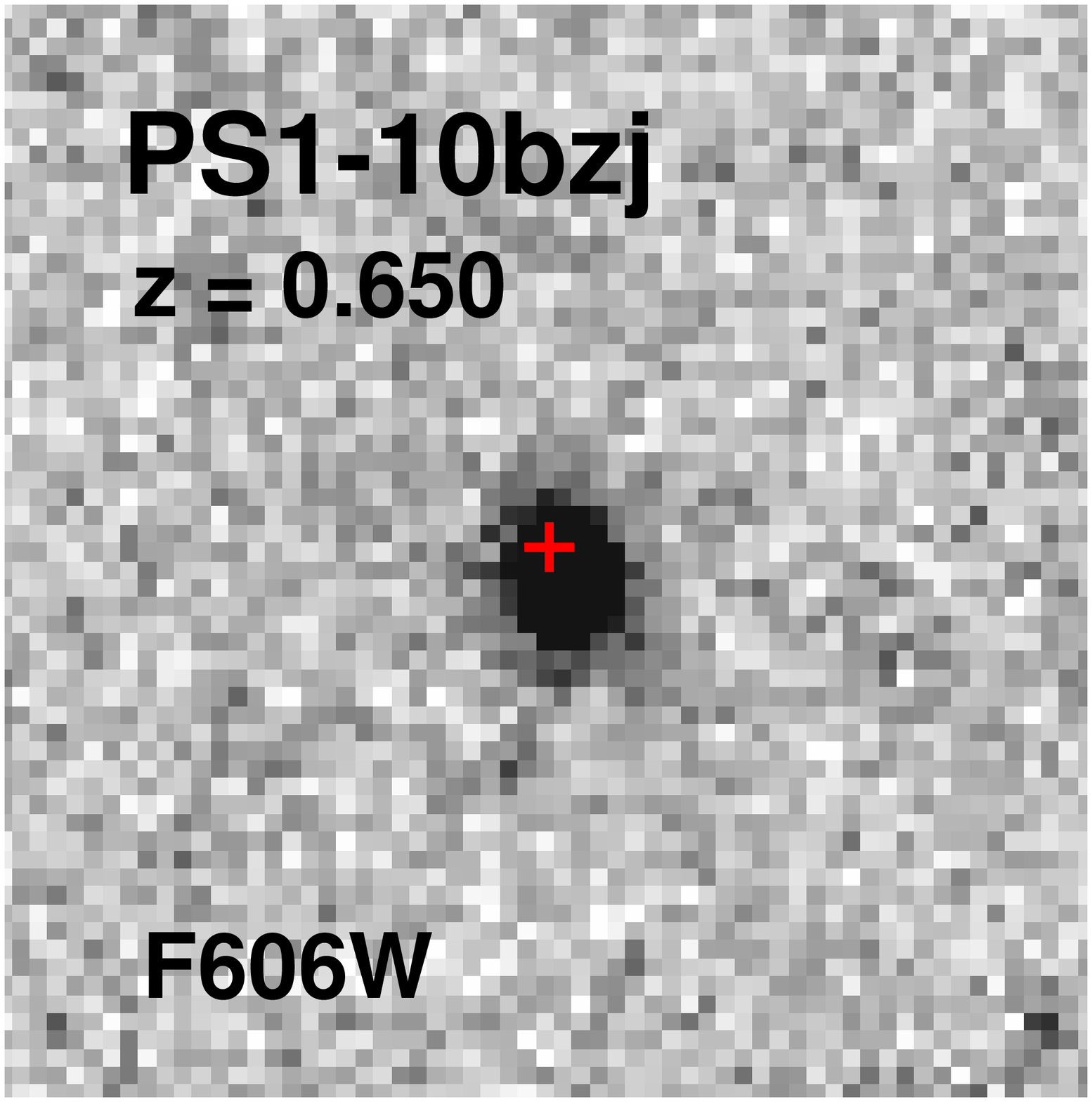}  &   \includegraphics[width=3.0cm]{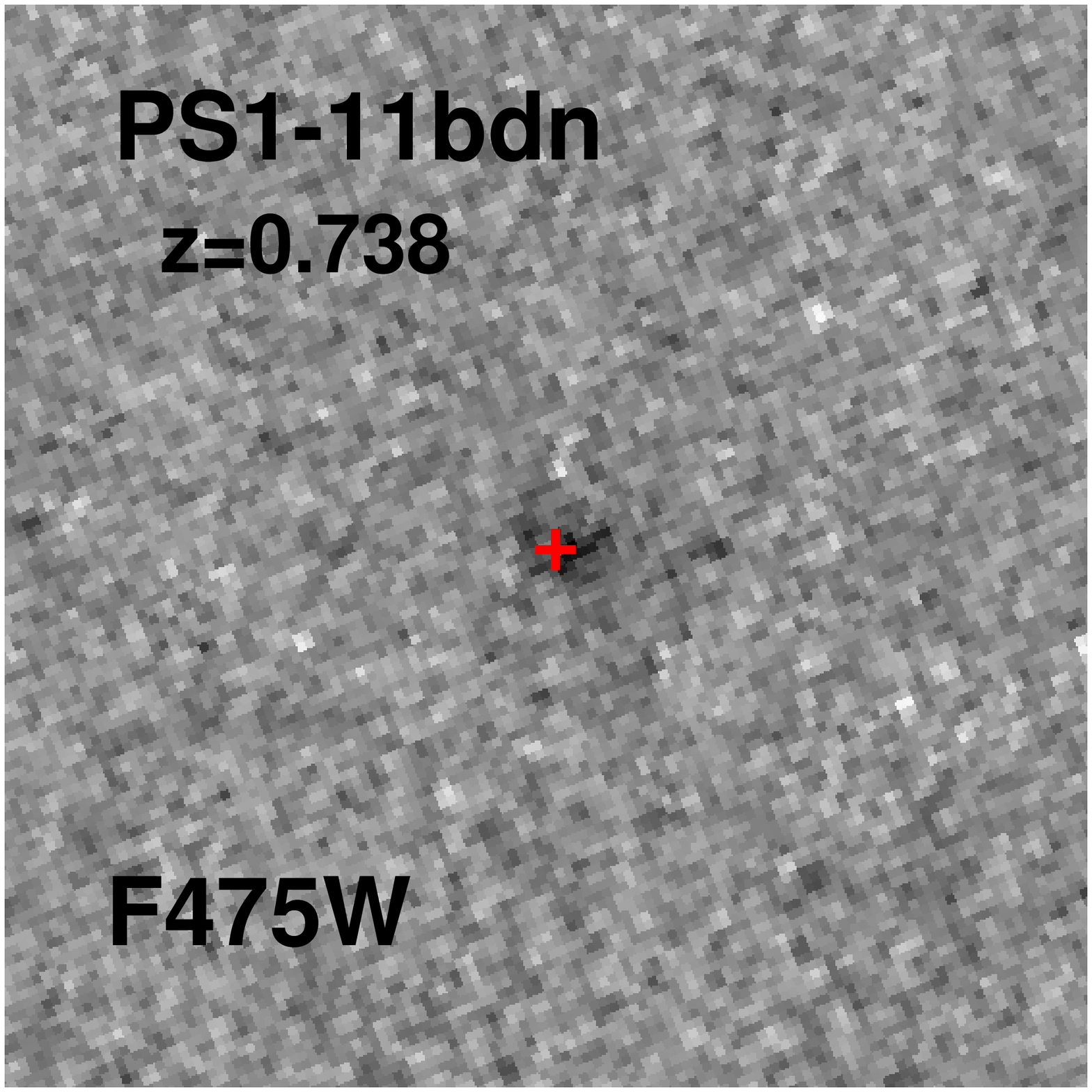}  & 
\includegraphics[width=3.0cm]{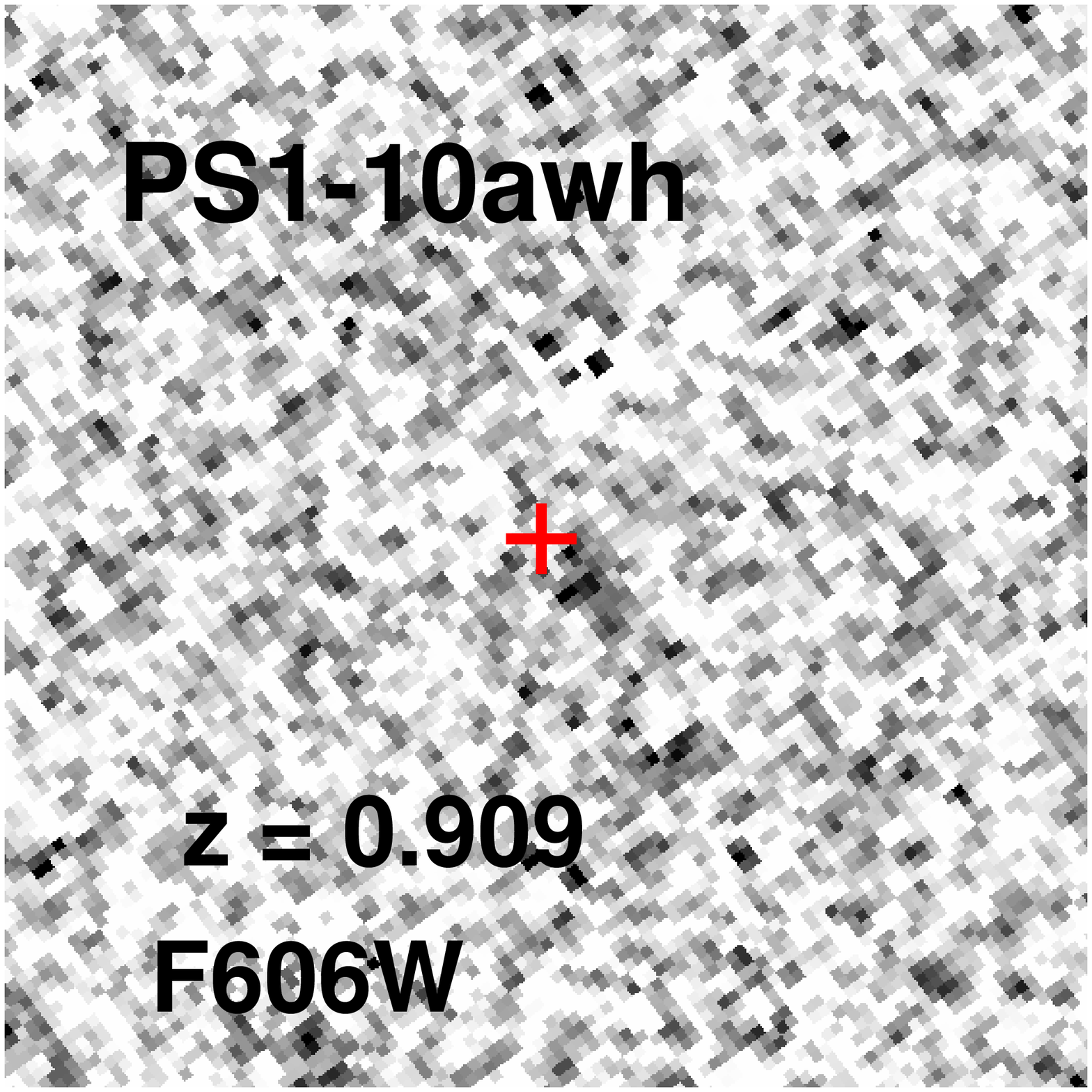}  \\ 
\includegraphics[width=3.0cm]{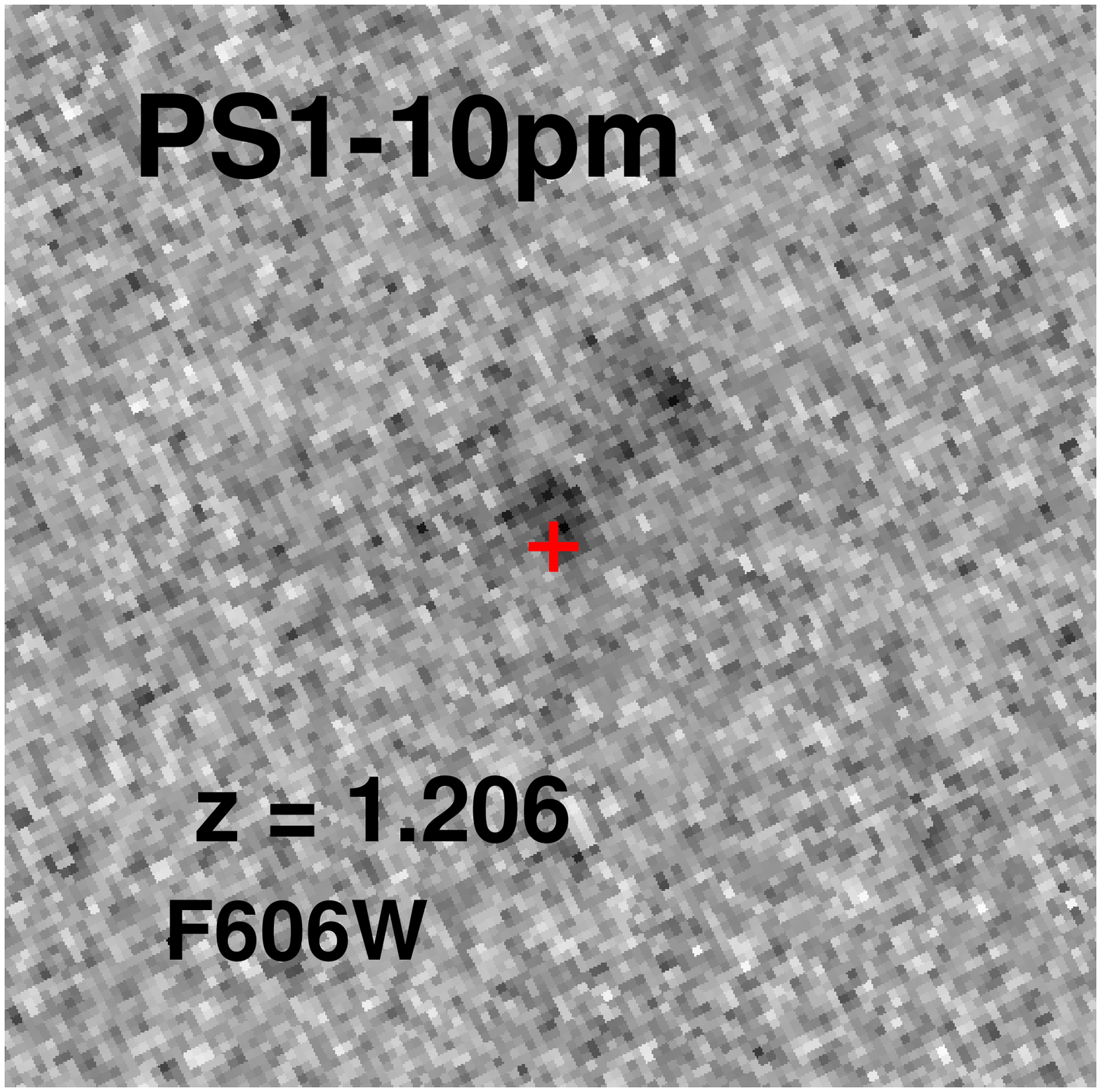}  &
 \includegraphics[width=3.0cm]{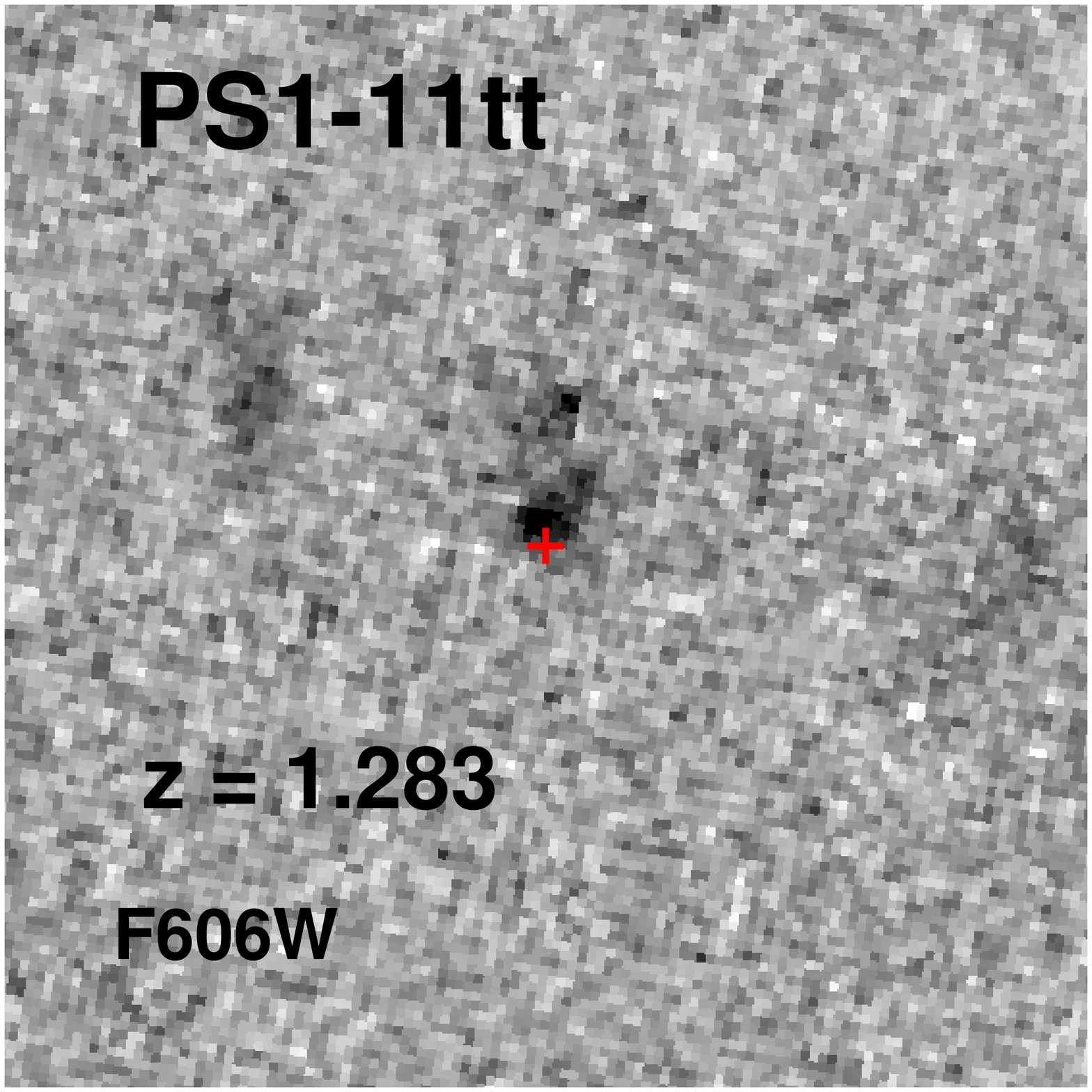} & \includegraphics[width=3.0cm]{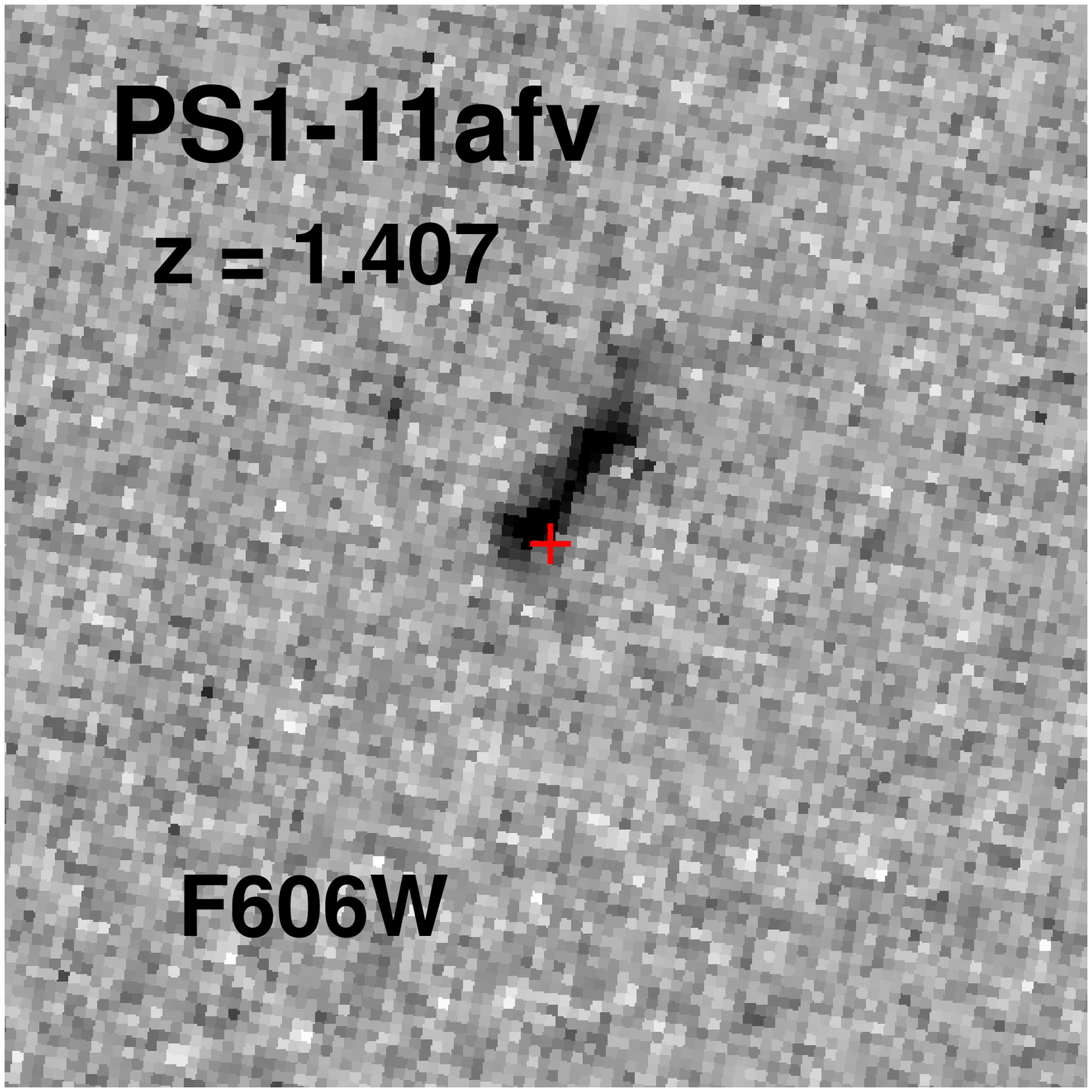}  &
\includegraphics[width=3.0cm]{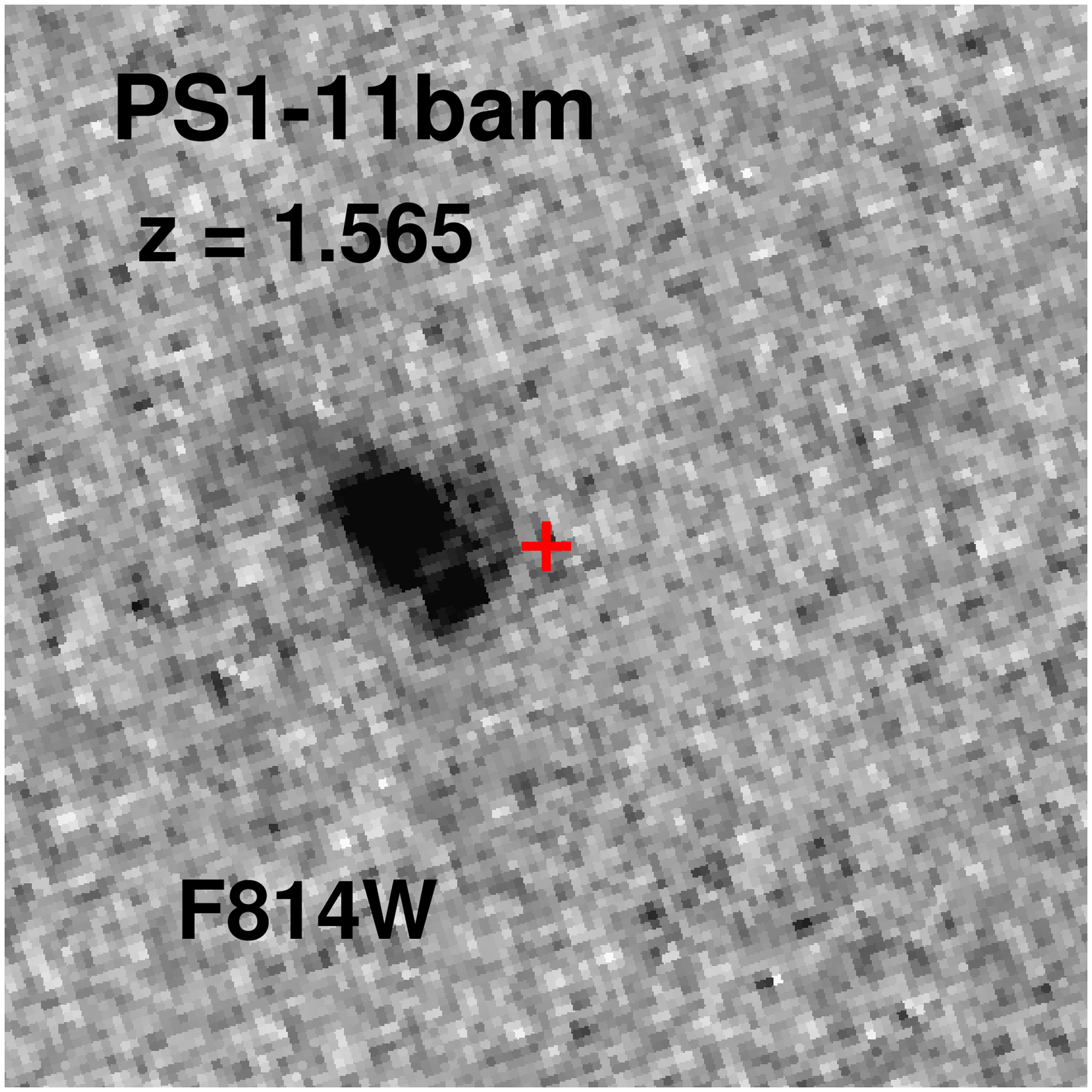} &  \includegraphics[width=3.0cm]{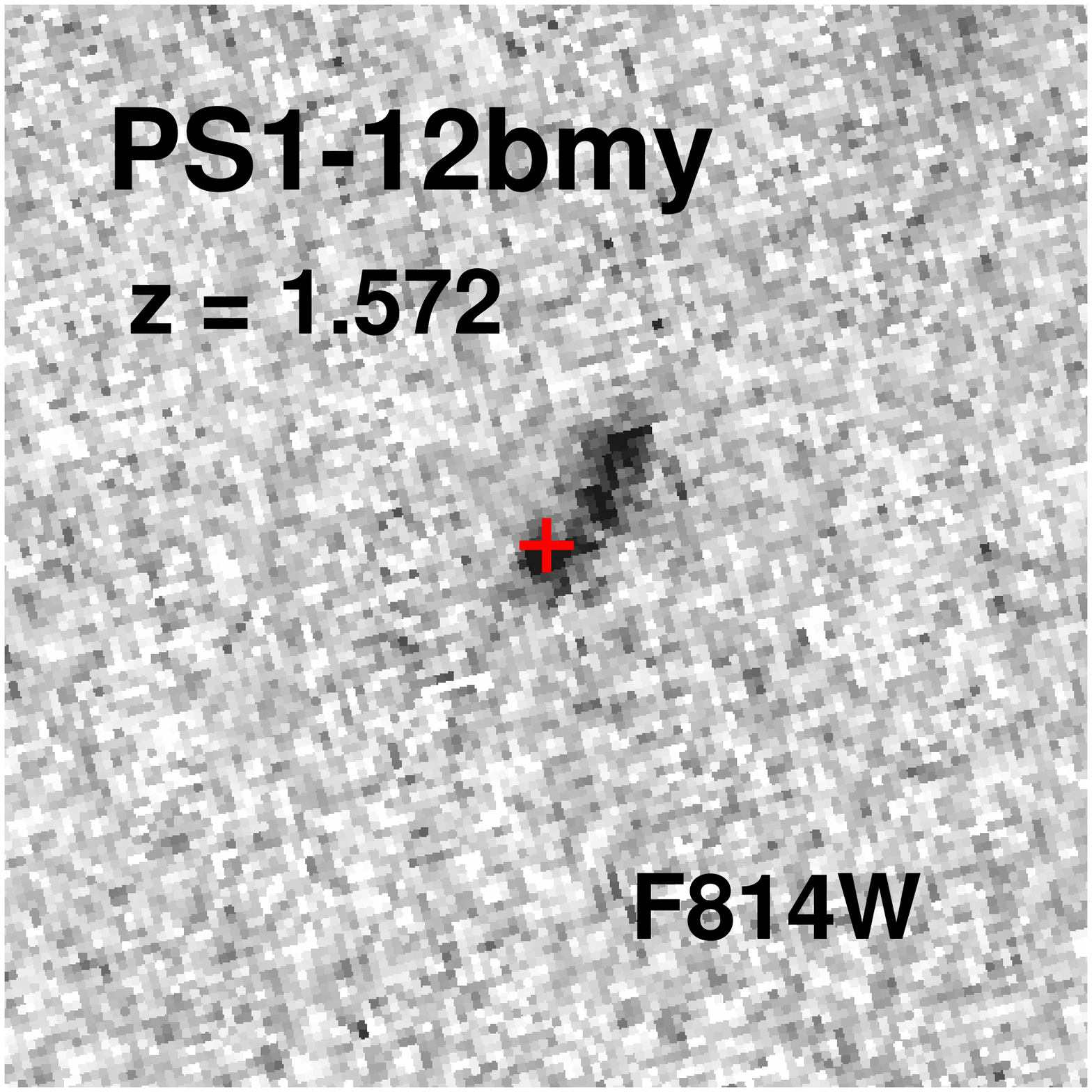} 
\end{tabular}
\caption{{\it HST} images of the remaining ten SLSN hosts in our sample (3\arcsec $\times$ 3\arcsec). The SN positions relative to the host are determined by astrometrically aligning the {\it HST} images with PS1 SN images, and shown as red crosses. }
\label{fig:hstpix}
\end{center}
\end{figure*}

\begin{figure}
\begin{center}
\includegraphics[width=3.5in]{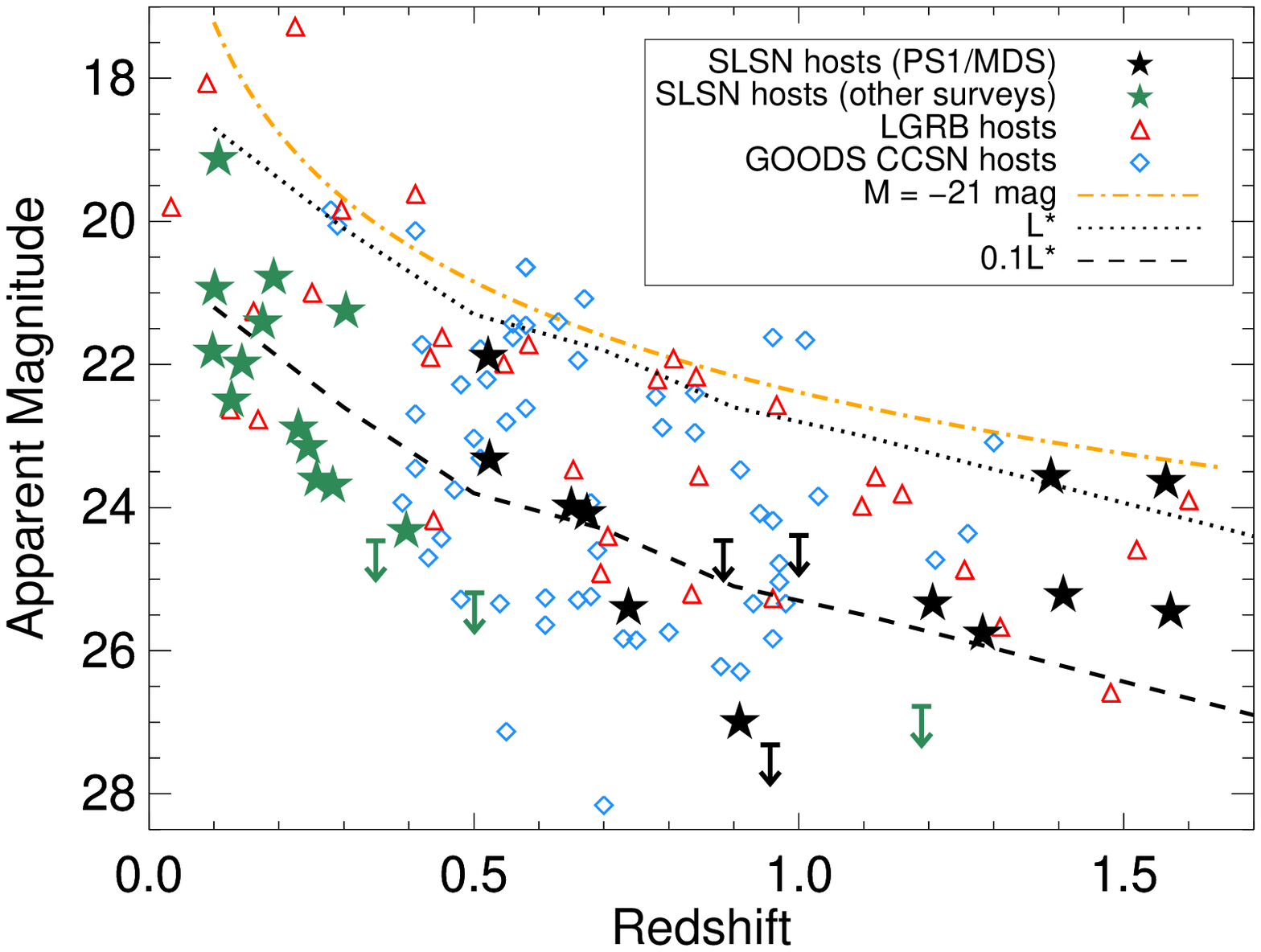}
\caption{Host apparent magnitude versus redshift for the SLSN host galaxies in our sample, with targets from PS1/MDS shown in black and other SLSN targets shown in green. Also shown are LGRB host galaxies, and core-collapse SN host galaxies from the GOODS survey, which we use as comparison samples (Section~\ref{sec:samples}). To guide the eye, the dotted and dashed black lines show tracks for $L_*$ and $0.1 L_*$. The SLSNe themselves generally peak above the dash-dotted orange line, which corresponds to an absolute magnitude $M = -21$~mag. $r$-band is plotted when available for the LGRB and SLSN host galaxies, though F606W is plotted instead for some SLSN hosts; $V$-band is plotted for the GOODS CCSN hosts.
\label{fig:rmag_z}}
\end{center}
\end{figure}

\subsubsection{Ground-Based NIR Photometry}
We obtained $J$- and $K_s$-band photometry for a subset of our targets using the FourStar Infrared Camera on the 6.5m Magellan/Baade telescope \citep{pms+13}. We used the IRAF/{\tt FSRED} package  (Andy Monson 2013, private communication) to calibrate, align, and co-add the Fourstar observations for each object and filter. We performed aperture photometry using standard packages in IRAF, using sources in common with 2MASS to determine the zeropoint.  All NIR photometry is listed in Table~\ref{tab:morephot}.

\subsubsection{{\it HST} Optical and NIR Photometry}
We obtained {\it HST} imaging of ten SLSN hosts from PS1 (programs GO-13022 and GO-13326; PI: Berger and Lunnan).  All hosts, with the exception of PS1-10ky were detected, and images corresponding to the rest-frame UV are shown in Figure~\ref{fig:hstpix}. The host of PS1-10bzj \citep{lcb+13} has serendipitous {\it HST} imaging from the GEMS survey \citep{rbb+04}, and is also shown in Figure~\ref{fig:hstpix}.

In addition to a filter covering the rest-frame UV, we imaged the hosts of PS1-11tt, PS1-11afv, PS1-10pm, PS1-10awh and PS1-10ky with a second filter covering the rest-frame optical (F850LP or F110W, depending on redshift). We processed and stacked all {\it HST} images using the AstroDrizzle software \citep{fh02, ghf+12}. As with our other photometry, we determined host galaxy fluxes using aperture photometry (Table~\ref{tab:morephot}).

\subsubsection{{\it Spitzer} Photometry}
Several PS1/MDS fields overlap with {\it Spitzer} survey coverage. Four of the lower-redshift PS1 host galaxies are detected in archival images from the \spitzer Extragalactic Representative Volume Survey (SERVS; \citealt{servs}), the COSMOS \spitzer survey (S-COSMOS; \citealt{scosmos}), or the \spitzer Extended Deep Survey (SEDS; \citealt{awf+13}). We use the catalog photometry for PS1-10bzj and PS1-12bqf (Table~\ref{tab:morephot}). The other two hosts (PS1-11ap and PS1-12zn) lack reliable catalog photometry, and so we downloaded the survey images and performed the photometry ourselves. 

At the depth of these observations, \textit{Spitzer} images are confusion-limited for faint sources. As a result, in several cases the region around the host galaxy is contaminated by light from nearby stars or galaxies. Prior to performing photometry, we used the \textsc{galfit} software package \citep{phi+02} to model and subtract these neighboring sources using the procedure described in \citet{lbc11}. We used a 3\arcsec~ aperture and a 3--7\arcsec~ background annulus and determined aperture corrections using the PSFs derived from the mosaics. We include the contribution of correlated noise from the mosaicking process in our estimate of the uncertainty on the derived fluxes following \citet{lbc11}.

Other PS1/MDS hosts also lie within the survey footprints, but were not detected. We find that the upper limits are too shallow to constrain the host spectral energy distributions (SED), so we do not consider them here. We also searched the {\it Spitzer} archive for observations of non-PS1 host galaxies; SN\,2005ap and SCP06F6 lie in areas of  {\it Spitzer} coverage, but the limits are not constraining.

\subsection{Astrometry}
To establish an absolute astrometry scale on the MMTCam and Magellan images, we download catalog images of the field (SDSS, PS1 or DSS) and use the IRAF routine {\tt ccmap} to align the images after identifying common point sources. For the non-PS1 objects, we do not have SN images available to precisely determine the location of the SN relative to its host galaxy, but we mark the absolute reported literature positions in Figure~\ref{fig:galpix}.

For the PS1/MDS objects we also have SN images that can be used to perform relative astrometry. We use these to determine the SN position relative to the host galaxies in the non-PS1 images, again by identifying common point sources in the two images and aligning them using the IRAF package {\tt ccmap}. Depending on the source density and depth of the PS1 image, the number of overlap sources varies from $\sim 10-100$, with a resulting uncertainty of the astrometric tie of $\approx 20-80~{\rm mas}$. The positions of the PS1/MDS SLSNe relative to their hosts are marked in Figures~\ref{fig:galpix} and \ref{fig:hstpix}.

\subsection{Spectroscopy}
We obtained spectra of 12 host galaxies at $z \lesssim 0.7$, using LDSS3, IMACS and BlueChannel. Beyond this redshift, our targets are generally too faint for spectroscopy, or at too high redshift to measure [\ion{O}{3}]$\lambda$ 5007, which is required for a metallicity determination (Figure~\ref{fig:rmag_z}). Table~\ref{tab:spec} summarizes the spectroscopic observations and observing setups. All spectra were taken at parallactic angle unless otherwise noted. Continuum and arc lamp exposures were obtained after each observation to provide a flatfield and wavelength calibration. Basic two-dimensional image processing tasks were performed using standard tasks in IRAF.  Observations of spectrophotometric standard stars were obtained on the same night, and we used our own IDL routines to apply a flux calibration and correct for telluric absorption bands.

Absolute flux calibration (to account for slit losses and/or non-photometric conditions) was determined by performing synthetic photometry on the observed spectra and applying an overall scaling factor to match the galaxy broadband photometry. We find that generally factors derived from different filters agree well, indicating that the standard star calibration reliably recovers the shape of the spectrum. In spectra where the continuum is not well detected, we do not make this correction, and the overall calibration is derived from the spectrophotometric standard stars only. The relative line fluxes and quantities that are derived by ratios (i.e., extinction and metallicity) are reliable, but the overall scaling and in particular a line flux derived star formation rate may be marginally affected. 

In cases where the galaxy continuum is well detected, we construct a stellar model spectrum using the FAST stellar population synthesis code \citep{kvl+09} by fitting the observed spectrum (with strong emission lines masked). We then subtract the model to correct for underlying stellar absorption in the Balmer lines before measuring line fluxes. In practice, we find that this correction only makes a significant difference ($\gtrsim 10\%$ correction in the H$\beta$ flux) for a few objects in our sample. Since the objects with weak continuum emission also exhibit the highest equvalent width (EW) emission lines, the correction for these objects is marginal.

We measure emission line fluxes by fitting Gaussian profiles, and list the results in Table~\ref{tab:lines}. In two objects, SN\,2006oz and PTF10hgi, low-precision redshifts were previously only known from cross-correlating SN features, but we now detect galaxy emission lines from both hosts and adjust the redshifts to $z=0.396$ for SN\,2006oz, and $z=0.098$ for PTF10hgi. These redshifts are consistent with the inferred SN redshifts ($0.376 \pm 0.014$ and $0.100 \pm 0.014$, respectively; \citealt{lcd+12,isj+13}).

All spectra are shown in Figure~\ref{fig:spec}. In addition to our spectra, 7 more objects have spectra with emission lines available in the literature: SN\,2010gx \citep{csb+13}, SN\,2007bi \citep{ysv+10}, PS1-11bam \citep{bcl+12}, PS1-10afx \citep{cbr+13}, PS1-10bzj \citep{lcb+13}, PTF09cnd \citep{qkk+11} and PS1-11ap \citep{msk+14}. Combining the spectra presented here with the literature data leads to line measurements for 19 hosts.

\begin{figure*}
\begin{center}
\includegraphics[width=7in]{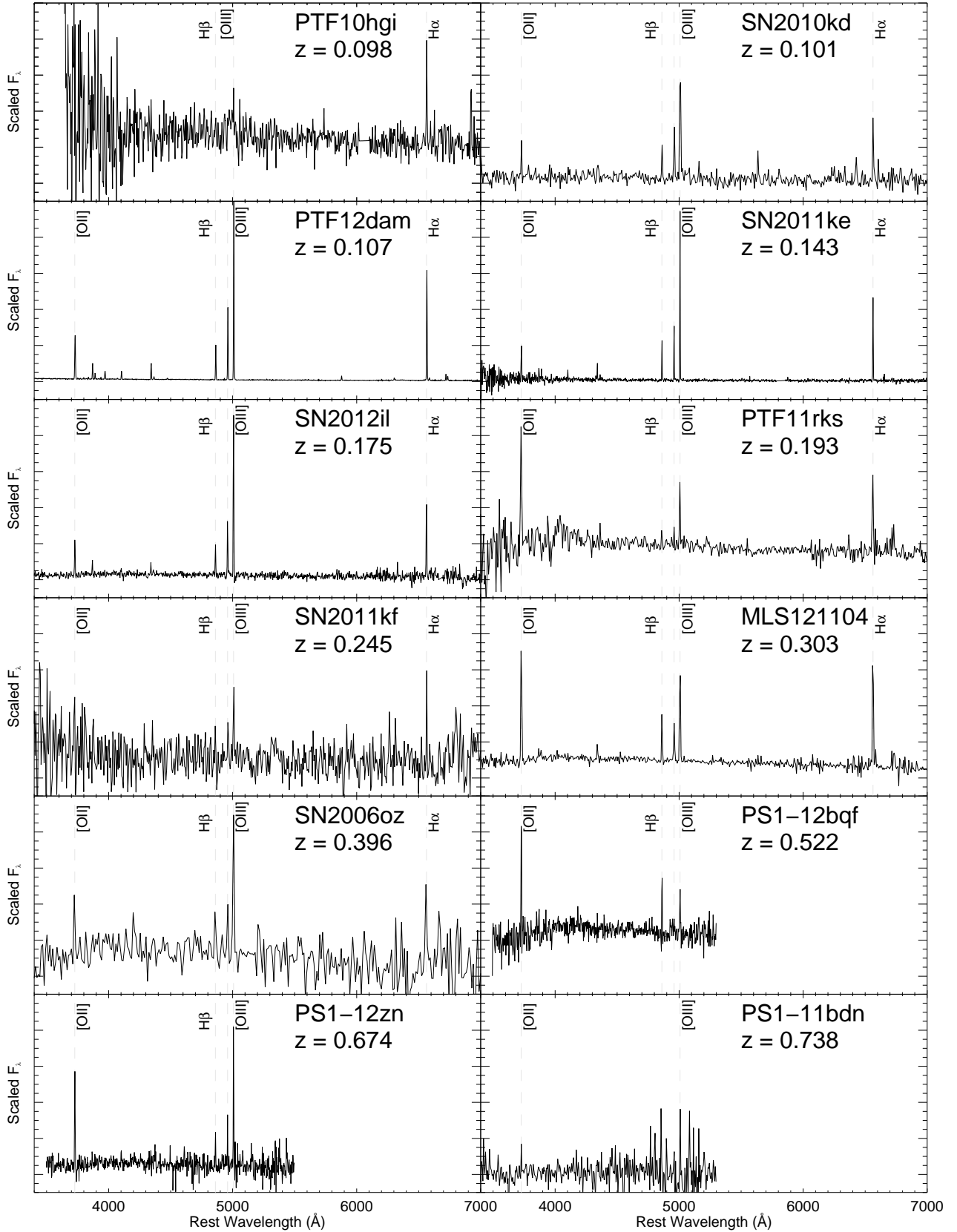}
\caption{Spectra of SLSN host galaxies at $z \lesssim 0.75$ (Table~\ref{tab:spec}). The main emission lines used for analyzing galaxy properties are marked. In addition to the 12 spectra shown here, an additional 7 hosts have emission line measurements available in the literature, providing spectroscopic information for more than half of our sample.
\label{fig:spec} }
\end{center}
\end{figure*}

\section{Comparison Samples}
\label{sec:samples}

We compare the SLSN host galaxies to galaxies hosting two other types of transients: LGRBs and core-collapse supernovae (CCSNe). For LGRB host galaxies, we use the sample from \citet{slt+10}, who provide luminosities, stellar masses and star formation rates of 34 hosts at $z \lesssim 1.2$ based on photometry reported in \citet{sgl09} and \citet{fls+06}. We supplement these data with spectroscopy from \citet{sgl09}, \citet{lbk+10,lkb+10} and \citet{gf13}, and also include any LGRB host galaxies at  $z \lesssim 1.7$ that are analyzed in these papers but which are not part of the sample in \citet{slt+10}. This leads to a sample of 44 LGRB hosts in the same redshift range as the SLSNe, of which 17 hosts also have metallicity measurements.

For core-collapse SN hosts, we use the GOODS sample \citep{fls+06,slt+10}. As GOODS was primarily searching for Type Ia SNe, only a subset of the SNe in this sample were spectroscopically typed, with the rest classified as core-collapse based on light curve properties \citep{srd+04} and so subtypes are not available. Still, this sample has two key advantages over local supernova hosts for our purposes: it is an untargeted sample, and it covers a similar redshift range as the SLSN hosts, thus minimizing effects due to galaxy redshift evolution. The GOODS sample includes luminosities, stellar masses and star formation rates derived from SED fits, but does not include metallicities.

In Figure~\ref{fig:zdist} we show the redshift distributions of the three samples, including separately the subsamples for which we have metallicity measurements, as well as the SLSNe from PS1/MDS . The redshift distributions are similar both for the full samples and the spectroscopic subsamples, thereby minimizing any potential galaxy evolution effects.

The SLSN host galaxy data contain both detections and upper limits. To include the information from the non-detections, we use techniques from survival analysis, as implemented in the \textsc{ASURV} statistics package \citep{asurv}. To estimate and display the distribution function of each quantity, we use the Kaplan-Meier estimator. This is a non-parametric estimator of the cumulative distribution function, where the weight of each upper limit is distributed uniformly among the detections at lower values. If there are no upper limits, the Kaplan-Meier estimator reduces to the usual empirical distribution function. For each detected value $x_i$ in the sample, $N_i$ is the number of objects (detected or undetected) with  $\geq x_i$, and $d_i$ is the number of objects at $x_i$. The Kaplan-Meier estimator is then given by

\begin{equation}
\widehat{S}_{KM} (x_i) = \prod_{x \geq x_i} \left(1 - \frac{d_i}{N_i}\right) .
\label{eq:km}
\end{equation}

In addition, the presence of upper limits means that common statistical tests for two-sample comparisons (e.g., the Kolmogorov-Smirnov test) cannot be applied. To test the null hypothesis that  two samples are drawn from the same underlying distribution, we instead use a generalized Wilcoxon rank-sum test (Peto-Prentice test), also available as part of the \textsc{ASURV} statistics package. The $p$-values we report are the probabilities for obtaining the calculated value of the test statistic, given the null hypothesis that the two samples are drawn from the same distribution. A $3\sigma$ significant difference thus corresponds to $p < 0.003$, whereas a $2\sigma$ significant difference corresponds to $p < 0.05$.

For two-sample comparisons where we do not have upper limits in the data, we use a Kolmogorov-Smirnov (K-S) test. The K-S test statistic $D$ is defined as $D = \sup_x | F_1(x) - F_2(x)| $, where $F_1$ and $F_2$ are the empirical distribution functions of the two samples. By comparing $D$ to the K-S distribution, we can calculate the probability ($p$-value) of obtaining a value $D$ under the null hypothesis that the two samples are drawn from the same underlying distribution.

\begin{figure}
\begin{center}
\includegraphics[width=3.5in]{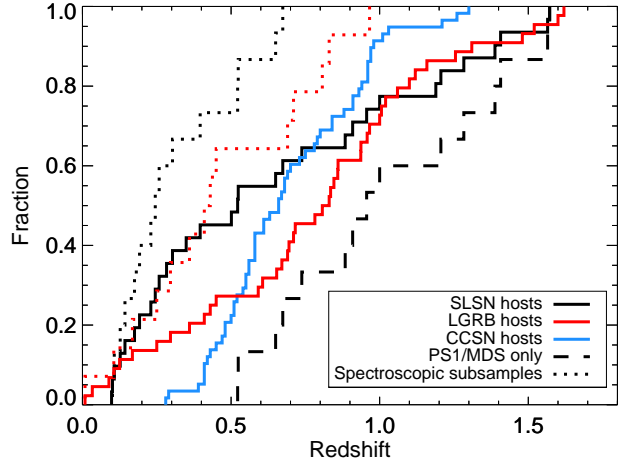}
\caption{Redshift distribution of the SLSN host sample (black) and the comparison samples: LGRB host galaxies (red) and the GOODS core-collapse SNe (blue). The dotted lines indicate the subsamples with metallicity measurements for the SLSN and LGRB host galaxies. The dashed black line shows the SLSNe from PS1/MDS only. 
\label{fig:zdist}}
\end{center}
\end{figure}

\section{Physical Properties of SLSN Host Galaxies}
\label{sec:galprop}

For the SLSN hosts with multi-band photometry, we construct galaxy models with the FAST stellar population synthesis code \citep{kvl+09}. FAST takes as inputs a choice of stellar population synthesis (SPS) models, IMF, reddening law, and a grid of stellar population properties (age, star formation timescale, dust content, metallicity and redshift), and computes model fluxes for each point in the grid. The best-fit parameters are determined by computing the $\chi^2$ at each point in the grid and finding the minimum. The confidence intervals are calibrated using Monte Carlo simulations, taking into account the uncertainties both in the observed fluxes and in the models \citep{bvc08}.

We fit the SLSN host galaxies using the \citet{mar05} stellar library, and assuming an exponential star formation history and a Salpeter IMF. We assume a metallicity of $Z = 0.5 Z_{\odot}$, unless the metallicity measured from spectroscopy (Section~\ref{sec:metal}) requires a library with $Z = 0.05 Z_{\odot}$ or $Z = 1 Z_{\odot}$. If we have a measurement of the extinction from spectroscopy (Section~\ref{sec:ext}), $A_V$ is restricted to that range but is otherwise allowed to vary freely.  In cases where our galaxy spectra show strong emission lines, the filter containing [\ion{O}{3}]$\lambda$5007 is typically excluded from the fit. The resulting best-fit galaxy SEDs are shown in Figure~\ref{fig:seds}.

\begin{figure*}
\begin{center}
\begin{tabular}{cccc}
\includegraphics[width=4cm]{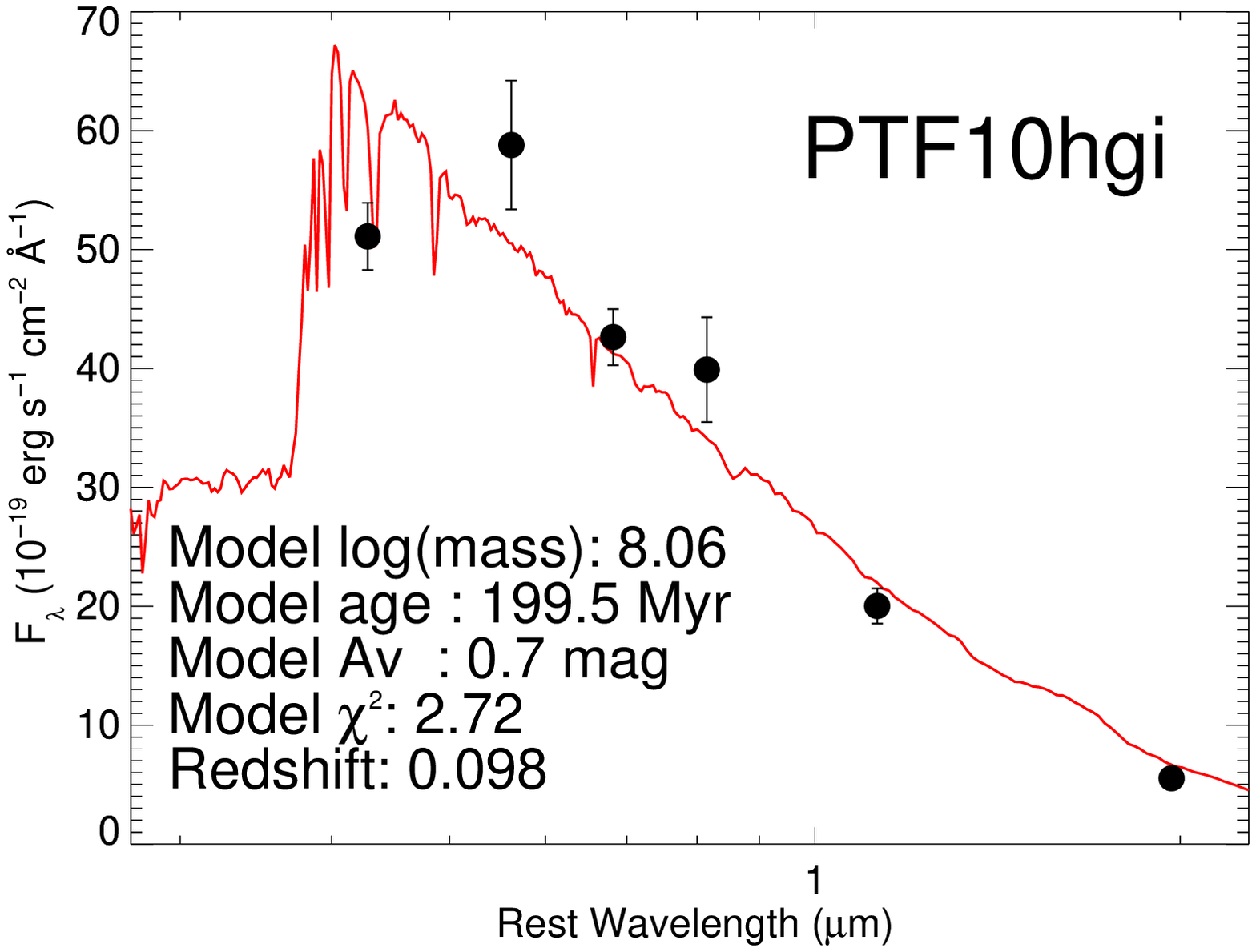} & \includegraphics[width=4cm]{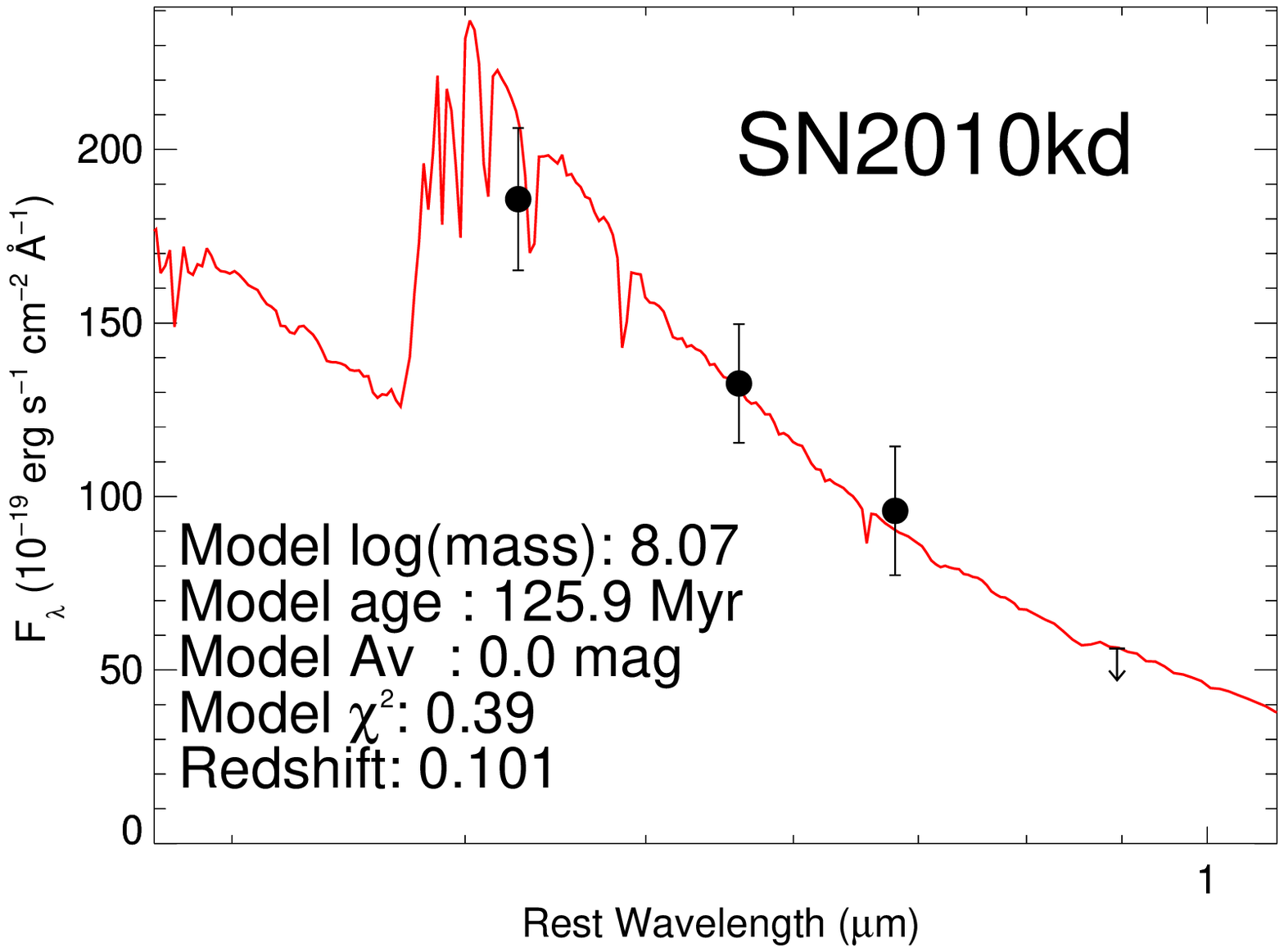} & \includegraphics[width=4cm]{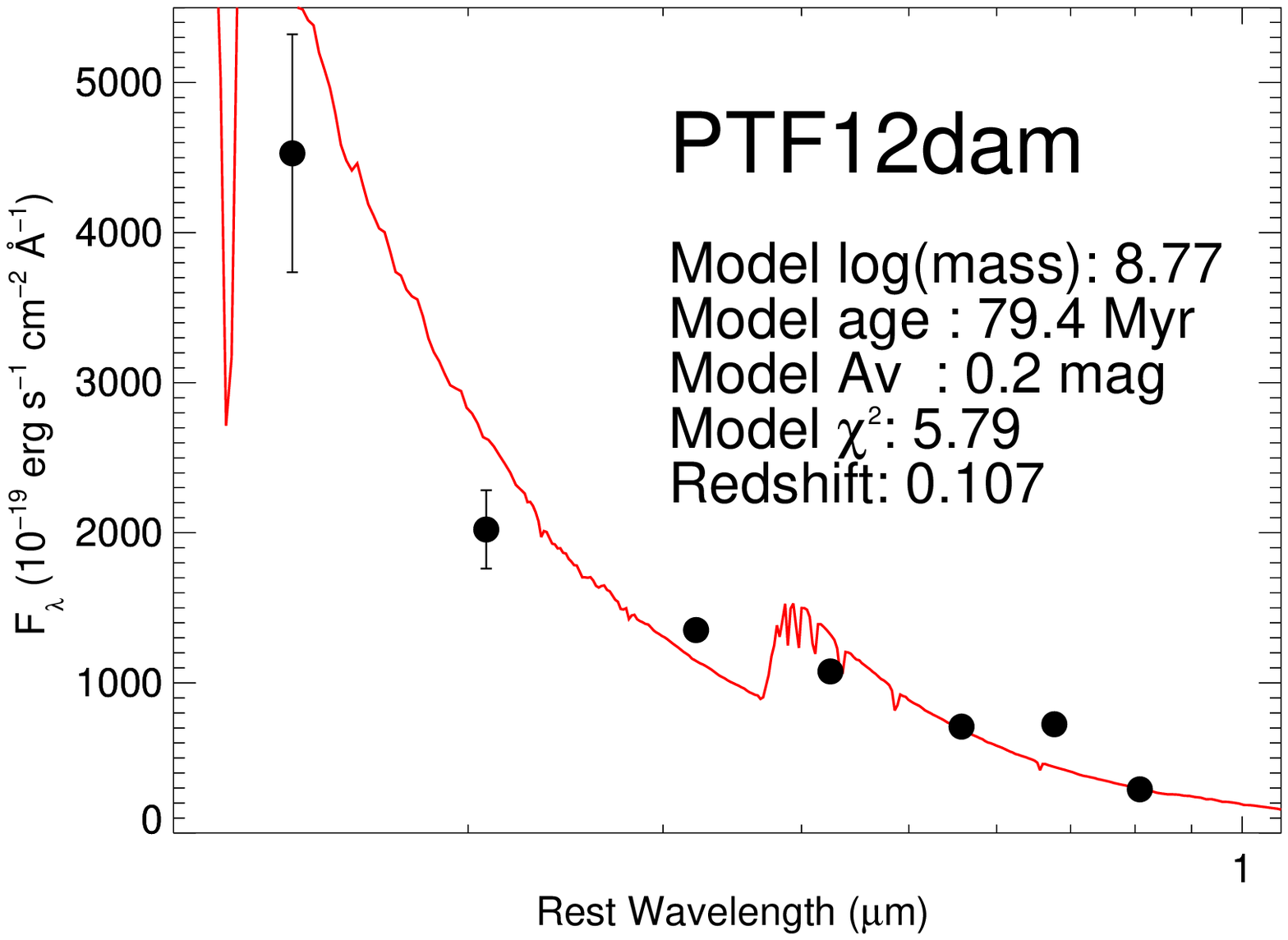}  & \includegraphics[width=4cm]{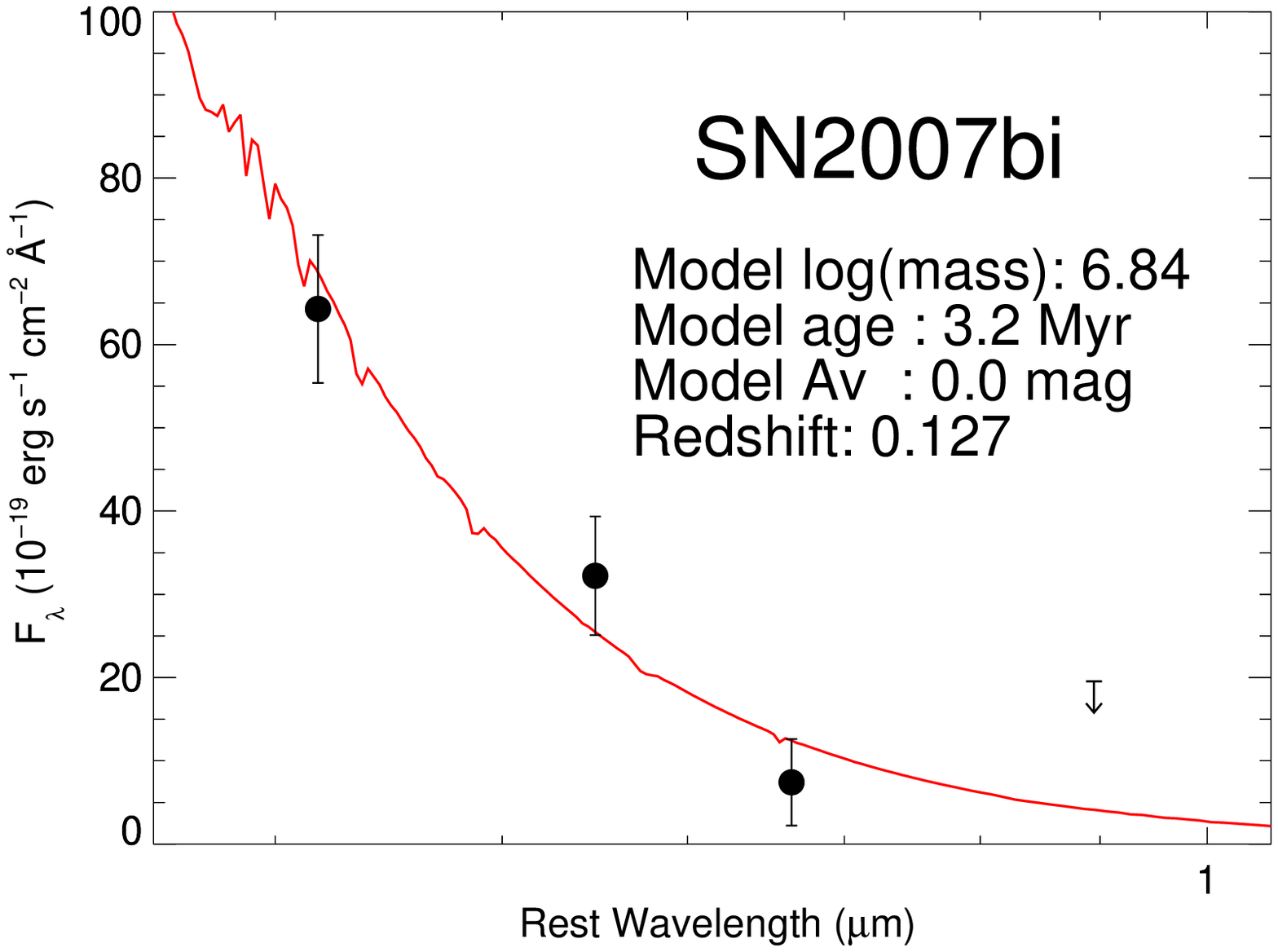} \\
\includegraphics[width=4cm]{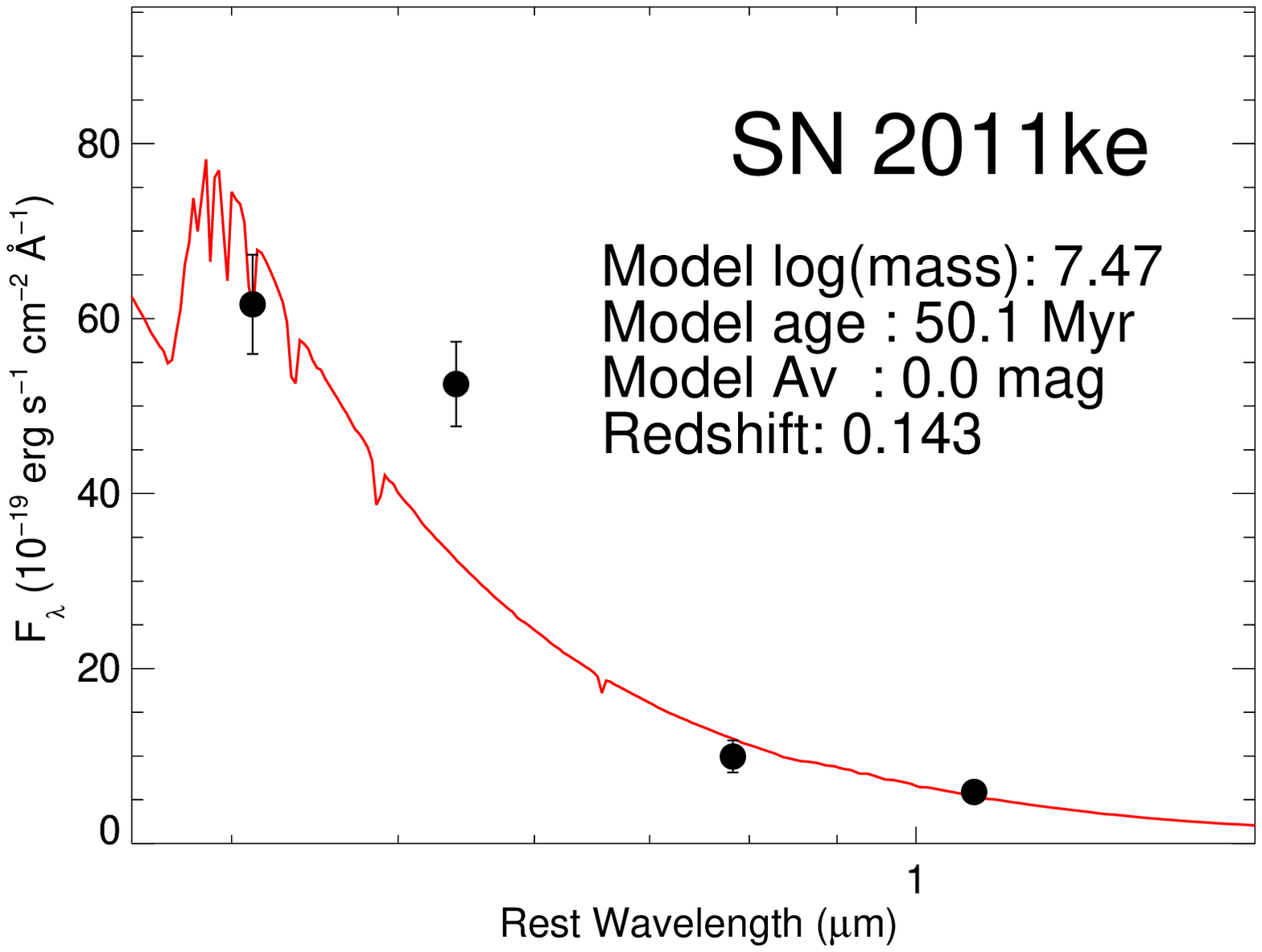} & \includegraphics[width=4cm]{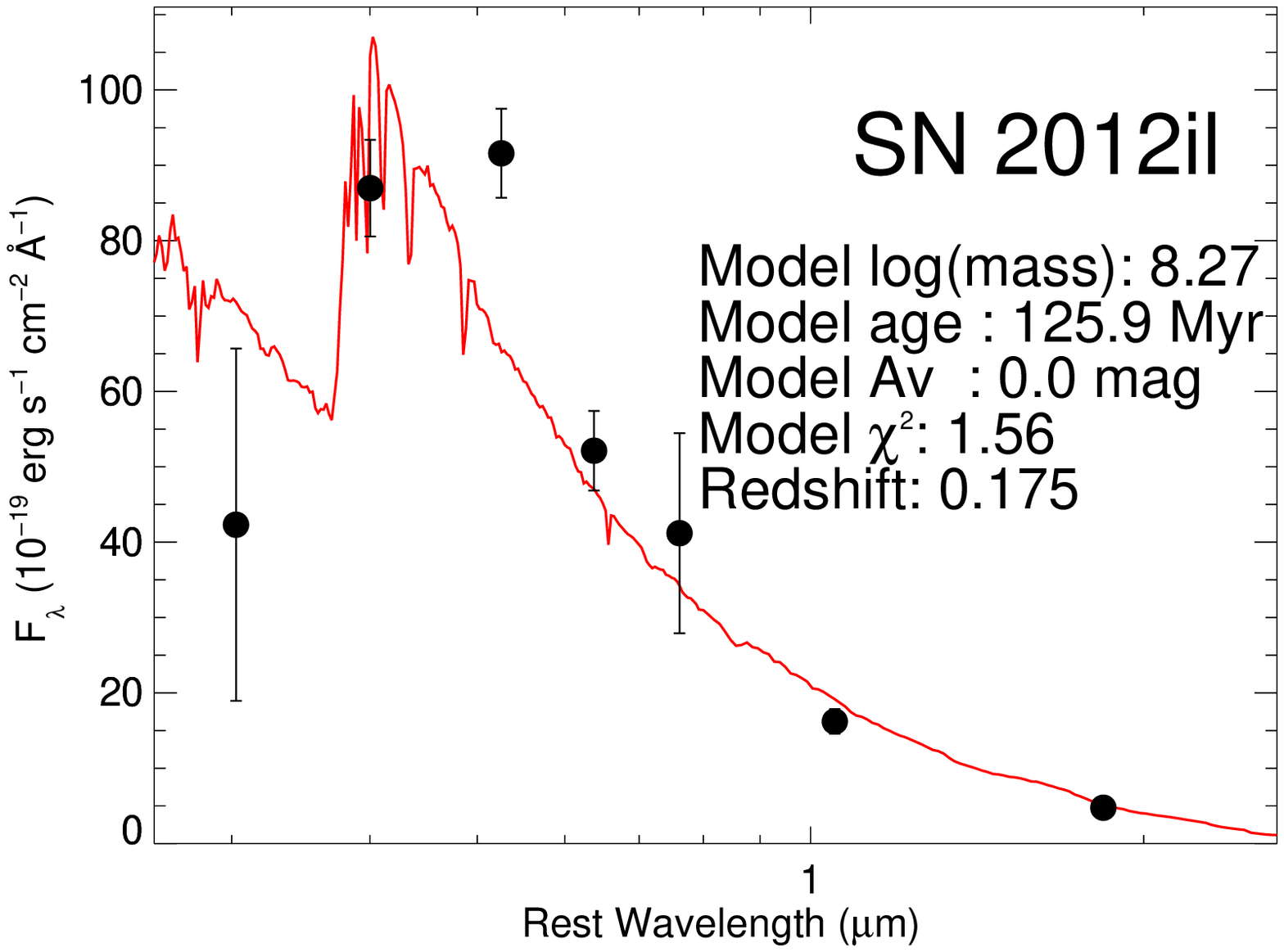} &
\includegraphics[width=4cm]{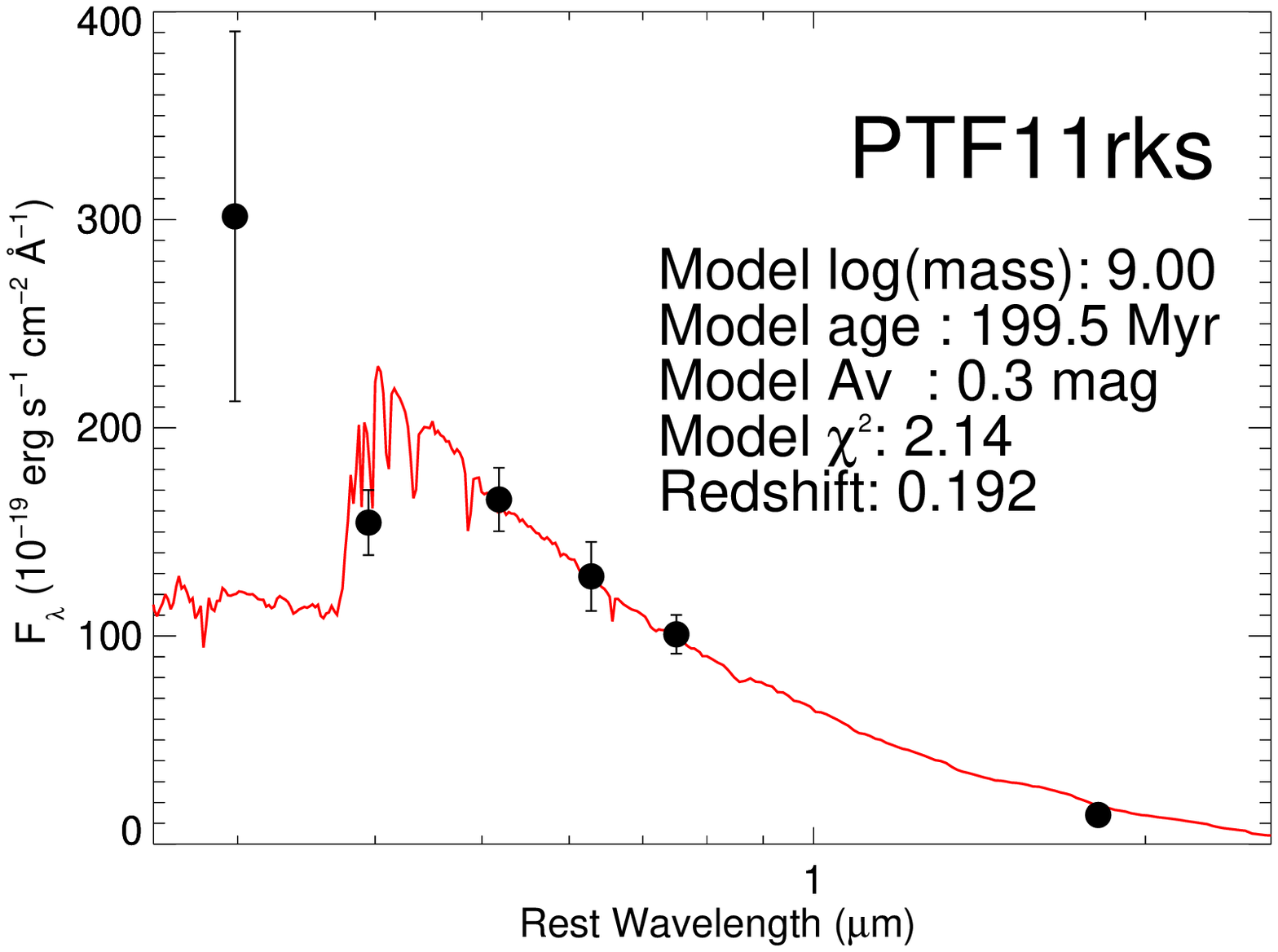} & \includegraphics[width=4cm]{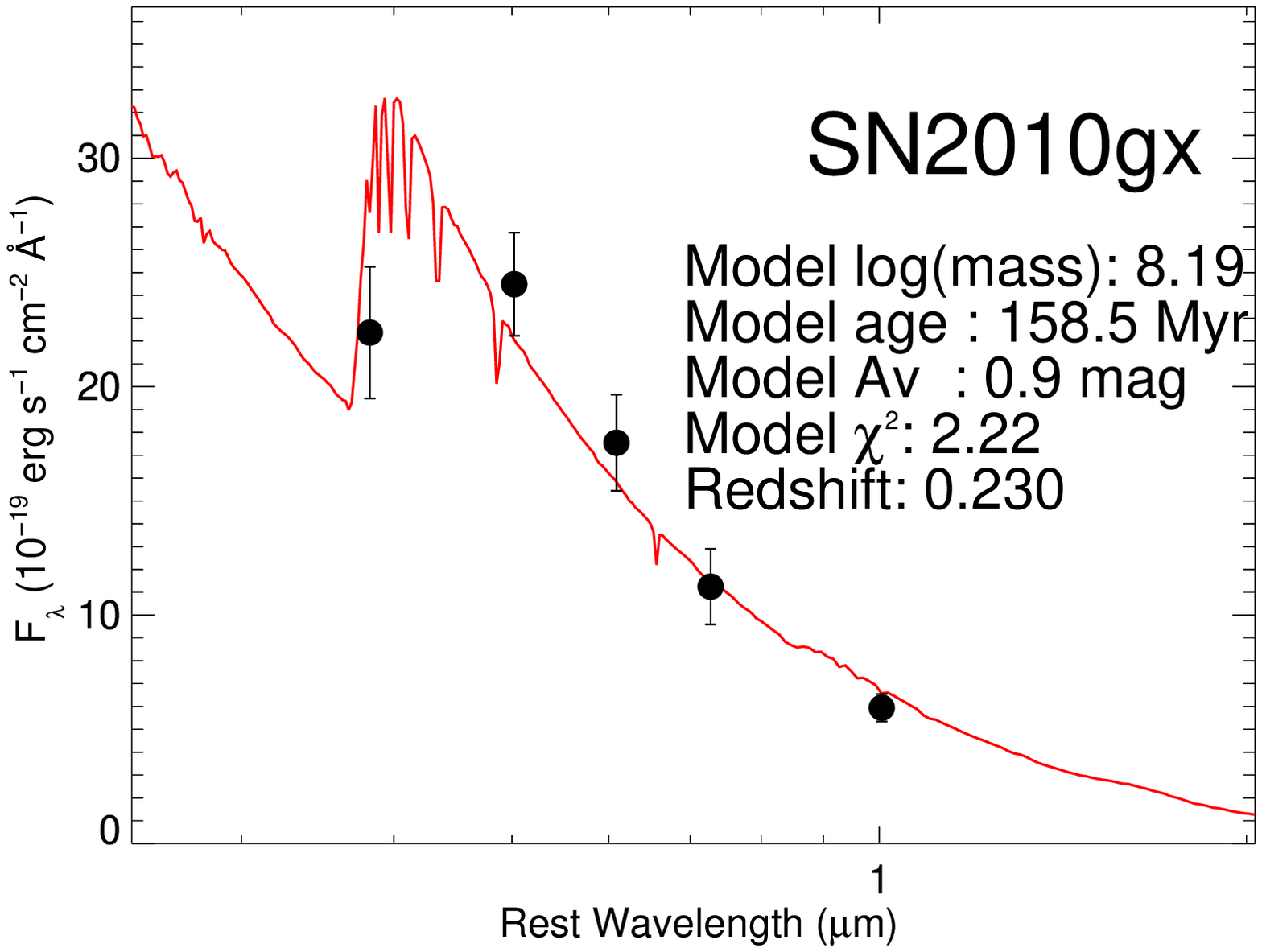} \\
 \includegraphics[width=4cm]{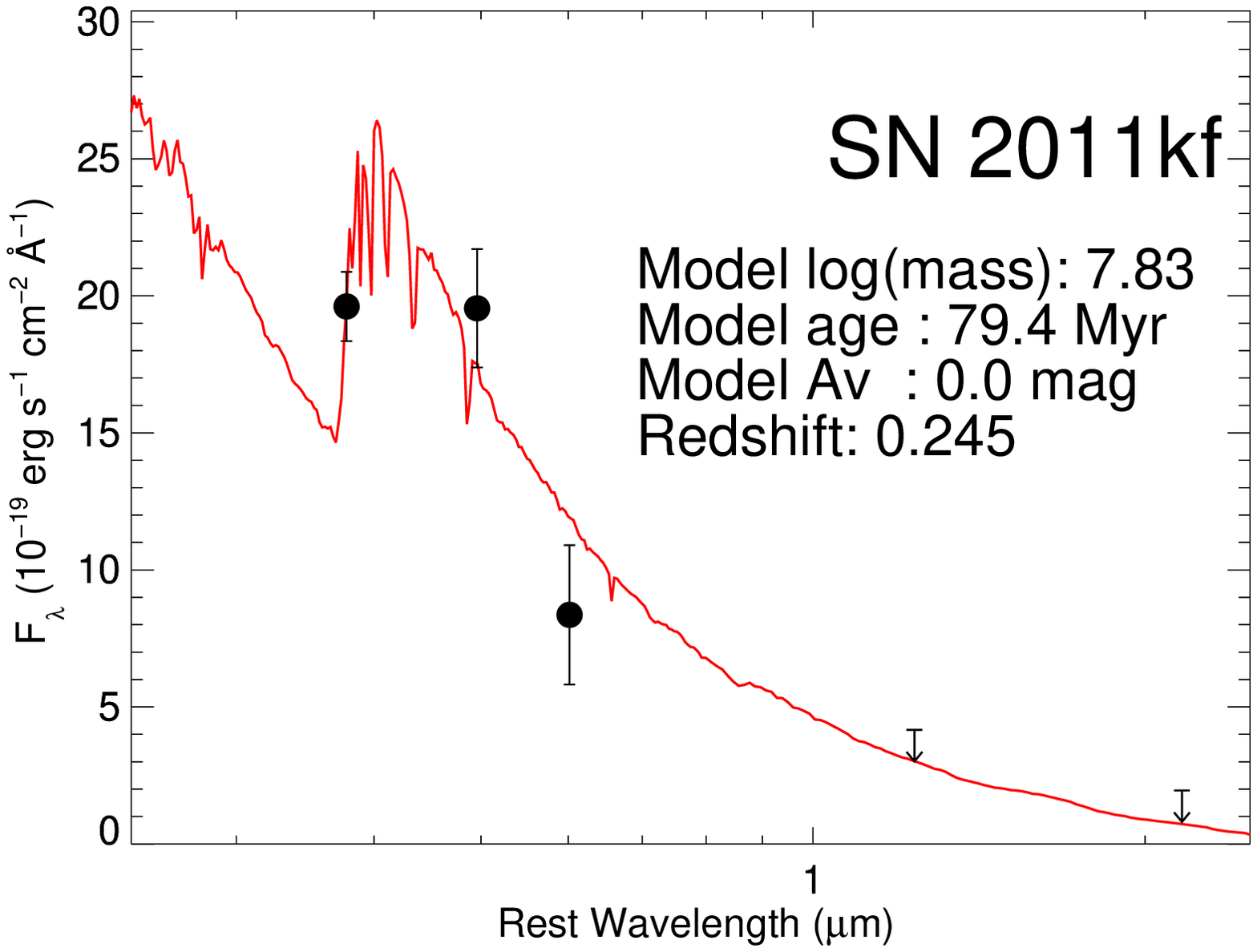} &
\includegraphics[width=4cm]{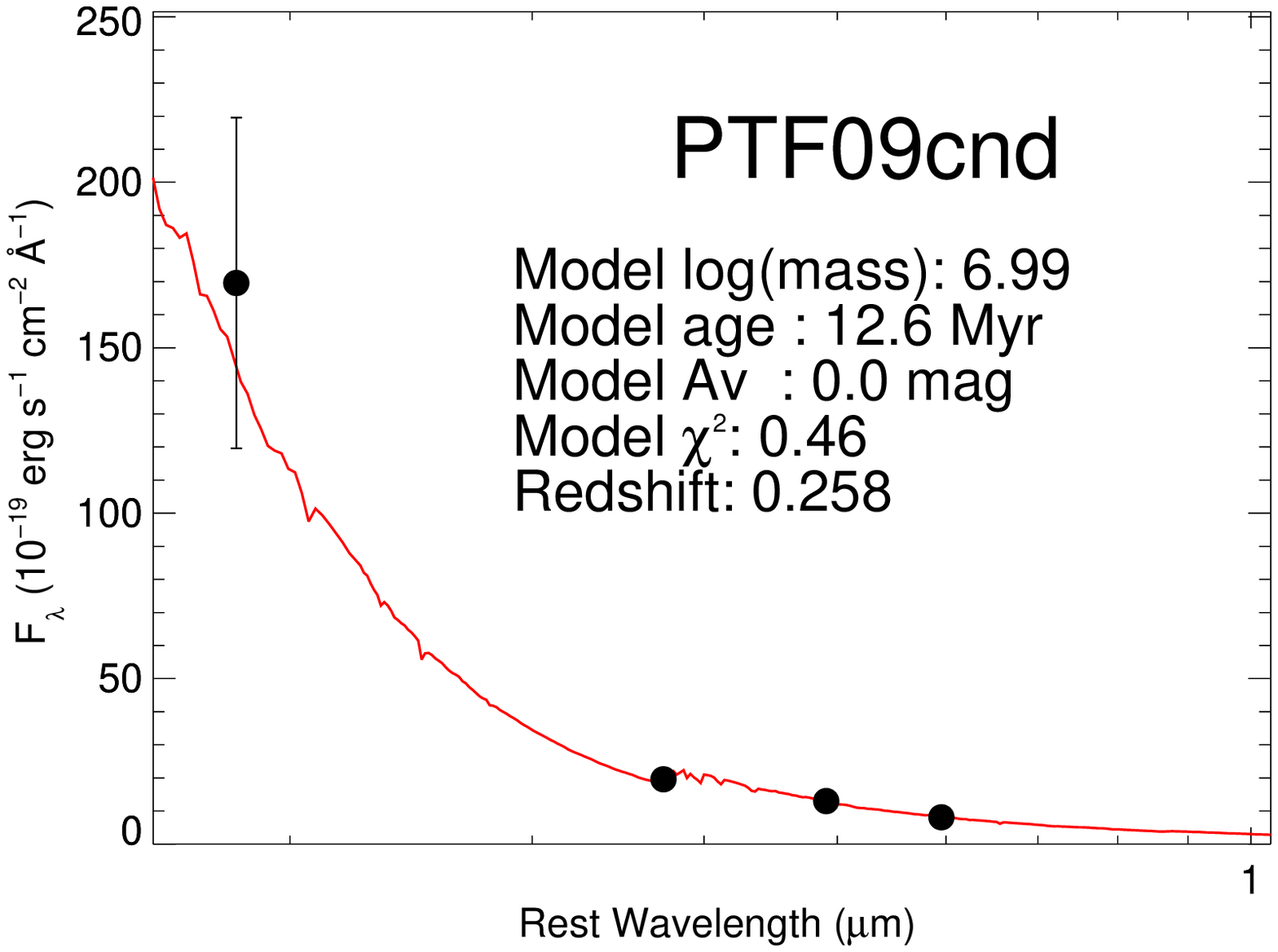} & \includegraphics[width=4cm]{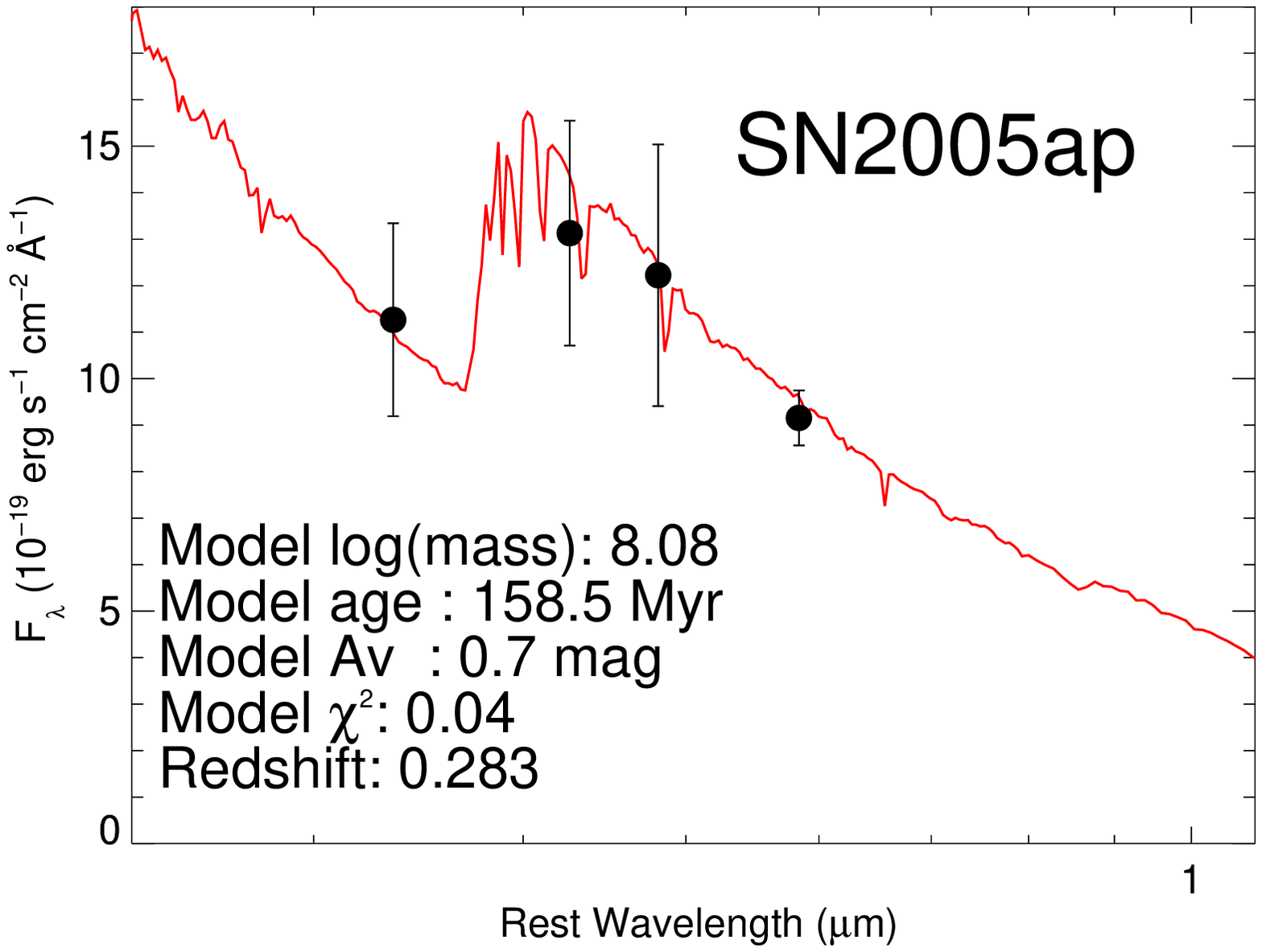} & \includegraphics[width=4cm]{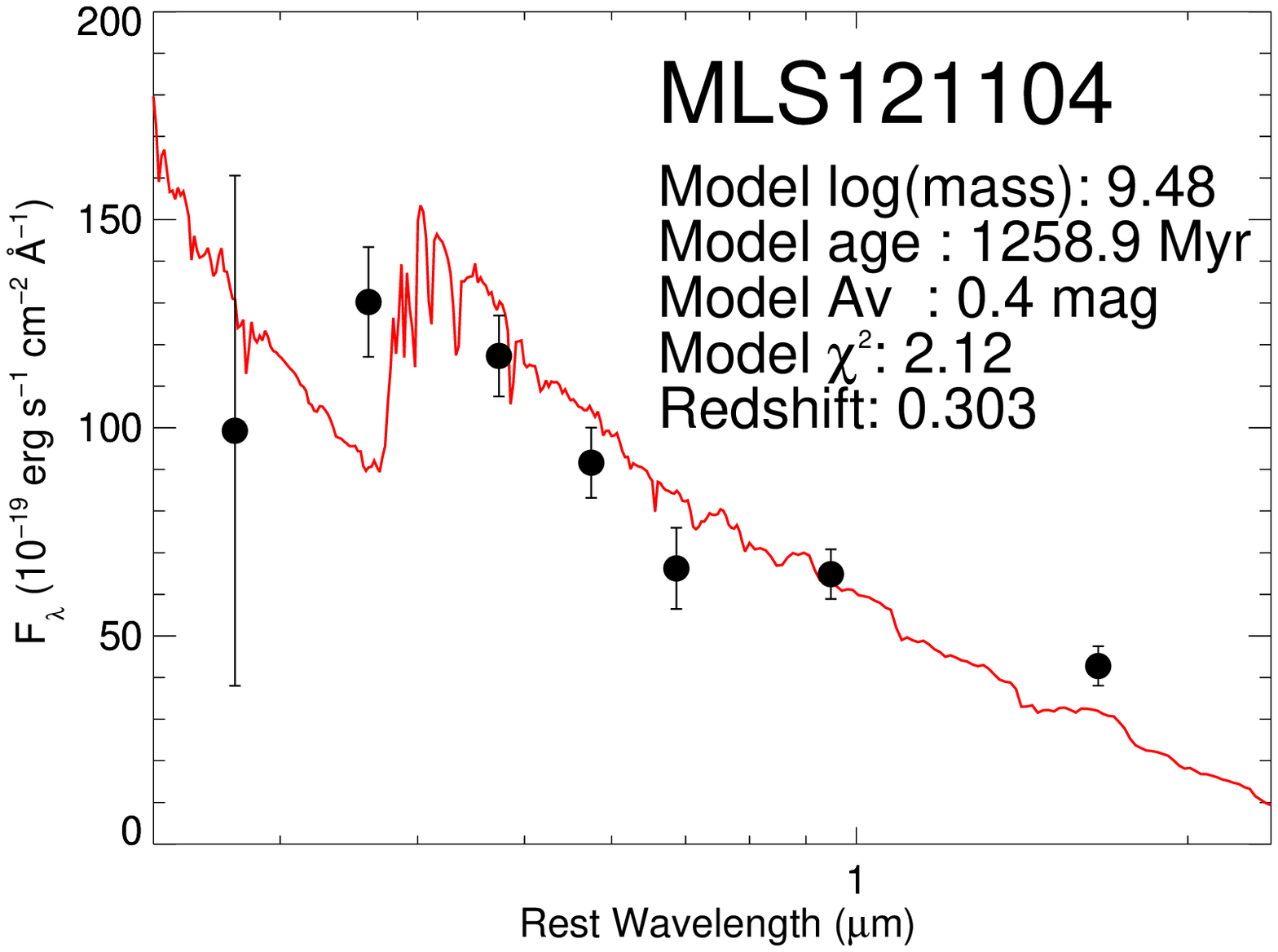} \\
\includegraphics[width=4cm]{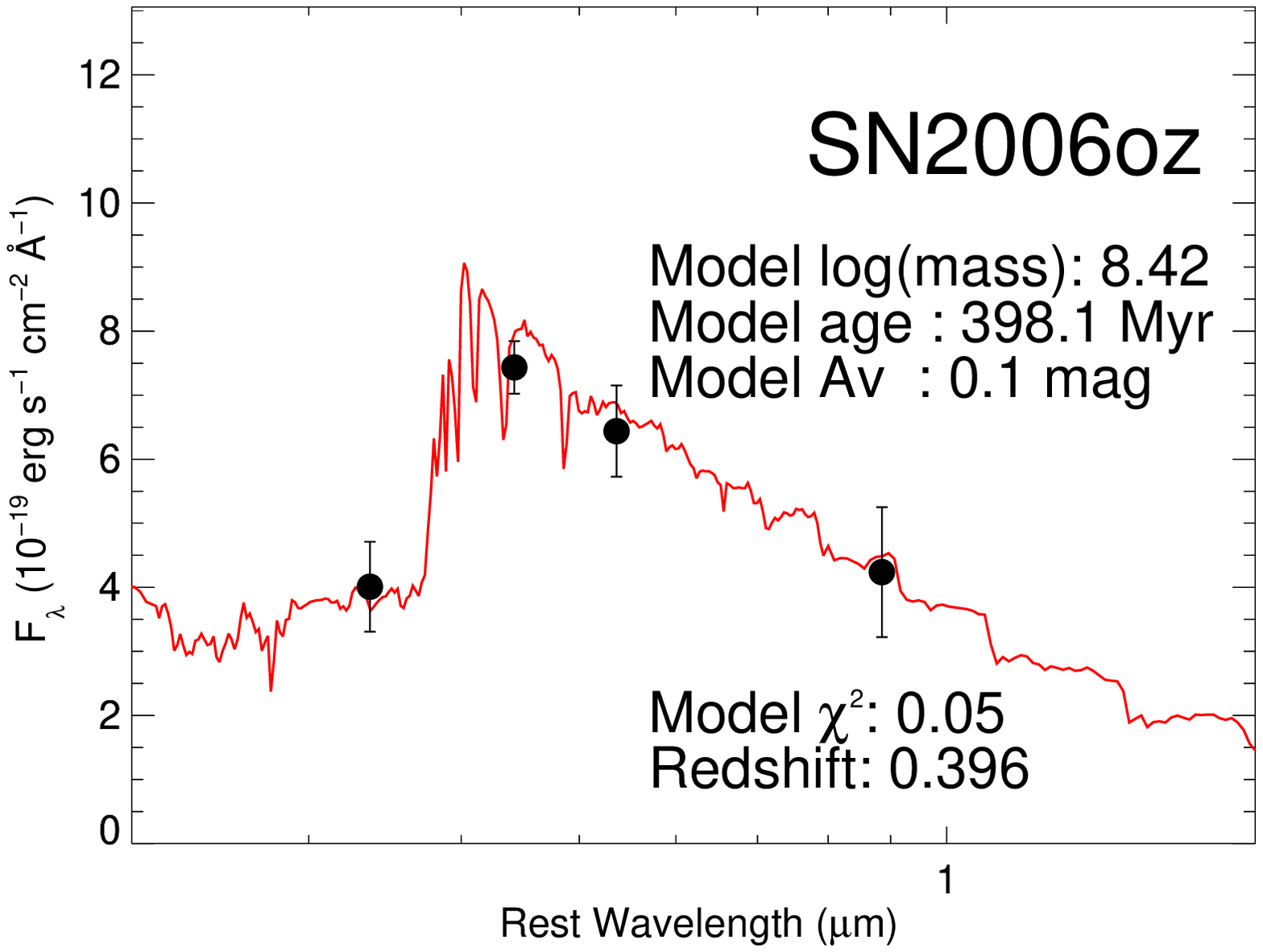} & \includegraphics[width=4cm]{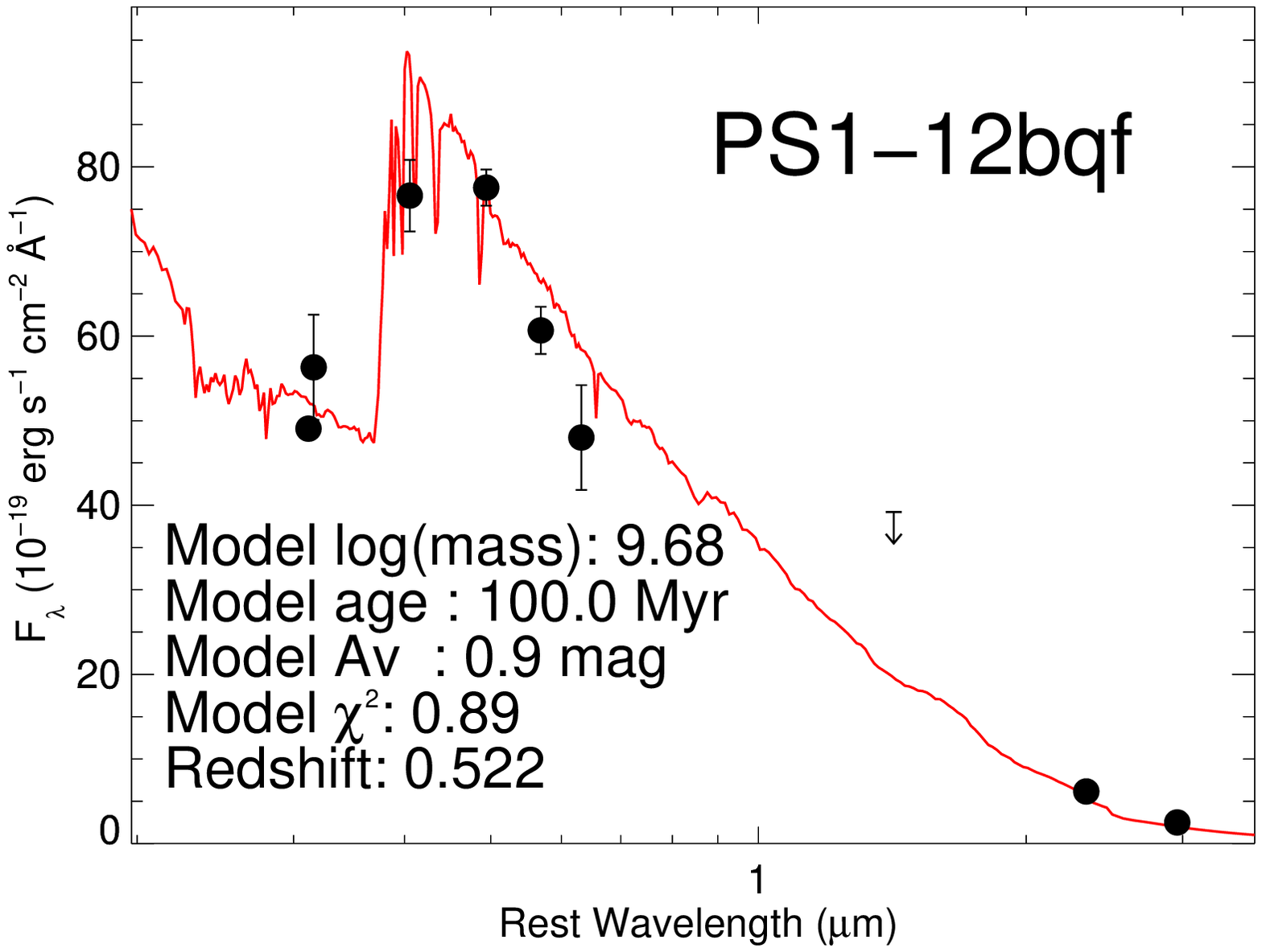} & \includegraphics[width=4cm]{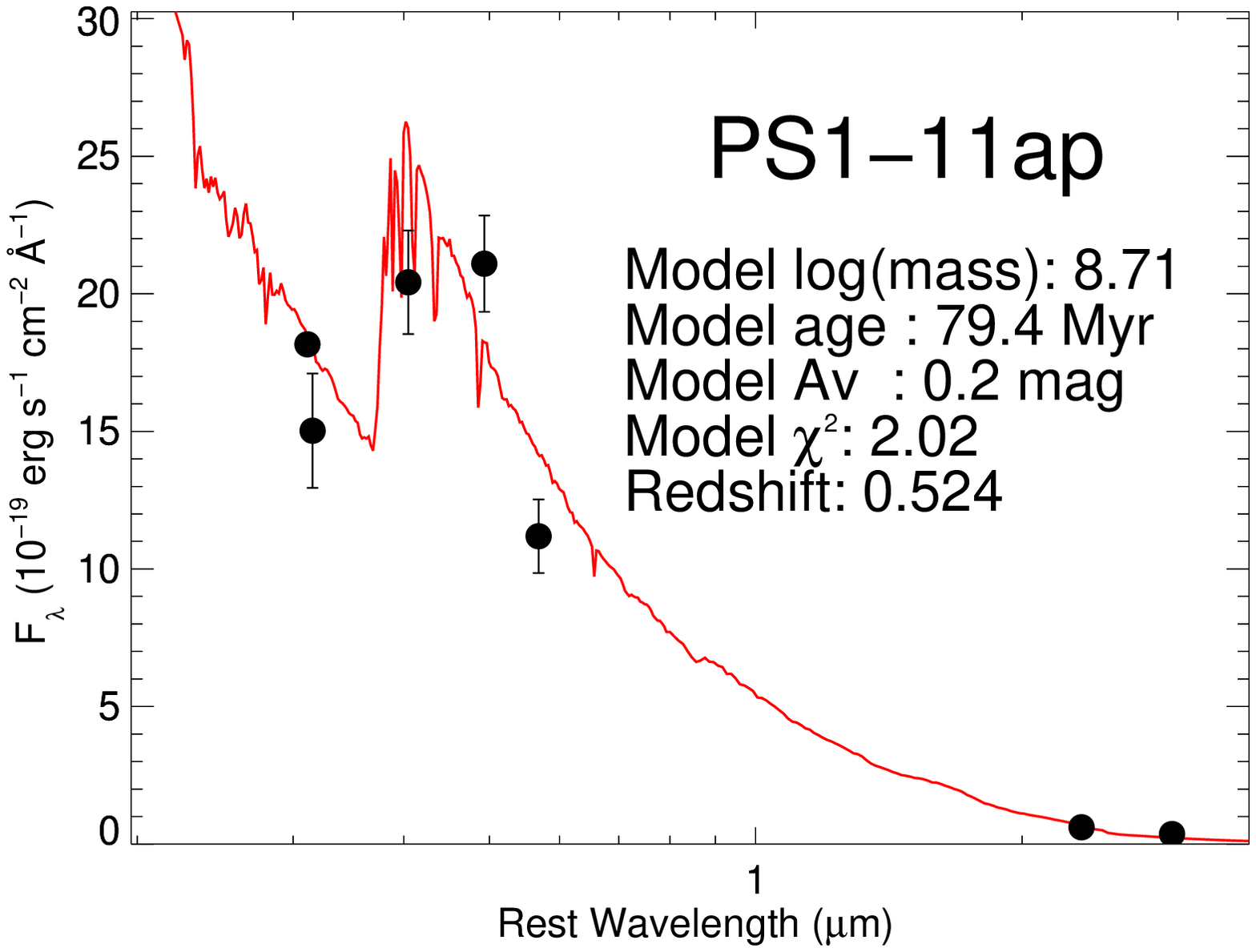} &
\includegraphics[width=4cm]{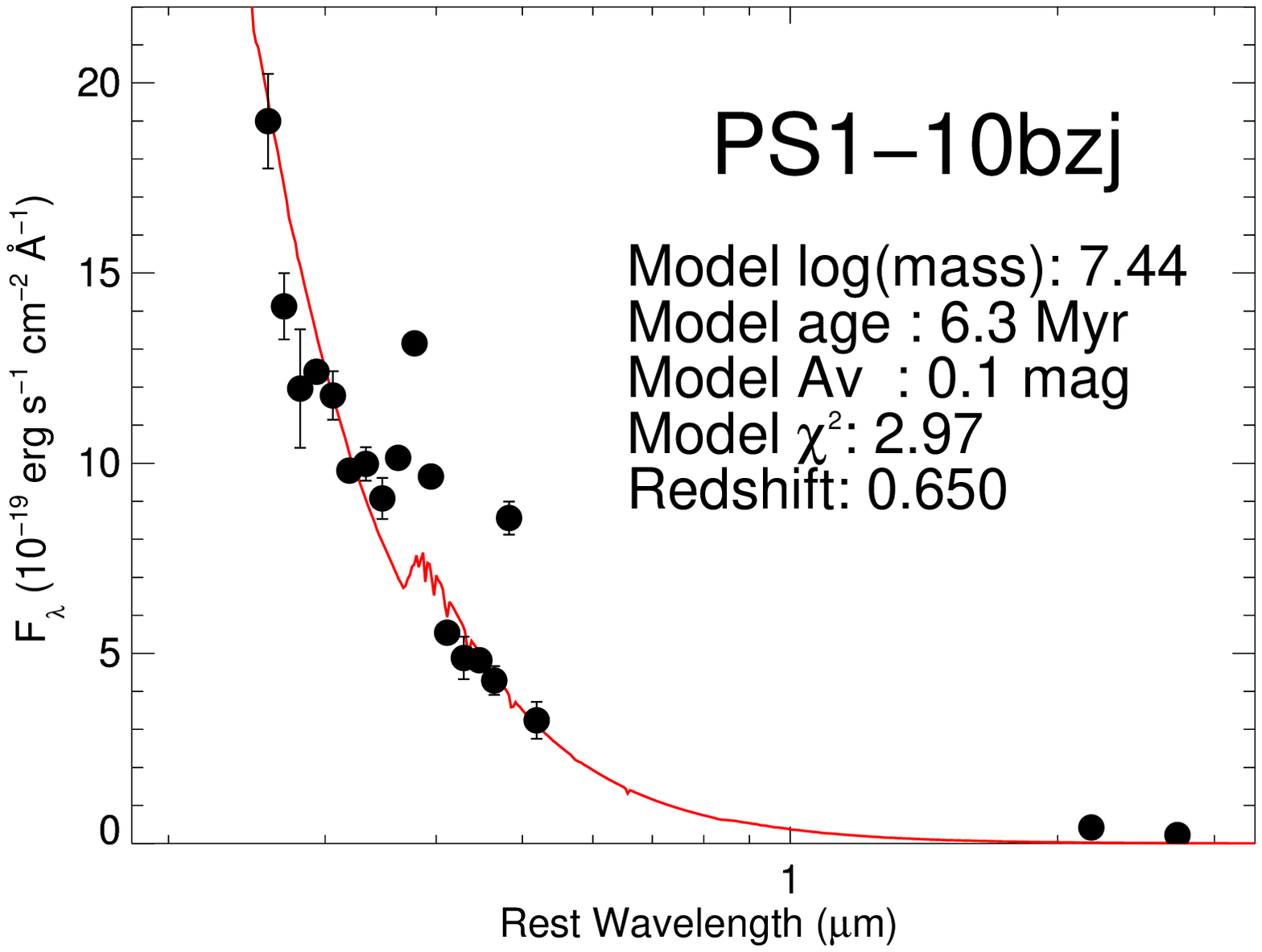} \\
 \includegraphics[width=4cm]{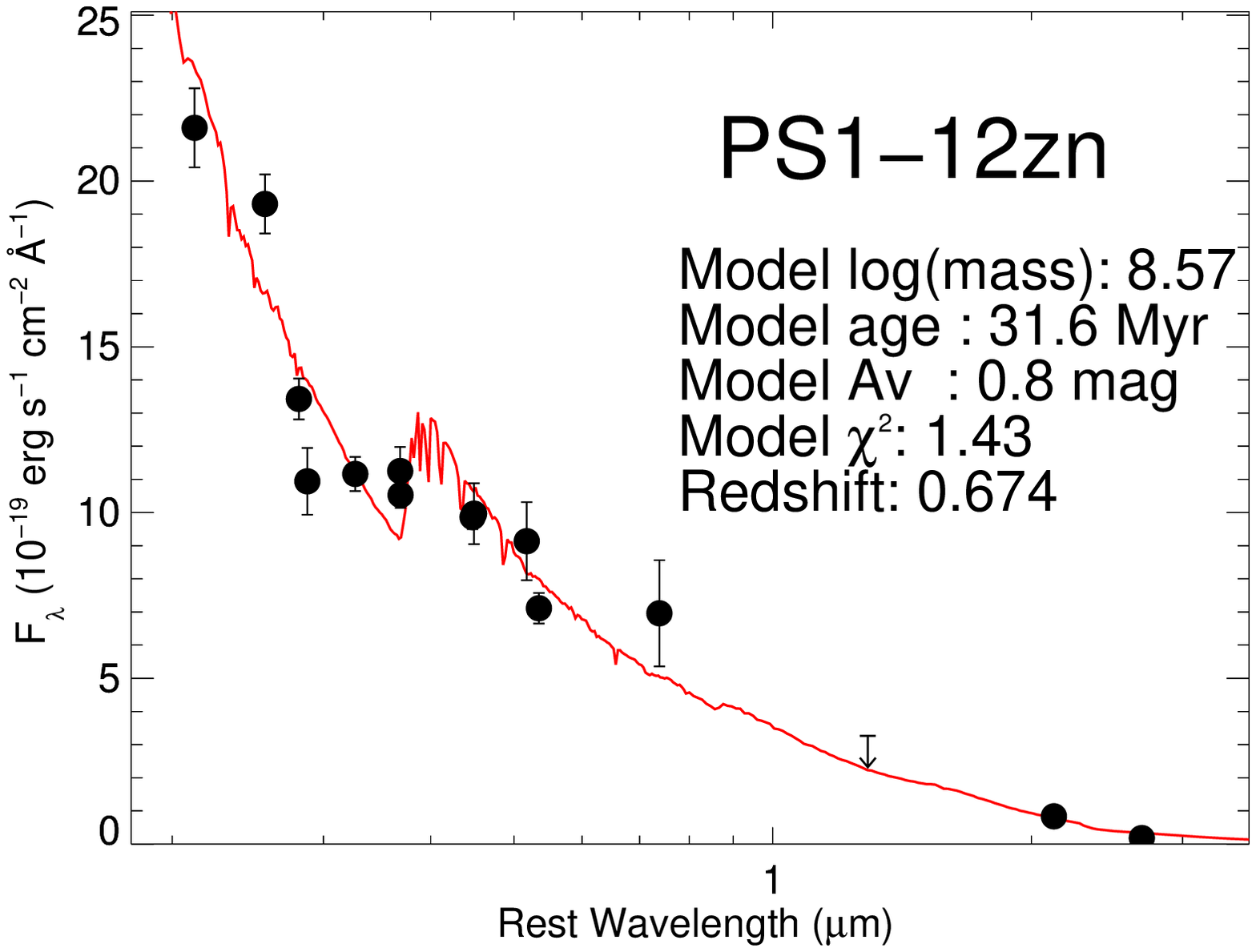} & \includegraphics[width=4cm]{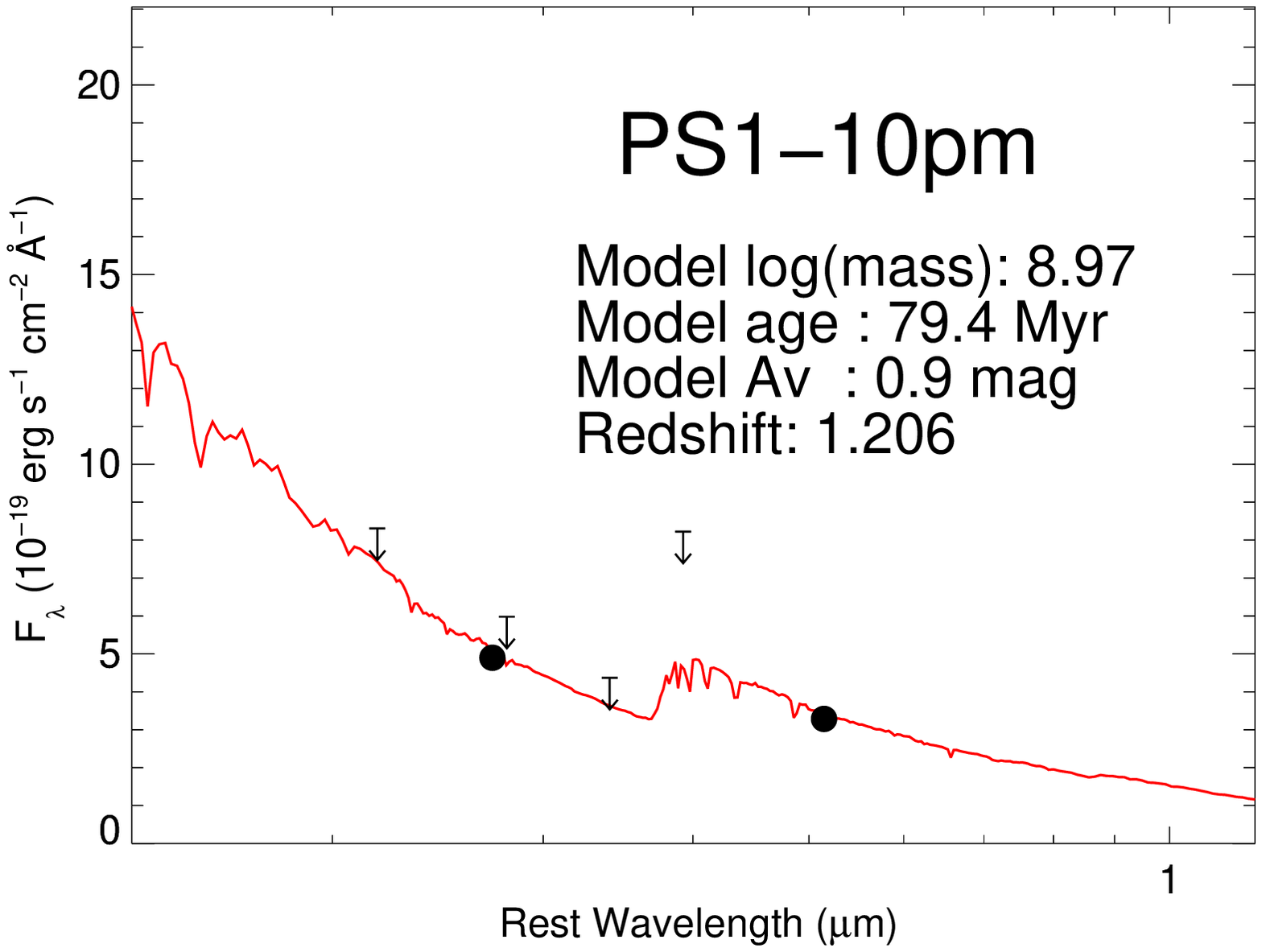} &
\includegraphics[width=4cm]{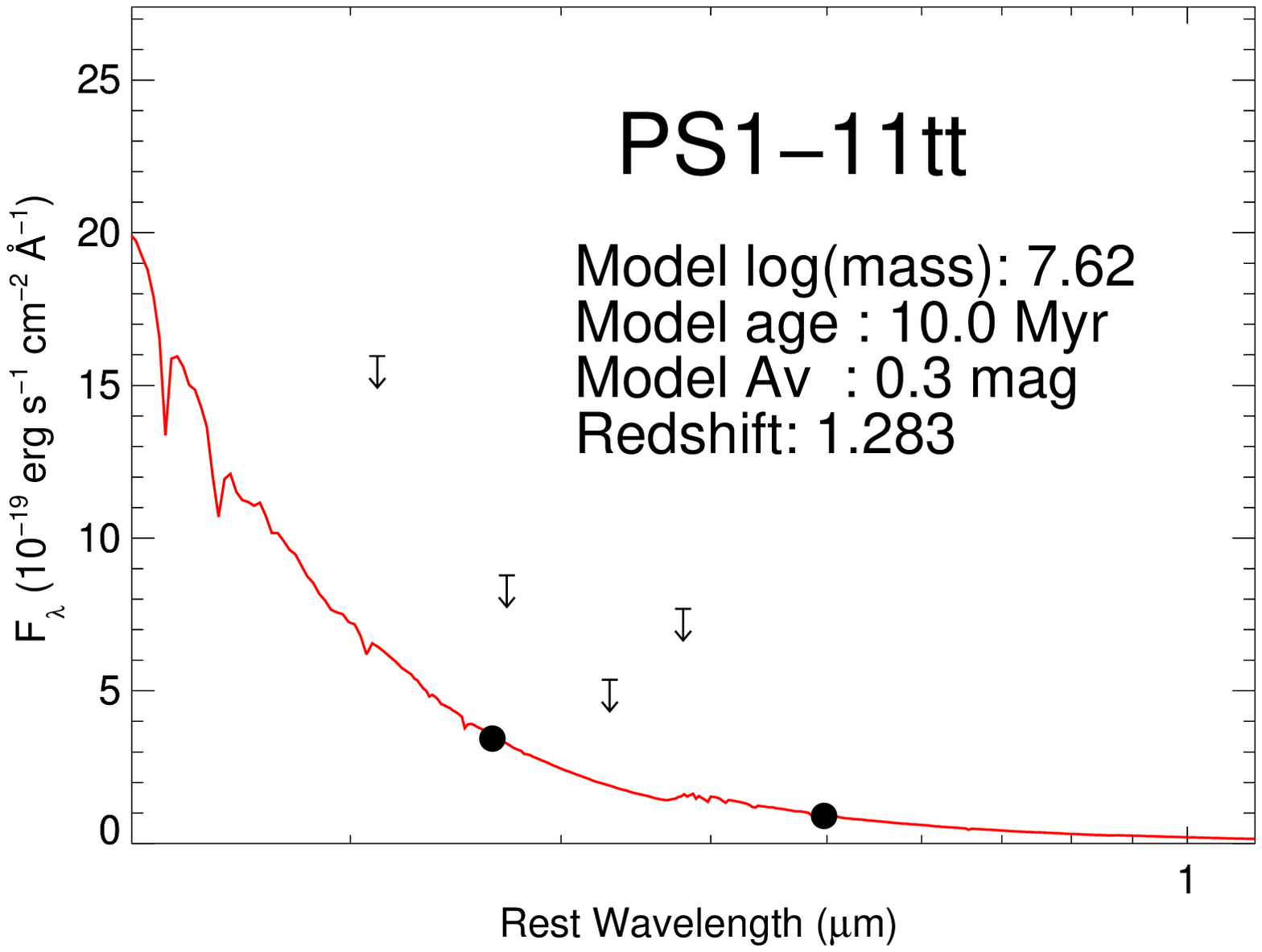} & \includegraphics[width=4cm]{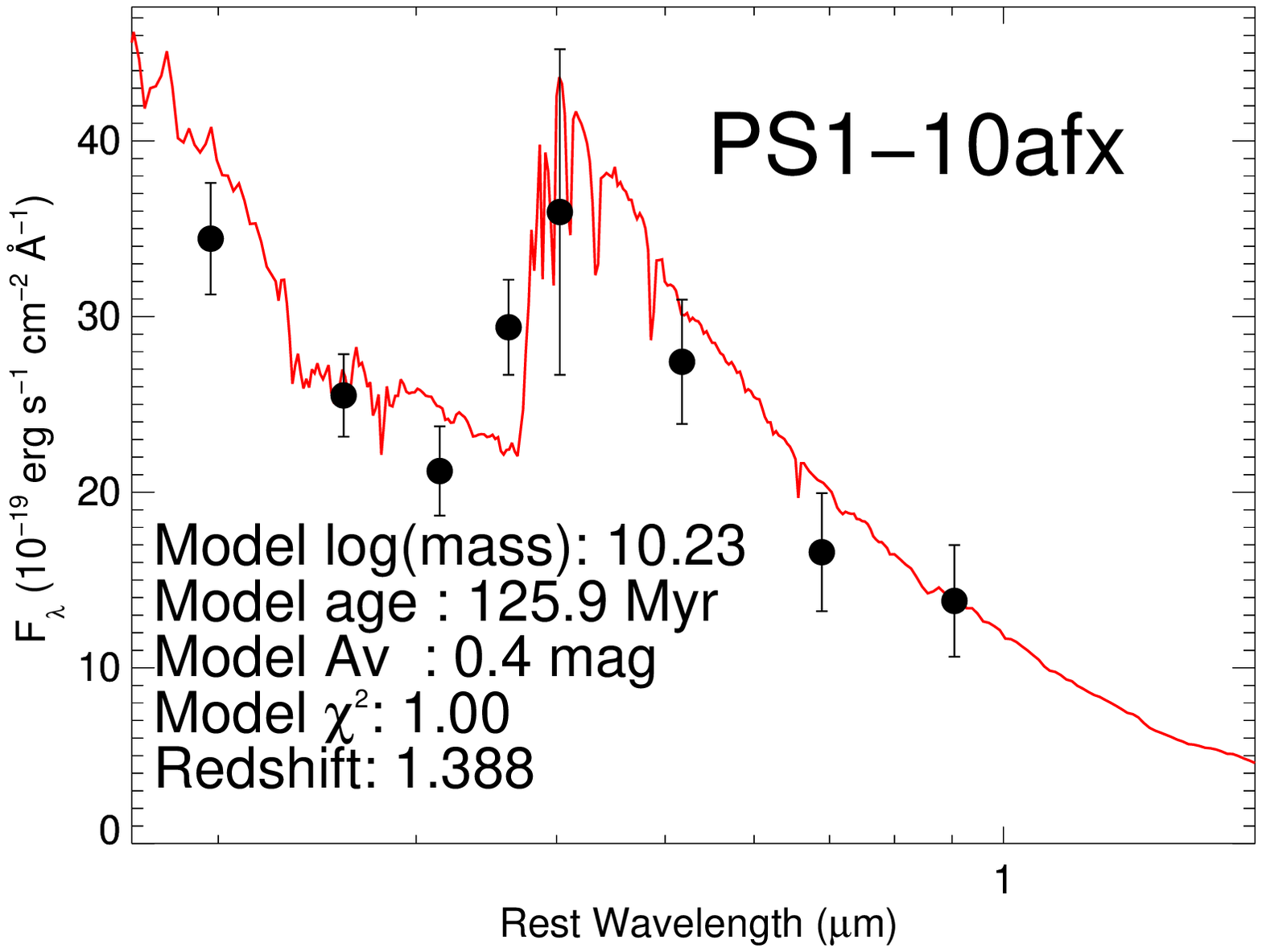} \\
 \includegraphics[width=4cm]{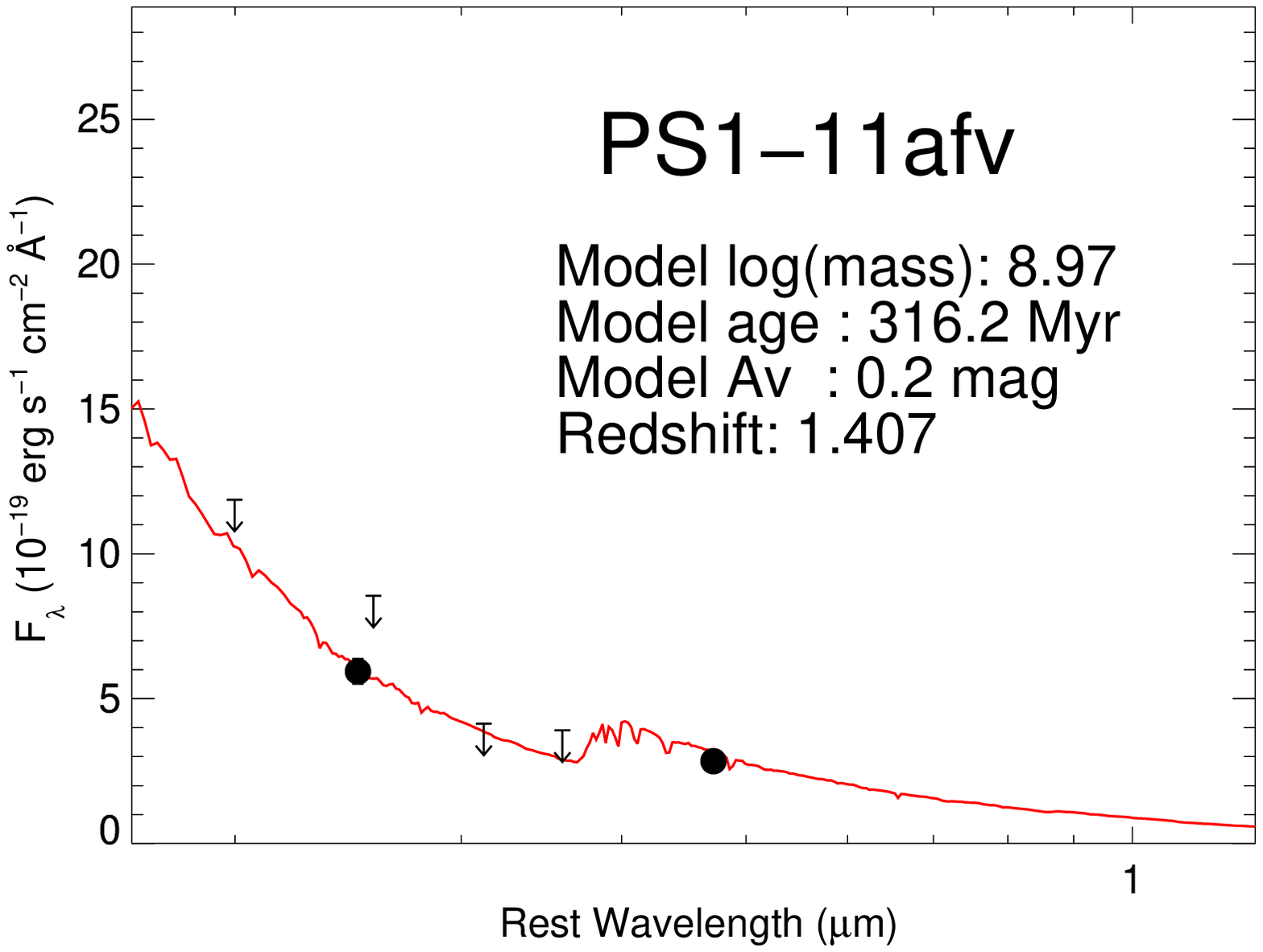} & \includegraphics[width=4cm]{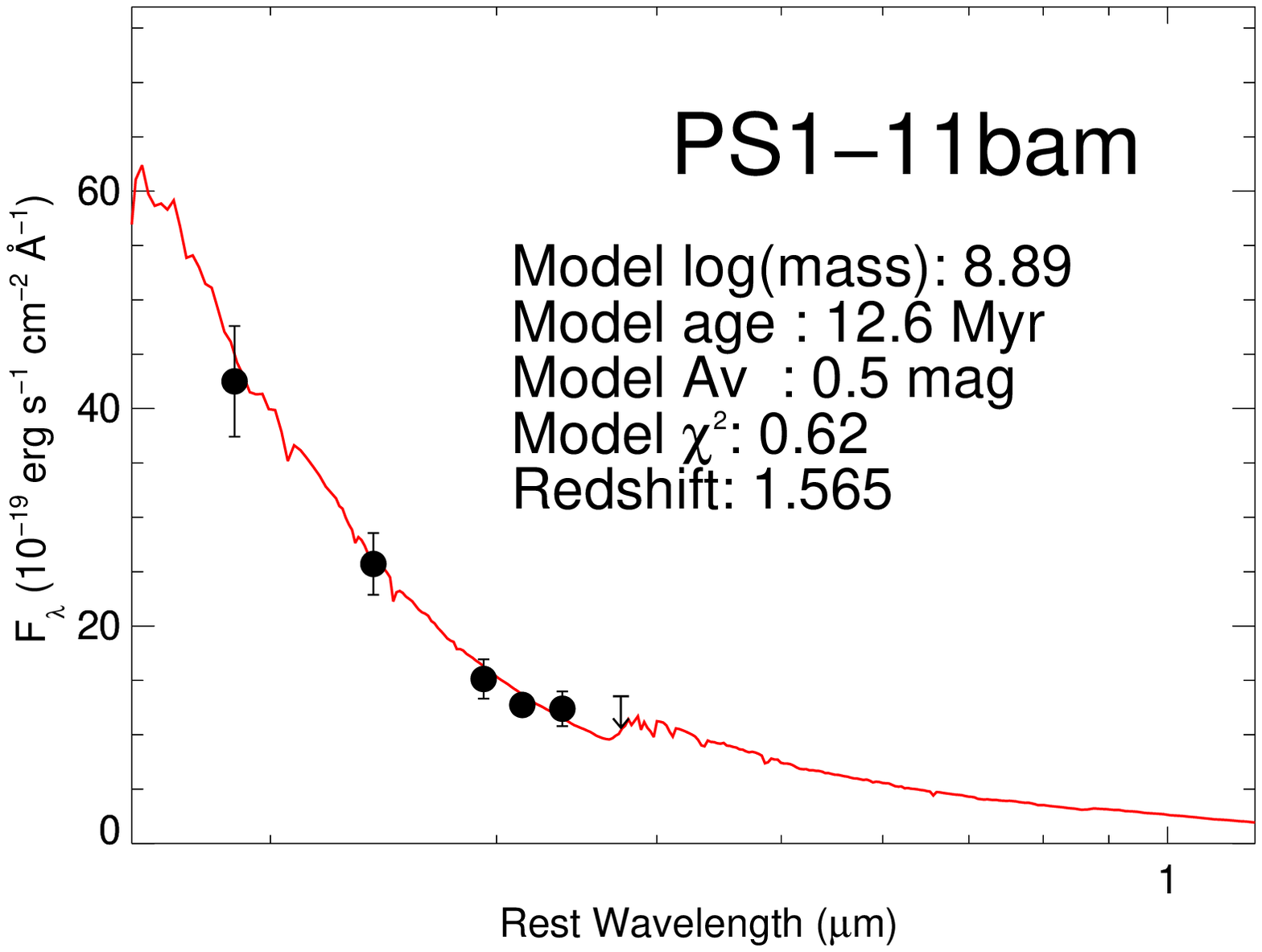} & 
\includegraphics[width=4cm]{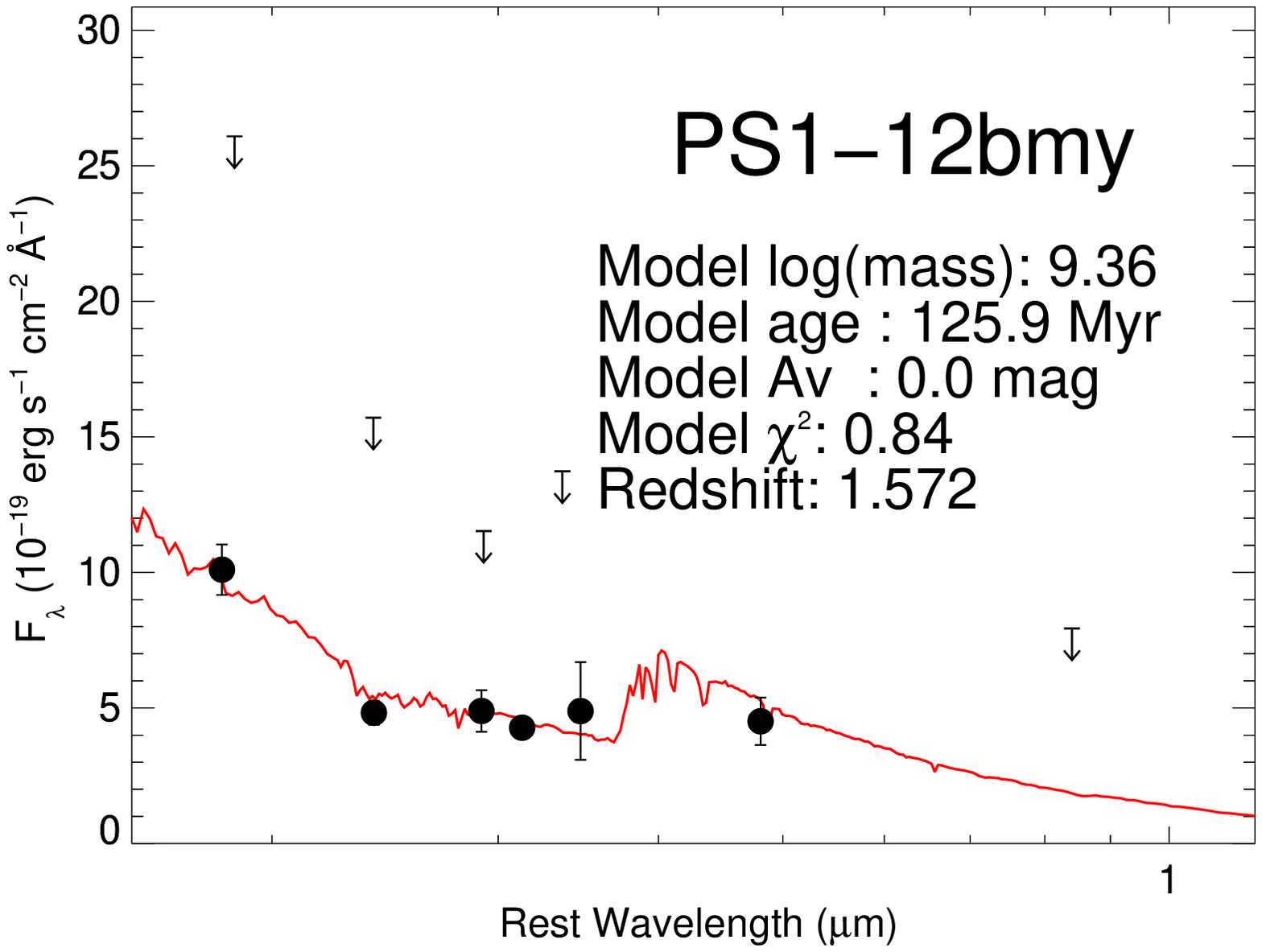} &  \\
\end{tabular}
\caption{Model fits to the SEDs of 23 host galaxies with multi-band photometry. The red lines show the model SEDs (calculated using FAST; \citealt{kvl+09}), while the black points with error bars show the photometry.  The main parameters of the model and fit are listed in each panel, and summarized in Table~\ref{tab:results}.
\label{fig:seds} }
\end{center}
\end{figure*}

\subsection{Extinction}
\label{sec:ext}
We estimate the reddening by measuring Balmer decrements, using the ratio of H$\alpha$ to H$\beta$ or H$\gamma$ to H$\beta$ (if H$\alpha$ is not available), assuming intrinsic ratios according to Case B recombination \citep{ost89}. The measured emission line fluxes are then corrected for reddening using the extinction curve of \citet{ccm89}. We estimate error bars on $A_V$ using the $1\sigma$ uncertainties in our line flux measurements. For galaxies where we can measure extinction from Balmer lines, the SED fit for the galaxy is constrained to the 1$\sigma$ uncertainty range from spectroscopy.

For hosts with no Balmer decrement measurements but with multi-band photometry, we do not restrict the range of allowed extinction in the SED fits. As with the stellar mass, the fitting procedure returns both a best-fit and a $1\sigma$ uncertainty range on the extinction. While the uncertainty from the SED fits is generally larger than from our Balmer decrements, we list the extinction estimates from SED modeling in Table~\ref{tab:results} for the galaxies where no estimates from Balmer lines are available.

While a wide range of $A_V$ is allowed by the SED fits, we find that with a few exceptions the data are consistent with zero extinction. We therefore also compute a set of galaxy models that assume zero extinction. Table~\ref{tab:results} also lists the stellar mass and population age for these fits.

\subsection{Absolute Magnitudes}
We calculate absolute $B$-band magnitudes by transforming the galaxy models to the rest frame and integrating over the $B$-band filter curve. We also use these models to calculate a mean $k$-correction as a function of redshift, and use this to determine absolute magnitudes or upper limits for the objects with only single-band photometry or non-detections. 

In Figure~\ref{fig:mbdist} we show the resulting $B$-band absolute magnitudes, both as a function of redshift and the cumulative distribution using the Kaplan-Meier estimator. The overall range is $-16$ to $-22~{\rm mag}$, but the population is strikingly low-luminosity with a median absolute magnitude of $\langle M_B \rangle \approx -17.6~{\rm mag}$ ($\approx 0.05 L_*$; \citealt{wfk+06}). A large fraction of the lowest-luminosity hosts are found at the low-redshift end: when we consider the PS1 sample + SCP06F6 separately ($z \gtrsim 0.5$), we find a median magnitude of $-18.8~{\rm mag}$ ($\approx 0.1 L_*$), whereas the sample at lower redshifts (all the non-PS1 hosts, and excluding SCP06F6) has a median magnitude of $-17.0~{\rm mag}$ ($\approx 0.04 L_*$). This may indicate that the typical host of a SLSN shifts to fainter galaxies at lower redshift, an effect one might expect if, for example, low metallicity is a driving ingredient for producing SLSNe. As the data at high and low redshift come from different surveys, however, this could also reflect different survey or follow-up strategies.

The highest-luminosity (and only $\sim L_*$) host galaxy in this sample is the host of PS1-10afx, which was also an outlier in terms of SN properties \citep{cbr+13}. It is debated whether this object was a true SLSN, or a normal Type Ia SN lensed by a foreground galaxy \citep{qwo+13}. However, we note that given our sample size this one outlier does not drive our results -- in fact, excluding it would lead us to find an even lower median luminosity. 

As is apparent from Figure~\ref{fig:mbdist}, the SLSN hosts are also significantly less luminous as a population than both LGRB hosts and CCSN hosts in the same redshift range. Applying the generalized Wilcoxon test, we find that the SLSN hosts are not consistent with being drawn from the same underlying distribution as either CCSN hosts or LGRB hosts in terms of their luminosities. The significance levels are listed in Table~\ref{tab:stats}. If we consider the PS1/MDS subsample separately, however, the luminosity distribution is consistent with both the LGRB host sample and the CCSN sample ($p = 0.26$ and $0.15$ respectively). The change in significance level results from both the PS1/MDS hosts being higher luminosity and the fact that the sample size is smaller.

\begin{figure*}
\begin{center}
\begin{tabular}{cc}
\includegraphics[width=3.5in]{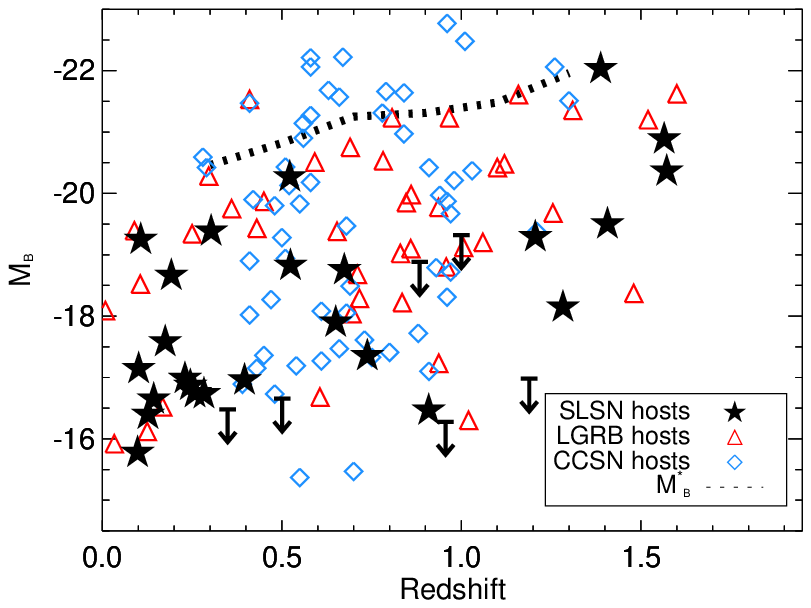} & \includegraphics[width=3.5in]{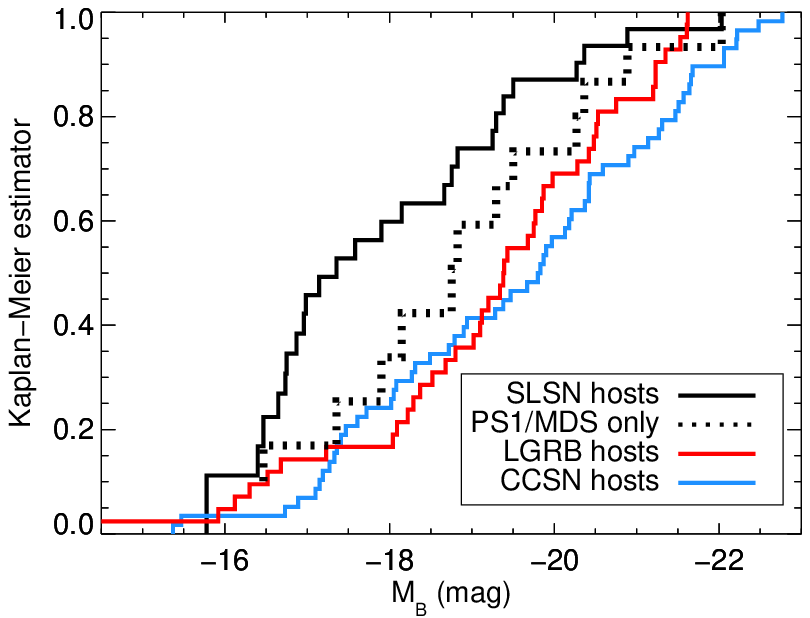}
\end{tabular}
\caption{Left: Absolute $B$-band magnitudes as a function of redshift for the SLSN host galaxies (black stars and arrows), LGRB host galaxies (red triangles) and GOODS CCSN hosts (blue diamonds). Also shown is the luminosity function parameter $M_B^*$ for blue galaxies as a function of redshift (dotted line; \citealt{wfk+06}). Right: The resulting distribution functions of the three populations, as calculated by the Kaplan-Meier estimator to include the information contained in upper limits. The dotted line shows only the hosts from the PS1/MDS subsample, illustrating how the difference between the SLSN hosts and the other populations is driven by the low-redshift end of the sample.
\label{fig:mbdist}}
\end{center}
\end{figure*}

\subsection{Stellar Masses}
The FAST SED fitting code provides the stellar mass of the best-fit model, and the $1\sigma$ confidence interval. The derived stellar masses and uncertainties are listed in Table~\ref{tab:results}. For host galaxies where we either only have upper limits, or detections in too few filters for an SED fit, we use the galaxy models to calculate a median mass-to-light ratio and use this to convert our single-band measurements into a mass estimate; in these cases the uncertainties quoted reflect the spread in possible mass-to-light ratios.

The resulting stellar masses are shown in Figure~\ref{fig:massdist}, both as a function of redshift and the cumulative distribution. As with the luminosities, the SLSNe are generally found in low-mass galaxies, with a median stellar mass of $\langle M_* \rangle \approx 2 \times 10^8~{\rm M}_{\odot}$. There is a range of three orders of magnitude in mass, from $10^7$ to $10^{10}~{\rm M}_{\odot}$, and the same trend towards smaller galaxies at lower redshift is also seen in the stellar masses. Again, the SLSN hosts are offset from both the CCSN hosts and the LGRB hosts, and the difference between the SLSN and CCSN host galaxies is significant both when comparing the full samples and when considering the full sample and the PS1/MDS subsample only ($p =  1.5 \times 10^{-7}$ and $9 \times 10^{-4}$, respectively). The difference between SLSN hosts and LGRB hosts is significant at the $2.7\sigma$ level ($p = 0.007$) when comparing to the full sample of SLSN hosts, but not significant when comparing to the PS1/MDS data only ($p = 0.17$).

\begin{figure*}
\begin{center}
\begin{tabular}{cc}
\includegraphics[width=3.5in]{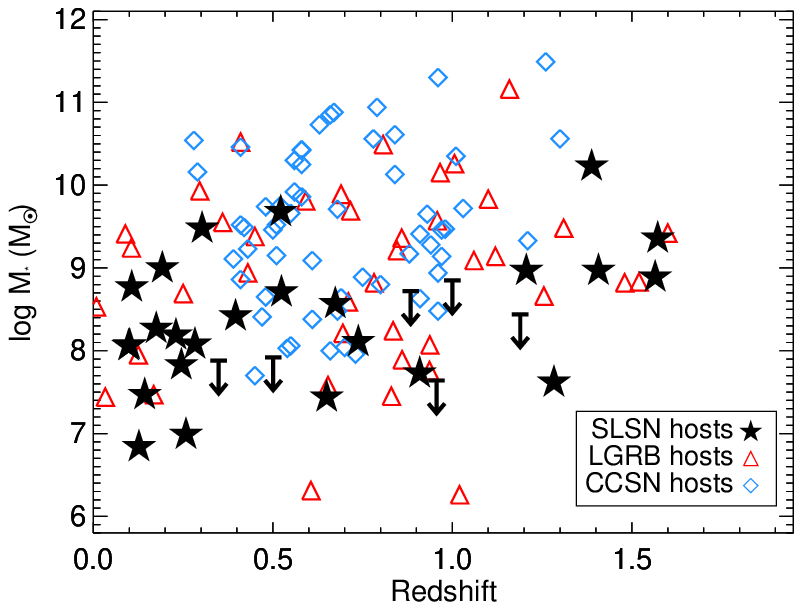} & \includegraphics[width=3.5in]{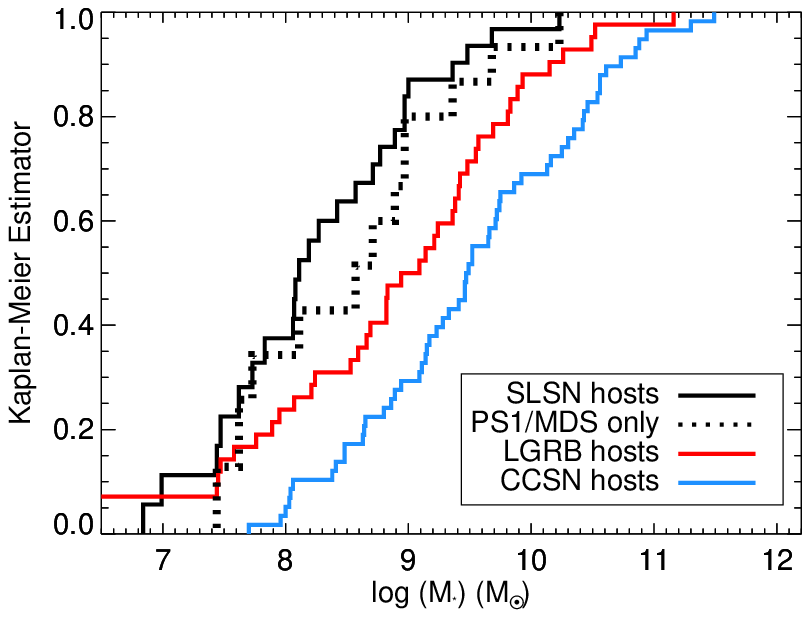}
\end{tabular}
\caption{Left: Stellar mass as a function of redshift for the SLSN host galaxies (black stars and arrows), LGRB host galaxies (red triangles) and CCSN hosts (blue diamonds). Right: The resulting distribution functions of the three populations. The difference between the SLSN and CCSN hosts is statistically significant, both when considering the full SLSN sample and the PS1/MDS subsample only. While having a lower median mass, the SLSN hosts are marginally consistent with being drawn from the same distribution as the LGRB hosts.
\label{fig:massdist}}
\end{center}
\end{figure*}

\subsection{Star Formation Rates}
Star formation rates (SFR) are derived using a variety of methods, depending on the available data for each host galaxy. If available, we calculate the SFR using the H$\alpha$ emission line flux, according to the relation ${\rm SFR} = 7.9 \times 10^{-42}L_{{\rm H}\alpha}({\rm erg~s^{-1}})$ \citep{ken98}. If H$\alpha$ is not available but H$\beta$ and H$\gamma$ are both detected, we calculate the expected H$\alpha$ flux using the measured reddening and H$\beta$ flux, assuming Case B recombination. For some galaxies at higher redshift no Balmer lines are available, but we detect the [\ion{O}{2}]$\lambda$3727 emission line. For these galaxies, we use  ${\rm SFR} = 1.4 \times 10^{-41}L_{{\rm [OII]}}({\rm erg~s^{-1}})$ \citep{ken98}.

Finally, for galaxies where we do not have line-based SFR estimates, we calculate a SFR based on the rest-frame UV flux. For galaxies at redshift $z \gtrsim0.6$, $g$-band covers rest-frame UV, and we use ${\rm SFR} = 1.4 \times 10^{-28}L_{\nu}({\rm erg~s^{-1}~Hz^{-1}})$ \citep{ken98}. We use the observed fluxes without correcting for extinction for this calculation, since the extinction is not particularly well constrained by the SED fits and also consistent with zero in most galaxies (Section~\ref{sec:ext}; Table~\ref{tab:results}). We also use this relation to calculate upper limits for the galaxies with rest-frame UV non-detections. In general, we find that the different diagnostics agree within a factor of $2-3$. For four galaxies (PTF09cwl, PTF09atu, SN\,2005ap and SCP06F6) we have neither emission line measurements nor rest-frame UV data, and we therefore lack SFR estimates.

The resulting star formation rate distributions are plotted in Figure~\ref{fig:sfrdist}. The median value for the SLSN host sample is $\langle {\rm SFR} \rangle \approx 1~{\rm M}_{\odot}~{\rm yr}^{-1}$ and varies from $10^{-2} - 10~{\rm M}_{\odot}~{\rm yr}^{-1}$. Consistent with their lower luminosities and stellar masses, the SLSN hosts also have slightly lower absolute SFRs than the LGRB and CCSN hosts. Only the difference between the CCSN and SLSN hosts is statistically significant (Table~\ref{tab:stats}).

We also consider the specific star formation rate (${\rm sSFR} \equiv {\rm SFR} / M_*$), which is the inverse of the time required to double the stellar mass of a galaxy given its current SFR. Since we cannot constrain the sSFR if we only have upper limits on both the SFR and the stellar mass, only detected galaxies are considered for this analysis (23 hosts). The distributions are shown in Figure~\ref{fig:ssfrdist}. Both at high and low redshifts, the SLSNe show a wide range of sSFRs, with a median of $\sim 2~{\rm Gyr}^{-1}$, corresponding to a characteristic doubling time of $\sim$500~Myr. Again their distribution is statistically indistinguishable from that of the LGRB hosts: applying a Kolmogorov-Smirnov test, we find that the LGRB and SLSN distributions are consistent with each other ($p = 0.55$). The SLSN and CCSN distributions are not ($p = 0.004$), mainly due to the tail of high sSFRs that is not observed in the CCSN hosts.

\begin{figure*}
\begin{center}
\begin{tabular}{cc}
\includegraphics[width=3.5in]{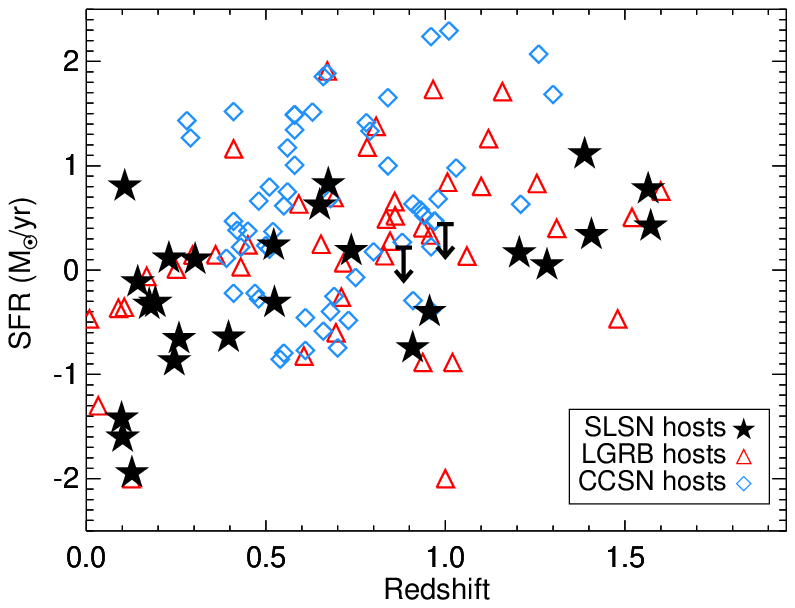} & \includegraphics[width=3.5in]{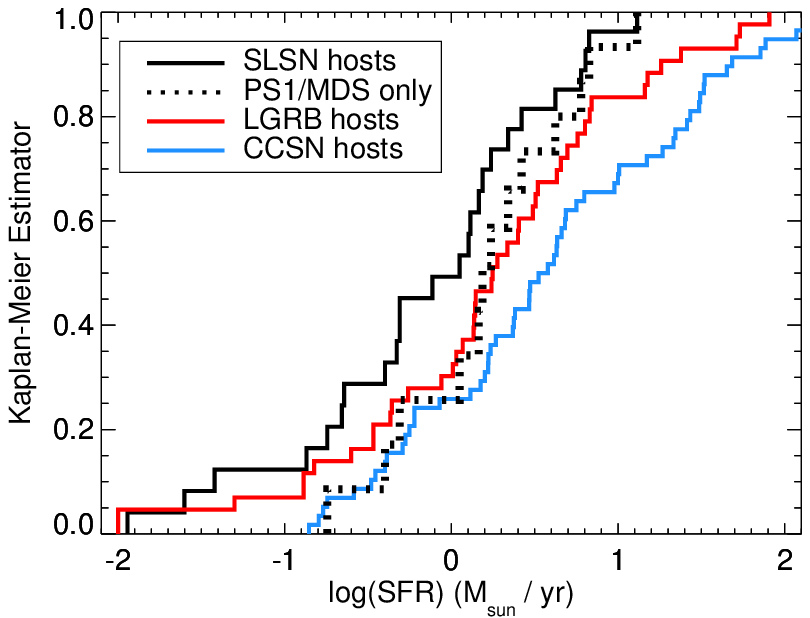}
\end{tabular}
\caption{Left: Star formation rates as a function of redshift for the SLSN host galaxies (black stars and arrows), LGRB host galaxies (red triangles) and GOODS CCSN hosts (blue diamonds). Right: The resulting distribution functions of the three populations. The difference between the SLSN and CCSN hosts is statistically significant, but the difference between LGRB and SLSN hosts is not.
\label{fig:sfrdist}}
\end{center}
\end{figure*}

\begin{figure*}
\begin{center}
\begin{tabular}{cc}
\includegraphics[width=3.5in]{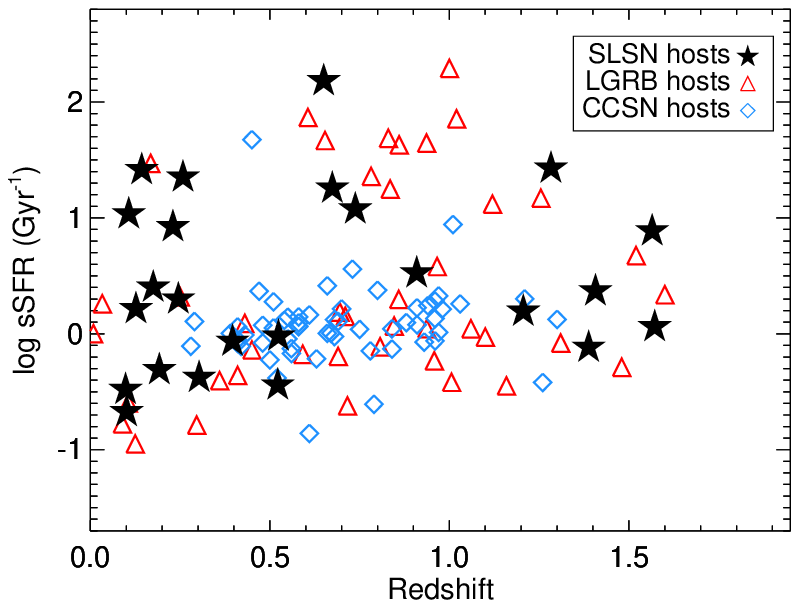} & \includegraphics[width=3.5in]{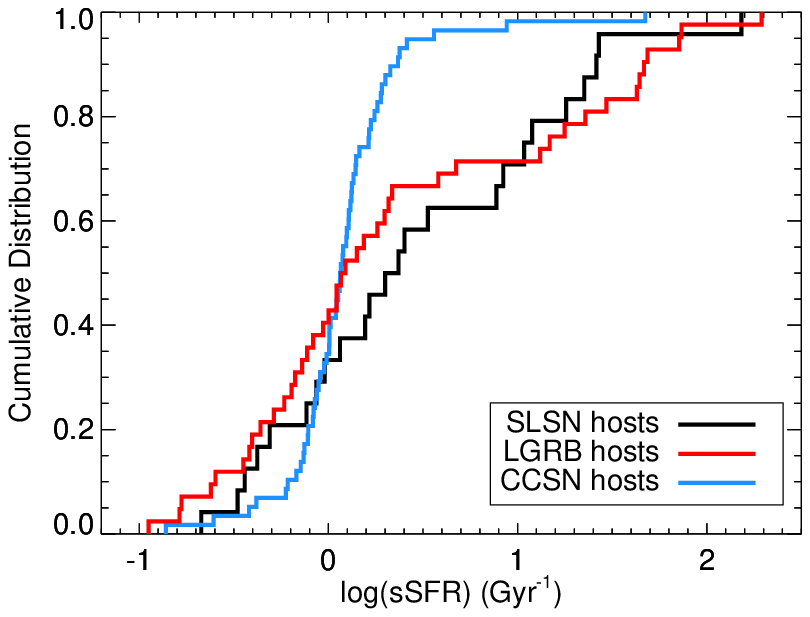}
\end{tabular}
\caption{Left: Specific star formation rates as a function of redshift for the SLSN host galaxies (black stars and arrows), LGRB host galaxies (red triangles) and GOODS CCSN hosts (blue diamonds). Right: The resulting distribution functions of the three populations. As we cannot place limits on the sSFR of undetected objects, only galaxies that are actually detected are plotted here. The three populations have similar medians, but both the LGRB hosts and SLSN hosts show a tail to high specific star formation rates that is not seen in the CCSN host population. 
\label{fig:ssfrdist}}
\end{center}
\end{figure*}

\subsection{H$\beta$ and [\ion{O}{3}] Equivalent Widths}
One striking characteristic of our SLSN host spectra (Figure~\ref{fig:spec}) is the strong nebular emission lines. The equivalent width of H$\beta$ is of particular interest, as it generally decreases monotonically with the age of the young stellar population \citep{cpd86,sv98}. We show the distribution of H$\beta$ EWs in our sample in Figure~\ref{fig:ews}, compared to LGRB hosts \citep{lbk+10,lkb+10} and a sample of star-forming field galaxies at $z \approx 0.3 - 1.0$ from the Team Keck Redshift Survey (TKRS; \citealt{wwa+04,kk04}). The SLSN and LGRB host distributions are similar (a KS test yields $p = 0.75$), indicating the presence of similar-age young stellar populations in the two groups. The comparison to TKRS shows that the  H$\beta$ EWs of the SLSN hosts are also higher than what would be expected if they were drawn from the general field galaxy population, and this remains true also if we weight the field galaxy distribution by star formation rate.

We also note that several of the SLSN hosts exhibit particularly strong [\ion{O}{3}]$\lambda$5007 emission. While the strength of this line is sensitive to a number of physical parameters, including ionization parameter and metallicity, it serves to illustrate how the SLSN host galaxies are different from the normal star-forming field galaxies. The right panel of Figure~\ref{fig:ews} shows the [\ion{O}{3}] equivalent widths measured, compared to the TKRS sample and a sample of Green Pea galaxies from SDSS. The Green Peas are a class of compact, intensely starforming galaxies, originally selected by their unusual colors that is due to extreme [\ion{O}{3}] emission \citep{css+09}. We see that the distribution of [\ion{O}{3}] EWs for SLSN hosts galaxies is clearly skewed towards higher values than what would be expected simply drawing from the star-forming population over this redshift range, with about one third of the SLSN sample showing [\ion{O}{3}] EWs comparable to what is seen in the lower range of Green Pea galaxies.

\begin{figure*}
\begin{center}
\begin{tabular}{cc}
\includegraphics[width=3.5in]{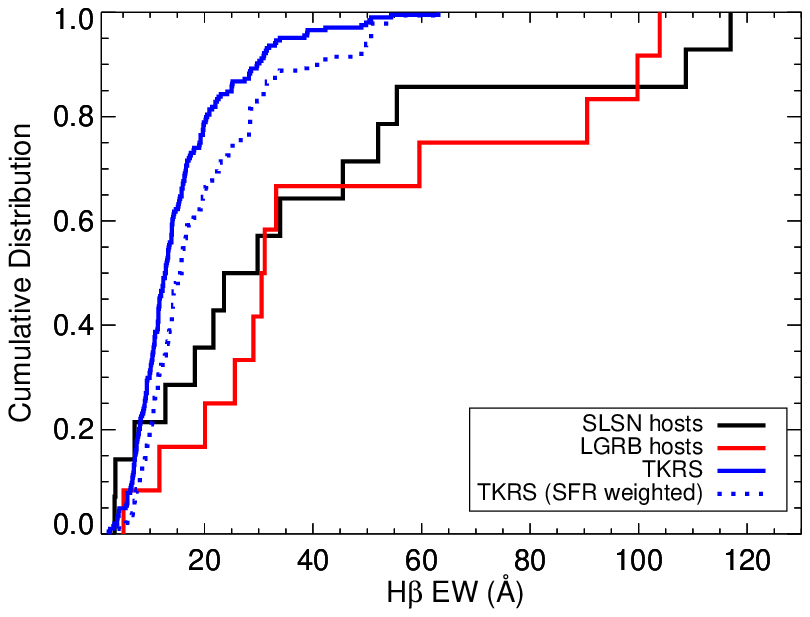} & \includegraphics[width=3.5in]{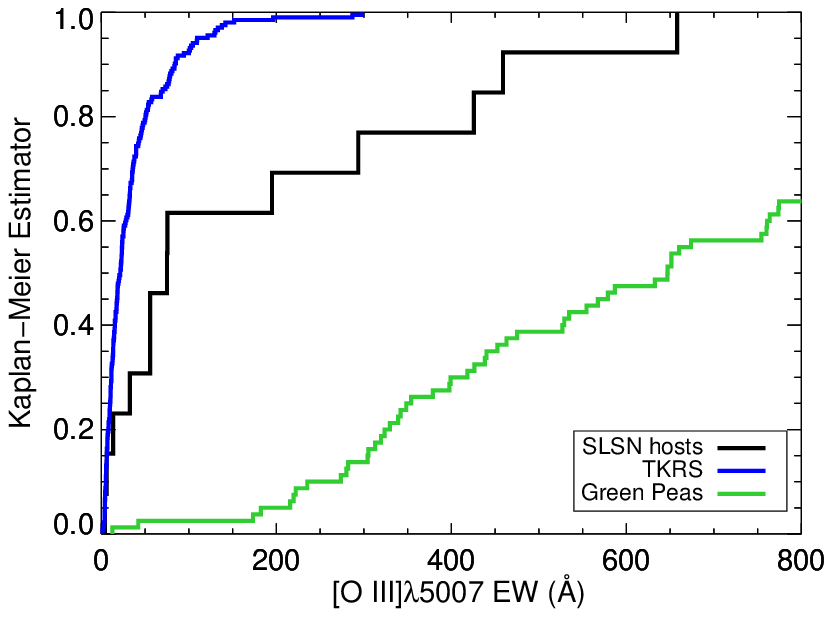}
\end{tabular}
\caption{Left: H$\beta$ equivalent widths for the SLSN host sample (black), a sample of LGRB hosts (red; \citealt{lbk+10,lkb+10}), and 
the Team Keck Redshift Survey (TKRS) sample of field galaxies at $z \approx 0.3 - 1.0$ (blue; \citealt{wwa+04,kk04}). The SLSN host sample and LGRB host sample have a similar H$\beta$ EW distribution, suggesting similar young stellar population ages. Right: [\ion{O}{3}]$\lambda$5007 equivalent widths for our SLSN host sample (black), the TKRS sample, and ``Green Pea'' galaxies from SDSS, a class of compact, intensely star-forming galaxies characterized by extreme [\ion{O}{3}]$\lambda$5007 emission \citep{css+09}. The SLSN hosts generally show much stronger [\ion{O}{3}]$\lambda$5007 emission than the field star-forming galaxies, with about one third of the sample within the Green Pea regime.
\label{fig:ews}}
\end{center}
\end{figure*}

\subsection{Metallicity}
\label{sec:metal}
There are a number of metallicity indicators available in the literature, depending on redshift range and the detected emission lines. However, there are known systematic offsets between them (e.g., \citealt{ke08}). We therefore focus on the R$_{23}$ diagnostic, which is available over the entire redshift range of interest, and we use the calibration in \citet{kk04}. This ensures consistent comparison within the SLSN host sample, and to other galaxy samples using the same diagnostics.

R$_{23}$ is a double-valued diagnostic, and additional information is needed to break the degeneracy between the high-Z and low-Z branches. We accomplish this in either of the following ways. First, if the [\ion{O}{3}]$\lambda$4363 line is detected, we assume the lower-metallicity branch, as this temperature-sensitive line is not present at high metallicities. Second, when detected, we use the ratio of [\ion{N}{2}]$\lambda$6584 to [\ion{O}{2}]$\lambda$3727 (or [\ion{N}{2}]$\lambda$6584 to H$\alpha$, if the reddening is not well constrained) to break the degeneracy. In some cases, [\ion{N}{2}]$\lambda$6584 is not detected, but the upper limit on this ratio  is sufficiently low to allow us to place the host galaxy on the lower metallicity branch. Finally, in some cases the value of R$_{23}$ falls in the turnover region, and either branch gives a value in the range $12 + \log(\rm{O / H}) \sim 8.3-8.6$.  If we cannot formally break the degeneracy, both possible values are listed. However, we note that the low masses of most SLSN host galaxies ($\sim 10^8~{\rm M}_{\odot}$) suggest that the lower branch solution is more likely over the supersolar metallicity given by the upper-branch solution. Indeed, of the eight galaxies where we can robustly break the degeneracy, only one (MLS121104) is found to lie on the upper branch. 

In galaxies where the auroral [\ion{O}{3}]$\lambda$4363 line is detected we can also calculate a ``direct'' metallicity through the electron temperature ($T_e$) method. We use the {\tt temden} task in the IRAF {\tt nebular} package \citep{sd94} to determine the temperature of O$^{++}$ and the electron density ($n_e$), from the ratio of the [\ion{O}{3}] lines and [\ion{S}{2}] lines respectively. The O$^+$ temperature is then calculated assuming the relation from \citet{sta82}. Finally, we determine O$^+/$H and O$^{++}/$H using the relations in \citet{skc06}. Four galaxies in our sample have detected [\ion{O}{3}]$\lambda$4363 emission: PS1-10bzj \citep{lcb+13}, SN\,2010gx \citep{csb+13}, SN\,2011ke and PTF12dam. The host of PTF12dam exhibits both auroral [\ion{O}{3}] and [\ion{O}{2}] lines; a detailed analysis of this host will be presented in Chen et al. (in preparation). 

The distribution of R$_{23}$ metallicities is plotted in Figure~\ref{fig:metal_dist}. As we cannot formally break the R$_{23}$ degeneracy in a number of cases, the dotted and dashed lines show what the distribution would be if we assumed all upper-branch or all lower-branch solutions for these galaxies. The solid lines assume the lower branch solution for host galaxies with a stellar mass lower than $10^8~{\rm M}_{\odot}$, and an equal probability of lower/upper branch solutions for the remaining objects. Taking this as the best estimate of the true distribution, we find a median metallicity of 8.35 ($\approx 0.45 Z_{\odot}$). Also shown in Figure~\ref{fig:metal_dist} are LGRB hosts, and hosts of Type Ib/c and Ic-BL (broad-lined) SNe from untargeted surveys \citep{ssl+12}. The SLSN host metallicity distribution is statistically consistent with that of the LGRB hosts and inconsistent with the SN Ib/c hosts, which are generally found at higher metallicities. We note that the SN samples shown here are local (median redshift $\langle z \rangle \approx 0.036$), as the GOODS CCSN sample does not have metallicity measurements.

Figure~\ref{fig:mz_r23} shows the SLSN hosts with metallicity measurements on a mass-metallicity (M-Z) diagram, compared to LGRB host galaxies, the local M-Z relation from SDSS \citep{thk+04}, local core-collapse SN host galaxies from the compilation of \citet{kk12}, and a sample of emission-line selected galaxies at redshift $z \sim 0.6-0.7$ \citep{hmf+13}. The SLSN hosts are predominantly found at low masses and metallicities, although there is clearly a wide range -- the host of MLS121104, for example, has well detected [\ion{N}{2}] lines that place it on the upper branch in the R$_{23}$ diagnostic, at approximately solar metallicity. This shows that any metallicity preference in producing SLSNe does not take the form of an absolute cutoff; the same is true for LGRBs (e.g., \citealt{lkg+10}).

It is also interesting to note that two of the host galaxies with the highest measured metallicities (MLS121104 and PTF11rks) also exhibit some of the largest offsets from the galaxy center to the SN explosion site (Figure~\ref{fig:galpix}). If there are metallicity gradients present in these hosts (e.g., \citealt{zkh94, mkb+11}), it is still possible that the SNe exploded in an environment with a lower metallicity, closer to the median of the SLSN host sample. Indeed, when comparing line ratios along the slit in the host of PTF11rks, we do find indications of a decreasing R$_{23}$ ratio in an extraction region in the outskirts compared to at the center of the galaxy. However, the poor signal-to-noise ratio in the H$\beta$ line prevents us from making a more quantitative statement. We also note that the majority of our galaxies are of such a small angular size (Figure~\ref{fig:galpix}) that in most cases there is little practical difference between metallicity determined for the galaxy as a whole compared to the explosion site.

\begin{figure}
\begin{center}
\includegraphics[width=3.5in]{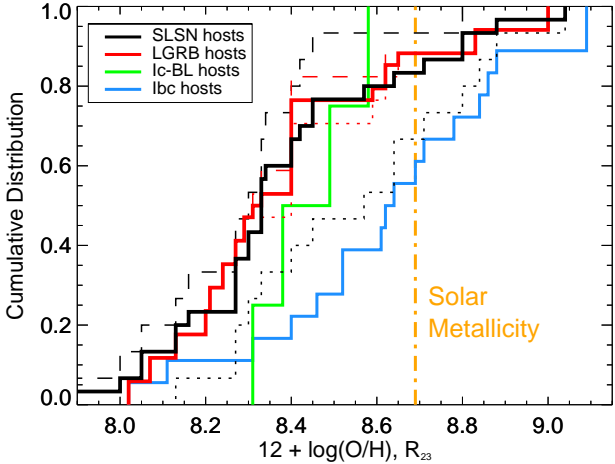}
\caption{Metallicity distribution of the SLSN host galaxies (black) and LGRB host galaxies (red). Also shown are hosts of Type Ic-BL (green) and Ib/c (blue) SNe from untargeted surveys \citep{ssl+12}. For a number of the SLSN hosts, we cannot formally break the R$_{23}$ degeneracy; the dashed and dotted line shows the resulting distributions if we assume that all of the hosts reside on the lower or upper branches, respectively. The solid line is the resulting distribution when assuming hosts with a stellar mass $\lesssim 10^8~ {\rm M}_{\odot}$ fall on the low-metallicity branch, and assigning equal probability to the upper/lower branch solutions for the rest. This distribution is statistically consistent with the LGRB host galaxies, but not with the Type Ib/c SN hosts.
\label{fig:metal_dist}}
\end{center}
\end{figure}

\begin{figure}
\begin{center}
\includegraphics[width=3.5in]{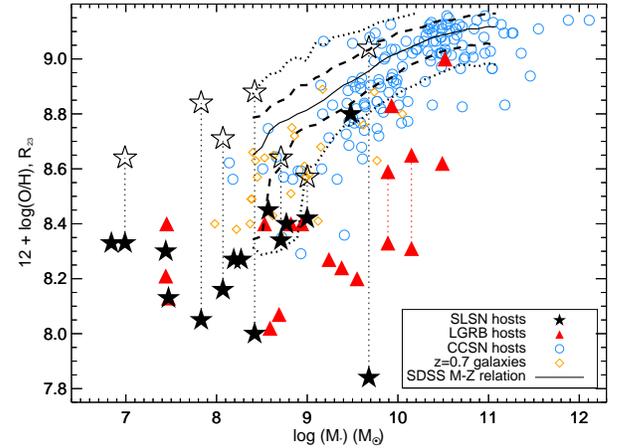}
\caption{Mass-metallicity diagram comparing the SLSN hosts (black stars) to LGRB hosts (red triangles), local CCSN hosts (blue circles), the SDSS M-Z relation (black lines), and a galaxy sample at redshift $z \sim 0.7$ (orange diamonds). All metallicities are on the \citet{kk04} scale to facilitate comparison. Points joined by lines represent cases where the R$_{23}$ degeneracy could not be formally resolved, and so both the upper- and lower-branch solutions are plotted.
\label{fig:mz_r23}}
\end{center}
\end{figure}

\section{Implications for SLSN progenitors}
\label{sec:disc}

We have shown that the H-poor SLSNe are preferentially found in low-luminosity, low-mass, low-metallicity hosts with high sSFR and evidence for very young stellar populations based on line EWs. However, as these properties are found to be correlated in the general galaxy population, it is not clear which is the driving factor in producing SLSNe. This is an on-going debate regarding LGRBs, and many of the same arguments are relevant to the SLSNe.

A number of factors point toward metallicity being a key ingredient in producing both H-poor SLSNe and LGRBs. They overall show a preference for low-metallicity environments compared to CCSNe as well as a preference for faint, blue irregular galaxies \citep{fls+06, sgb+06,mkk+08}. If star formation were the only factor required for producing SLSNe, we would expect them to also occur in star-forming regions of more massive galaxies, and so their galaxy distribution to be more similar to the GOODS CCSN sample.  We also note that the potential redshift evolution we see in our SLSN host sample is consistent with a metallicity-based selection: since the mass-metallicity relation evolves with redshift, shifting to lower metallicities for a given stellar mass at higher redshift (e.g., \citealt{zgk+13} and references therein), we expect a trend toward lower-mass galaxies at lower redshift for a given metallicity. This is indeed what we observe for the SLSN hosts (Figures~\ref{fig:mbdist} and \ref{fig:massdist}).

On the other hand, we do observe a range of metallicities in the SLSN host galaxies, and we do not find evidence of a metallicity cutoff; the same is true for LGRB hosts (e.g., \citealt{lkb+10, lkg+10}). While there are LGRB hosts at higher metallicities, as a population they tend to fall below the local M-Z relation (Figure~\ref{fig:mz_r23}). It has been argued that this could be a result of a proposed anticorrelation between SFR and metallicity at a given stellar mass; the driving factor then would be star formation rather than metallicity \citep{mcm+10,msc11,kw11}. However, even when taking into account the LGRBs in heavily dust-obscured galaxies, the number of LGRBs in massive galaxies still falls short of what would be expected in a purely star formation-selected sample \citep{plt+13}, suggesting that the LGRB rate is also a function of metallicity. A similar argument can be made for SLSNe: while it is not clear whether they fall below the M-Z relation (Figure~\ref{fig:mz_r23}), they do exclusively populate the low-mass end of this diagram. Regardless of whether they are low-metallicity for their mass, then, they are clearly not simply following the star-forming population.

In terms of progenitor models, a low-metallicity environment preference could be linked to a requirement for high angular momentum in the core. Rotation is thought to be the link between LGRBs and metallicity from the theoretical side, where the GRB is a result of accretion onto a newly formed black hole, following the collapse of a rapidly rotating, massive star (e.g., \citealt{mw99}). Higher metallicities are associated with increased mass loss through stellar winds \citep{vd05} which strips the core of angular momentum, and so it has been proposed that the observed preference for low-metallicity environments for LGRBs is linked to the need to maintain high rotation \citep{yl05, ln06}. A similar argument can be applied to the H-poor SLSNe in the scenario where the energy source is a magnetar: in order to reproduce the observed timescales and luminosities rapid initial neutron star spin is required, as well as a strong magnetic field \citep{kb10,ccs+11,lcb+13,isj+13}. Alternatively, the central engine could also be a black hole, where large core angular momentum could allow material which remains bound in the explosion to form an accretion disk and inject energy into the supernova \citep{dk13}. However, this line of reasoning does not explain how the SLSN progenitors shed their hydrogen envelopes and why that mechanism would not remove angular momentum; this is a puzzle also regarding LGRBs, which are associated with Ic-BL SNe.

We note that while SLSNe and LGRBs seem to be found in similar environments, it does not follow that their progenitors must share common properties, but rather that the environmental causes for producing massive stars that end their lives as a LGRB or SLSN are likely similar. For example, if the high-mass end of the initial mass function (IMF) varies with environment (e.g., \citealt{bcm10, kwp+12} and references therein), that could potentially explain the trends we see without needing to invoke a metallicity dependence. Another possibility is dynamical effects: \citet{vp13} speculate that both LGRBs and SLSNe are end products of different dynamical processes in young, dense star clusters, with SLSNe being the result of runaway stellar collisions -- our findings at least support their premise that both SLSNe and LGRBs are associated with young star-forming regions. 

Due to the expected suppression of stellar winds in low-metallicity progenitors, one might initially expect that an interaction model would be harder to explain in a low-metallicity context. However, the mass loss required to explain the observed light curves of SLSNe is too large to be explained by line-driven stellar winds \citep{ci11, ccs+11, lcb+13}. A proposed alternative mechanism for ejecting the necessary mass shells is a pulsational pair-instability (e.g., \citealt{wbh07,cw12}), a phenomenon that may be sensitive to both the rotation and metallicity of the progenitor. A binary star channel has also been proposed \citep{che12}, where the mass loss is driven by common envelope evolution of a compact object within the envelope of a massive star, and the SN itself is triggered by inspiral of the compact object to the core of the companion star, though it is not clear why such a channel would be environment-dependent as we are finding in this work. We note that whatever the scenario, at the very least our findings suggest that if H-poor SLSNe are powered by strong circumstellar interaction, the mechanism that causes the mass loss is likely to be operating preferentially in low-metallicity environments.

\subsection{Possible Selection Effects}
\label{sec:sel}

\subsubsection{Extinction}
While we have taken care to compare events from untargeted surveys over a similar redshift range, one might worry that selection effects could still be driving the differences we see between the SLSN hosts and the other galaxy populations. One such effect is that the SLSN host sample is likely to be biased against host galaxies with high extinction, since it is selected for hosting a population of blue optical transients. This is consistent with what we find in our SED fits, in that virtually all the host galaxies in our sample are consistent with zero or moderate extinction ($\lesssim 0.5~{\rm mag}$). This may partially explain the marginally significant difference in galaxy luminosities seen between the SLSN and LGRB hosts, since LGRBs are selected via gamma-rays and therefore much less sensitive to dust extinction. The mid-IR transient SDWFS-MT-1 was proposed to be a dust-enshrouded SLSN (though of unknown type; \citealt{kks+10}), suggesting that there may exist a population of these objects in obscured environments that current optical surveys are missing. This would only impact our result if such a population was hosted in significantly different galaxies, however; in this one known case, the host was still a low-metallicity dwarf galaxy.

Extinction is unlikely to explain the difference between the SLSN hosts and the CCSN sample though, as this sample is also selected optically and would suffer from a similar extinction bias.  Conversely, if the SLSN hosts were indeed drawn from the same population as CCSN hosts, it would mean that current surveys are only detecting a small fraction of the SLSNe -- only $\sim 15$\% of the GOODS CCSNe were found in galaxies fainter than $M_B = -17.3~{\rm mag}$, the median of the SLSN host galaxy sample. As explored in Section~\ref{sec:followup}, we do not consider this to be a likely scenario.

\subsubsection{Incomplete Follow-Up}
\label{sec:followup}
We also note that while all the samples we are comparing come from untargeted surveys, the spectroscopic follow-up is not complete. In selecting the subsample of objects that will be followed up and confirmed spectroscopically, both light curve and host galaxy properties are typically considered, and this could therefore introduce biases based on host galaxy properties. Here, we can only address how the PS1/MDS sample was selected. The targets included in this paper were primarily chosen for spectroscopic follow-up by some combination of long rise times and/or being significantly brighter than any apparent host. The former effect arises both due to the intrinsically long rise of many SLSNe, as well as due to time dilation since the redshift distribution of the PS1/MDS sample peaks at $z \approx 1$. This selection could bias us toward lower-luminosity galaxies if faster-rising SLSNe were preferentially found in brighter host galaxies; exploring any such correlations is, however,  outside the scope of this paper.

The second effect of preferentially following up faint-host/high-contrast objects is potentially more problematic for the results presented here. We know that our sample is not complete -- the question is whether we missed objects systematically due to galaxy properties, and if so, whether the effect is large enough to influence our results. To quantify this, we carried out a number of tests. First, if we assume that the true distribution of SLSN host galaxies was that of the GOODS CCSN host galaxies, we can simulate the effects of a (crude) selection bias by excluding all the CCSN host galaxies brighter than a given magnitude, and ask what fraction of objects we must be missing. We find that to make the GOODS CCSN host galaxy mass distribution marginally consistent with that of the SLSN host galaxies, we must set the cut at $V = 23.5~{\rm mag}$, excluding half of the GOODS sample. This implies that if selection effects were the only driver behind our result, we should have systematically missed about half of the SLSNe in PS1/MDS. 

To quantify how many such potential bright-host SLSNe we could have missed, we searched the entire PS1/MDS photometric database for transients that had good-quality light curves (bright enough to be considered for spectroscopy) and long observed rise times, but that were not selected for spectroscopic follow-up. We excluded from this sample any object that had a light curve consistent with a Type Ia SN, as determined by PSNID \citep{sbc+11}. We tuned our cuts such that the search would let through all the PS1 SLSNe where the rise is observed, with the exception of the fastest-rising objects such as PS1-10bzj \citep{lcb+13}. Applying the same light curve cuts to the spectroscopic subsample, we found that in addition to SLSNe the main group of objects making it through these cuts are Type II SNe at lower redshifts ($z < 0.5$).

For the time period we considered, this left us with 17 long-rising transients without spectroscopic classification. 13 of these have host galaxies that are well-detected in SDSS, and thus have photometric redshifts available \citep{olc+08}. If we adopt these redshifts, the median implied peak absolute magnitude of these transients is $-18~{\rm mag}$, consistent with being the counterparts to the slow-rising Type II SNe seen in the spectroscopic control sample, and in particular unlikely to be missed SLSNe. This leaves us with only four candidates of unknown type/redshift, which already rules out having missed a considerable number of bright-host SLSNe simply due to a bias in the follow-up. Two of these four objects were actually targeted for spectroscopy, but the spectra were inconclusive, showing a blue and featureless continuum. It is unlikely that these objects belong to the subclass studied in this paper - if they were indeed superluminous, they would be at sufficiently high redshift that we would expect to detect the characteristic broad UV absorption features seen in most H-poor SLSNe \citep{qkk+11,ccs+11}.

In addition, these four remaining objects are all found in host galaxies fainter than $22.0~{\rm mag}$ (two have undetected hosts) and so would not be bright-host SLSNe: if we assign them redshifts by assuming the transients were indeed SLSNe (i.e., that the transient light curves peaked at $M = -22.5~{\rm mag}$), adding them to the sample considered in this paper does not change any of our conclusions.  Therefore, we are confident that our results are due to a real effect, rather than a bias towards preferentally following up transients with faint host galaxies.

We note that the preference for low-luminosity hosts is even stronger in the low-redshift non-PS1/MDS sample. This is reassuring, in the sense that the same general trend is found independently by more than one survey, which is certainly a necessary condition for it being a real physical effect. The stronger preference for low-luminosity galaxies at lower redshifts can be interpreted as an evolutionary effect, that may come about if for example metallicity affects the SLSN rate. Without a better understanding of how the different surveys select targets for follow-up, disentangling any selection effects from redshift evolution will be difficult, however.

\section{Conclusions}
\label{sec:conc}

We have presented the first comprehensive study of the host galaxy environments of H-poor SLSNe, with 31 objects over the redshift range $z \approx 0.1-1.6$. This is the first study to look at the hosts of this subclass of SLSNe specifically, and the largest study of SLSN hosts so far: previous studies \citep{nsg+11,sps+11} mixed both H-rich and H-poor SLSNe and only detected a few hosts of H-poor SLSNe. Our main findings can be summarized as follows:

\begin{itemize}
\item H-poor SLSNe are generally found in low-luminosity galaxies. In our sample, we find the following median properties: $B$-band luminosity of $-17.3$~mag, stellar mass of $\sim 2 \times 10^8~{\rm M}_{\odot}$, star formation rate of $\sim 1~ {\rm M}_{\odot}~{\rm yr}^{-1}$ and specific star formation rate of $\sim 2~{\rm Gyr}^{-1}$.

\item Compared to the hosts of core-collapse SNe over the same redshift range, the SLSNe occur systematically in lower-luminosity, lower-mass, lower-metallicity and higher sSFR galaxies. These results are statistically significant at the $> 3\sigma$ level.

\item Compared to the hosts of LGRBs over the same redshift range, the SLSNe are consistent with being drawn from the same galaxy population as GRBs in terms of stellar mass, SFR, sSFR, and metallicity; we do, however, find them in lower-luminosity and lower-mass galaxies particularly at low redshift.

\item The SLSNe predominantly occur in low-metallicity galaxies, with a median value of $12 + \log({\rm O/H})\approx 8.35$ and four galaxies in the sample having a detected [\ion{O}{3}]$\lambda 4363$ emission line. However, we do find a range of metallicities, including a host galaxy at solar metallicity, and so there is no evidence for a strict metallicity cutoff.

\item The preference for low-luminosity galaxies is strongest in the low-redshift ($z \lesssim 0.5$) sample, suggesting that there could be redshift evolution in the host population. A better understanding of how this sample was selected is necessary to disentangle evolution effects and potential selection effects, however.

\end{itemize}

We have shown that SLSNe select host environments that are similar to those selected by LGRBs over the same redshift range, though seem to prefer even lower-luminosity galaxies. As is the case with LGRBs, the implications in terms of SLSN progenitors are not straightforward. However, if interpreted as a preference for low-metallicity environments as the effect driving the selection, this could lend support to a millisecond magnetar being the energy source powering SLSNe. A key component of this progenitor model is that the magnetar must initially be spinning at close to breakup speeds, and maintaining fast rotation in the core is thought to be more effective at low metallicities since less angular momentum is lost to line-driven stellar winds. 

It is less clear how our findings could be interpreted in the context of an interaction model for powering SLSNe, but our results at least indicate that the mechanism responsible for mass loss is likely to be environment-dependent. It would be interesting to compare the results to our study to the host galaxies of hydrogen-rich (Type IIn) SLSNe, since these SLSNe do show clear signs of interaction in their SN spectra. If their host population is found to be similar to the H-poor SLSN hosts, this could point to a similar progenitor population for the two classes. 

\acknowledgements
We thank the staffs at PS1, MMT and Magellan for their assistance with performing these observations, and Andy Monson for help with processing the FourStar data. The Pan-STARRS1 Surveys (PS1) have been made possible through contributions of the Institute for Astronomy, the University of Hawaii, the Pan-STARRS Project Office, the Max-Planck Society and its participating institutes, the Max Planck Institute for Astronomy, Heidelberg and the Max Planck Institute for Extraterrestrial Physics, Garching, The Johns Hopkins University, Durham University, the University of Edinburgh, Queen's University Belfast, the Harvard-Smithsonian Center for Astrophysics, the Las Cumbres Observatory Global Telescope Network Incorporated, the National Central University of Taiwan, the Space Telescope Science Institute, the National Aeronautics and Space Administration under Grant No. NNX08AR22G issued through the Planetary Science Division of the NASA Science Mission Directorate, the National Science Foundation under Grant No. AST-1238877, the University of Maryland, and Eotvos Lorand University (ELTE). Support for programs number GO-13022 and GO-13326 was provided by NASA through a grant from the Space Telescope Science Institute, which is operated by the Association of Universities for Research in Astronomy, Inc., under NASA contract NAS5-26555. This paper includes data gathered with the 6.5 m Magellan Telescopes located at Las Campanas Observatory, Chile. Some observations reported here were obtained at the MMT Observatory, a joint facility between the Smithsonian Institution and the University of Arizona. This paper includes data based on observations made with the NASA/ESA {\it Hubble Space Telescope} and obtained from the Hubble Legacy Archive, which is a collaboration between the Space Telescope Science Institute (STScI/NASA), the Space Telescope European Coordinating Facility (ST-ECF/ESA) and the Canadian Astronomy Data Centre (CADC/NRC/CSA). This work is based in part on observations made with the {\it Spitzer Space Telescope}, which is operated by the Jet Propulsion Laboratory, California Institute of Technology under a contract with NASA.This research used the facilities of the Canadian Astronomy Data Centre operated by the National Research Council of Canada with the support of the Canadian Space Agency. Some of the computations in this paper were run on the Odyssey cluster supported by the FAS Science Division Research Computing Group at Harvard University. Partial support for this work was provided by National Science Foundation grants AST-1009749 and AST-1211196.

{\it Facilities:} \facility{PS1}, \facility{MMT}, \facility{Magellan:Baade}, \facility{Magellan:Clay}, \facility{HST}.

\begin{deluxetable*}{lccccr}
\tablewidth{0pt}
\tablecaption{H-Poor SLSN Sample\label{tab:gallist}}
\tablehead{
   \colhead{SN Name} & \colhead{Redshift} & \colhead{RA} &
   \colhead{Dec} & \colhead{E(B-V)\tablenotemark{a}} & \colhead{Reference}  \\
   \colhead{} & \colhead{} & \colhead{(J2000)} & \colhead{(J2000)} &
   \colhead{(mag)} & \colhead{} }

\startdata
PTF10hgi   &  $0.098$   & \ra{16}{37}{47.04}   & \dec{+06}{12}{32.3}  & 0.074 &  1,2 \\
SN\,2010kd &  $0.101$   & \ra{12}{08}{00.89}   & \dec{+49}{13}{32.88} & 0.021 & 3,4  \\
PTF12dam   &  $0.107$   & \ra{14}{24}{46.20}   & \dec{+46}{13}{48.3}  & 0.010 & 5,6 \\
SN\,2007bi &  $0.127$   & \ra{13}{19}{20.14}   & \dec{+08}{55}{43.7}  & 0.024 & 7     \\
SN\,2011ke &  $0.143$   & \ra{13}{50}{57.77}   & \dec{+26}{16}{42.8}  & 0.011 & 2,8 \\  
SN\,2012il &  $0.175$   & \ra{09}{46}{12.91}   & \dec{+19}{50}{28.7}  & 0.019 & 2,9   \\
PTF11rks   &  $0.192$   & \ra{01}{39}{45.64}   & \dec{+29}{55}{27.0}  & 0.038 & 2,10  \\
SN\,2010gx &  $0.23 $   & \ra{11}{25}{46.71}   & \dec{-08}{49}{41.4}  & 0.035 & 11,12  \\
SN\,2011kf &  $0.245$   & \ra{14}{36}{57.34}   & \dec{+16}{30}{57.14} & 0.020 & 2,8,13 \\
PTF09cnd   &  $0.258$   & \ra{16}{12}{08.94}   & \dec{+51}{29}{16.1}  & 0.021 & 12  \\
SN\,2005ap &  $0.283$   & \ra{13}{01}{14.83}   & \dec{+27}{43}{32.3}  & 0.008 & 14  \\
MLS121104:021643+204009\tablenotemark{b} & $0.303$  & \ra{02}{16}{42.51} & \dec{+20}{40}{08.47} & 0.150 & 15,16  \\
PTF09cwl   &  $0.349$   & \ra{14}{49}{10.08}   & \dec{+29}{25}{11.4}  & 0.014 & 12  \\
SN\,2006oz &  $0.396$   & \ra{22}{08}{53.56}   & \dec{+00}{53}{50.4}  & 0.041 & 17  \\
PTF09atu   &  $0.501$   & \ra{16}{30}{24.55}   & \dec{+23}{38}{25.0}  & 0.042 & 12  \\
PS1-12bqf  &  $0.522$   & \ra{02}{24}{54.621}  & \dec{-04}{50}{22.72} & 0.025 & 18 \\
PS1-11ap   &  $0.524$   & \ra{10}{48}{27.752}  & \dec{+57}{09}{09.32} & 0.007 & 19 \\
PS1-10bzj  &  $0.650$   & \ra{03}{31}{39.826}  & \dec{-27}{47}{42.17} & 0.007 & 20 \\
PS1-12zn   &  $0.674$   & \ra{09}{59}{49.615}  & \dec{+02}{51}{31.85} & 0.019 & 18 \\
PS1-11bdn  &  $0.738$   & \ra{02}{25}{46.292}  & \dec{-05}{06}{56.57} & 0.025 & 18 \\
PS1-13gt   &  $0.884$   & \ra{12}{18}{02.035}  & \dec{+47}{34}{45.95} & 0.015 & 18 \\
PS1-10awh  &  $0.909$   & \ra{22}{14}{29.831}  & \dec{-00}{04}{03.62} & 0.070 & 21 \\
PS1-10ky   &  $0.956$   & \ra{22}{13}{37.851}  & \dec{+01}{14}{23.57} & 0.031 & 21 \\
PS1-11aib  &  $0.997$    & \ra{22}{18}{12.217}  & \dec{+01}{33}{32.01} & 0.044 & 18 \\
SCP\,06F6  &  $1.189$   & \ra{14}{32}{27.395}  & \dec{+33}{32}{24.83} & 0.009 & 22  \\
PS1-10pm   &  $1.206$   & \ra{12}{12}{42.200}  & \dec{+46}{59}{29.48} & 0.016 & 23 \\
PS1-11tt   &  $1.283$   & \ra{16}{12}{45.778}  & \dec{+54}{04}{16.96} & 0.008 & 18 \\
PS1-10afx  &  $1.388$   & \ra{22}{11}{24.160}  & \dec{+00}{09}{43.49} & 0.048 & 24 \\
PS1-11afv  &  $1.407$   & \ra{12}{15}{37.770}  & \dec{+48}{10}{48.62} & 0.014 & 18 \\
PS1-11bam  &  $1.565$   & \ra{08}{41}{14.192}  & \dec{+44}{01}{56.95} & 0.024 & 25 \\ 
PS1-12bmy  &  $1.572$   & \ra{03}{34}{13.123}  & \dec{-26}{31}{17.21} & 0.015 & 18
\enddata
\tablenotetext{a}{Foreground extinction \citep{sfd98, sf11}.}
\tablenotetext{b}{Referred to as MLS121104 throughout the paper.}
\tablecomments{References: (1) \citet{qko+10}, (2) \citet{isj+13}, (3) \citet{vzr+10}, (4) \citet{qya+13}, (5) \citet{qas+12}, (6) \citet{nsj+13},(7) \citet{gmo+09}, (8) \citet{ddm+11a}, (9) \citet{ddm+12}, (10) \citet{qga+11}, (11) \citet{psb+10}, (12) \citet{qkk+11}, (13) \citet{pdm+12}, (14) \citet{qaw+07}, (15) \citet{ddm+12b}, (16) \citet{fg12}, (17) \citet{lcd+12}, (18) Lunnan et al., in prep., (19) \citet{msk+14}, (20) \citet{lcb+13}, (21) \citet{ccs+11}, (22) \citet{bdt+09}, (23) \citet{msr+14}, (24) \citet{cbr+13}, (25) \citet{bcl+12}.     }
\end{deluxetable*}

\begin{deluxetable}{lccccc}
\tablewidth{0pt}
\tablecaption{Host Galaxy Photometry \& Limits from PS1/MDS Stacks \label{tab:ps1phot}}
\tablehead{
   \colhead{SN Name} & \colhead{\gps} & \colhead{\rps} &
   \colhead{\ips} & \colhead{\zps} & \colhead{\yps} }

\startdata
PS1-10ky   &   $ > 24.7$      &  $ > 24.6$        & $ > 24.5$         & $ > 24.0$        &  $ > 22.1 $  \\
PS1-10awh  &   $ > 25.0$      &  $ > 25.1$	    & $ > 25.3$	 & $ > 24.7$	     & $ > 22.7 $  \\
PS1-11bam  &  $23.63 \pm 0.13$ & $ 23.64 \pm 0.12$  & $ 23.78 \pm 0.13$	 & $ 23.69 \pm 0.14$ & $ > 23.4 $  \\
PS1-10afx  &  $23.84 \pm 0.10$ & $ 23.57 \pm 0.10$  & $ 23.34 \pm 0.13$	 & $ 22.68 \pm 0.10$ & 22.29 $\pm$ 0.28  \\
PS1-10bzj  &  $24.35 \pm 0.08$ & $ 23.98 \pm 0.12$  & $ 23.75 \pm 0.10$	 & $ 22.72 \pm 0.05$ & $ > 21.7 $  \\
PS1-10pm   &  $ > 25.2$       & $ > 25.1$	    & $ > 25.0$	 & $ > 24.0$	     &  $ > 23.0 $  \\
PS1-11ap   &  $24.20 \pm 0.15$ & $ 23.32 \pm 0.10$  & $ 22.86 \pm 0.09$	 & $ 23.24 \pm 0.13$ &  $ > 22.5 $  \\
PS1-11tt   &  $ > 24.6$       & $ > 24.7$	    & $ > 24.8$	 & $ > 24.1$	     &  $ > 23.0 $ \\
PS1-11aib  &  $ > 24.2$       & $ > 24.4$	    & $ > 24.7$	 & $ > 23.9$	     &  $ > 22.2 $ \\
PS1-11afv  &  $ > 24.9$       & $ > 24.8$	    & $ > 25.1$	 & $ > 24.9$	     & $ > 22.8 $  \\
PS1-11bdn  &  $ > 24.8$       & $ > 24.0$	    & $ > 24.9$	 & $ > 23.9$	     &  $ > 22.5 $  \\
PS1-12bmy  &  $ > 24.2$       & $ > 24.2$	    & $ > 24.1$	 & $ > 23.6$	     & $ > 22.3 $  \\
PS1-12zn   & $24.64 \pm 0.10$  & $ 24.07 \pm 0.07$  & $ 23.77 \pm 0.10$	 & $ 23.56 \pm 0.14$ &  $ > 22.5 $  \\
PS1-12bqf  & $22.76 \pm 0.12$  & $ 21.89 \pm 0.06$  & $ 21.44 \pm 0.03$	 & $ 21.40 \pm 0.05$ &  21.46 $\pm$ 0.14 \\
PS1-13gt   &  $ > 24.5$       & $ > 24.5$	    & $ > 24.7$	 & $ > 24.4$        & $ > 22.7 $  
\enddata
\tablecomments{Corrected for foreground extinction. Upper limits are 3$\sigma$.}
\end{deluxetable}

\begin{deluxetable}{lcccc}
\tablewidth{0pt}
\tablecaption{Additional Host Galaxy Photometry \label{tab:morephot}}
\tablehead{
   \colhead{SN Name} & \colhead{Filter} & \colhead{AB mag} & \colhead{Instrument}  & \colhead{UT date}  }
\startdata
PTF10hgi   & $g'$  & 22.56 $\pm$ 0.06 & IMACS  & 2013-05-07 \\
PTF10hgi   & $i'$  & 21.75 $\pm$ 0.06 & IMACS  & 2013-05-07 \\
PTF10hgi   & $z'$  & 21.43 $\pm$ 0.12 & IMACS  & 2013-04-11 \\
PTF10hgi  & $J$  & $ 21.48 \pm 0.08$  & FourStar & 2013-05-20 \\
PTF10hgi  & $K_s$  & $ 21.66 \pm 0.13$  & FourStar & 2013-05-20 \\
SN\,2011ke\tablenotemark{a} & $g'$  & 22.44 $\pm$ 0.10 & CFHT & \ldots \\
SN\,2011ke\tablenotemark{a} & $r'$  & 22.01 $\pm$ 0.10 & CFHT & \ldots \\
SN\,2011ke\tablenotemark{a} & $z'$  & 23.00 $\pm$ 0.30 & IMACS  & 2013-04-11 \\
SN\,2011ke\tablenotemark{a}  & $J$  & $ 22.86 \pm 0.15$  & FourStar & 2013-05-20 \\
SN\,2012il  & $J$  & $ 21.78 \pm 0.11$   & FourStar & 2013-05-19 \\
SN\,2012il  & $K_s$  & $ 21.90 \pm 0.20$   & FourStar & 2013-05-21 \\ 
PTF11rks   & $z'$  & 20.52 $\pm$ 0.10 & LDSS3 & 2013-10-04 \\
PTF11rks   & $K_S$ & 20.75 $\pm$ 0.34 & FourStar & 2013-12-18 \\
SN\,2010gx  & $J$  & $ 22.92 \pm 0.11$  & FourStar & 2012-12-04 \\
SN\,2011kf & $g'$  & 23.74 $\pm$ 0.07 & IMACS & 2013-05-07 \\
SN\,2011kf & $r'$  & 23.15 $\pm$ 0.12 & IMACS  & 2013-05-10 \\
SN\,2011kf & $i'$  & 23.65 $\pm$ 0.33 & MMTCam & 2013-04-29 \\
SN\,2011kf  & $J$  &  $ > 23.1$  & FourStar & 2013-05-22 \\
SN\,2011kf  & $K_s$&  $ > 22.7$  & FourStar & 2013-05-22 \\
PTF09cnd   & $g'$  & 23.75 $\pm$ 0.16 & MMTCam  & 2013-05-02 \\
PTF09cnd   & $r'$  & 23.60 $\pm$ 0.25 & MMTCam  & 2013-03-15 \\
PTF09cnd   & $i'$  & 23.70 $\pm$ 0.27 & MMTCam & 2013-05-02 \\
SN\,2005ap & $i'$  & 23.59 $\pm$ 0.07 & MMTCam & 2014-03-23 \\
MLS121104  & $J$   & 20.39 $\pm$ 0.10 & FourStar & 2013-12-18 \\
MLS121104  & $K_S$ & 19.63 $\pm$ 0.12 & FourStar & 2013-12-18 \\
PTF09cwl   & $r'$  & $ > 24.4 $      & MMTCam & 2013-03-13 \\
SN\,2006oz  & $J$  & $ 23.43 \pm 0.26$  & FourStar & 2012-12-04 \\
PTF09atu   & $r'$  & $ > 25.2 $      & IMACS  & 2013-05-07 \\
PS1-12bqf  & F475W & 22.94 $\pm$ 0.02 & {\it HST}/ACS & 2013-11-18 \\
PS1-12bqf  & $K_S$ & $ > 19.9 $     & FourStar &  2013-12-18   \\
PS1-12bqf   & 3.6$\mu$m & $20.82 \pm 0.06$ & {\it Spitzer}/IRAC  & \ldots \\
PS1-12bqf   & 4.5$\mu$m & $21.29 \pm 0.06$ & {\it Spitzer}/IRAC & \ldots \\ 
PS1-11ap  & F475W  & 24.02 $\pm$ 0.02 & {\it HST}/ACS & 2013-10-09 \\
PS1-11ap   & 3.6$\mu$m & $23.33 \pm 0.39$ & {\it Spitzer}/IRAC & \ldots \\
PS1-11ap   & 4.5$\mu$m & $23.38 \pm 0.29$ & {\it Spitzer}/IRAC & \ldots \\ 
PS1-10bzj   & 3.6$\mu$m & $23.79 \pm 0.16$ & {\it Spitzer}/IRAC & \ldots \\
PS1-10bzj   & 4.5$\mu$m & $24.00 \pm 0.18$ & {\it Spitzer}/IRAC & \ldots \\
PS1-12zn  & $J$  & $ 23.09 \pm 0.25$ & FourStar & 2013-05-20 \\
PS1-12zn  & $K_s$  &  $ > 22.7$ & FourStar & 2013-05-20 \\
PS1-12zn   & 3.6$\mu$m & $23.09 \pm 0.12$ & {\it Spitzer}/IRAC & \ldots \\
PS1-12zn   & 4.5$\mu$m & $24.24 \pm 0.57$ & {\it Spitzer}/IRAC & \ldots \\
PS1-11bdn  & F475W    & $26.09 \pm 0.10$ & {\it HST}/ACS & 2013-11-13 \\
PS1-11bdn  & $r'$  &  $> 25.5 $       & IMACS & 2012-07-19 \\ 
PS1-11bdn  & $i'$  &  $25.40 \pm 0.25$        & LDSS3  & 2013-10-05 \\
PS1-11bdn  & $z'$  &  $> 24.2 $       & LDSS3 & 2013-01-12 \\ 
PS1-11bdn & $J$  &  $ > 24.2$        & FourStar  & 2012-12-04  \\
PS1-10awh  & F606W  &  27.00 $\pm$ 0.20 & {\it HST}/ACS  & 2013-09-04 \\
PS1-10ky   & F606W &   $> 27.4$       & {\it HST}/ACS & 2012-12-13 \\
PS1-10ky   & F850LP &  $> 27.0 $      & {\it HST}/ACS & 2012-12-13 \\
PS1-10pm   & F606W  & $25.38 \pm 0.05$ &  {\it HST}/ACS & 2012-12-10 \\
PS1-10pm   & F110W  &  $24.40 \pm 0.08$ & {\it HST}/WFC3 & 2013-01-15 \\
PS1-11tt   & F606W  & $25.78 \pm 0.08$ &  {\it HST}/ACS & 2012-12-02 \\
PS1-11tt  & F110W  &  $25.83 \pm 0.05$ & {\it HST}/WFC3 & 2013-04-21 \\
PS1-11afv   & F606W  & $25.26 \pm 0.08$ &  {\it HST}/ACS & 2013-04-09 \\
PS1-11afv & F110W  &  $24.65 \pm 0.08$ & {\it HST}/WFC3 & 2012-11-24 \\ 
PS1-11bam  & F814W  & 23.82 $\pm$ 0.02 & {\it HST}/ACS & 2013-10-11 \\
PS1-12bmy  & $g'$  &  25.25 $\pm$ 0.10 & LDSS3  & 2013-10-05 \\
PS1-12bmy  & $r'$  &  25.46 $\pm$ 0.10 & LDSS3  & 2013-10-04 \\
PS1-12bmy  & $i'$  &  25.10 $\pm$ 0.16 & LDSS3  & 2013-10-05 \\
PS1-12bmy  & $z'$  &  24.64 $\pm$ 0.40 & LDSS3  & 2013-10-05 \\
PS1-12bmy  & F814W  & 25.01 $\pm$ 0.05 & {\it HST}/ACS & 2013-09-17 \\
PS1-12bmy  & $J$    & 24.02 $\pm$ 0.21 & FourStar  & 2013-12-18  \\
PS1-12bmy  & $K_s$    & $> 22.2  $       & FourStar  & 2013-12-18
\enddata
\tablenotetext{a}{Flux from dwarf galaxy host only; see Section~\ref{sec:11xk} for details.}
\tablecomments{Corrected for foreground extinction. Upper limits are 3$\sigma$.}
\end{deluxetable}

\begin{deluxetable*}{lcccccccc}
\tabletypesize{\scriptsize}
\tablecaption{Log of Host Galaxy Spectroscopic Observations}
\tablehead{
\colhead{Object}  &
\colhead{UT Date} &
\colhead{Instrument} &
\colhead{Wavelength Range} &
\colhead{Slit} &
\colhead{Grating} &
\colhead{Filter} &
\colhead{Exp. time} &
\colhead{Mean} \\
\colhead{} &
\colhead{(YYYY-MM-DD.D)} &
\colhead{} &
\colhead{(\AA)} &
\colhead{($\arcsec$)} &
\colhead{} &
\colhead{} &
\colhead{(s)} &
\colhead{Airmass}
}
\startdata
PTF10hgi     & 2013-04-11.3 & IMACS & 4000-10270  & 0.9 & 300-17.5     & none & 3600 & 1.22 \\ 
PTF10hgi     & 2013-05-07.4 & IMACS & 4000-10270  & 0.9 & 300-17.5     & none & 4200 & 1.75 \\ 
SN\,2010kd   & 2013-05-13.3 & BlueChannel & 3330-8550 & 1 & 300GPM   & none & 1800 & 1.23 \\
PTF12dam     & 2013-07-13.3 & BlueChannel &  3300-8530  & 1   & 300GPM     & none & 1800 & 1.57   \\
SN\,2011ke\tablenotemark{a}   & 2013-04-11.3 & IMACS & 4000-10270  & 0.9 & 300-17.5     & none & 2400 & 1.80 \\ 
SN\,2012il   & 2013-04-15.2 & BlueChannel & 3350-8570   & 1   & 300GPM     & none & 1800 & 1.15   \\
SN\,2011kf   & 2013-05-10.2 & IMACS & 4000-10270  & 0.9    &  300-17.5    & none & 1800 & 1.44 \\
SN\,2011kf   & 2013-06-03.1 & IMACS & 4000-10270  & 0.9 & 300-17.5   & none & 5400 & 1.43 \\
PTF11rks     & 2013-10-05.3 & LDSS3 &  3900-10000     &  1 &  VPH-All  & none   &  1800 & 1.95 \\ 
MLS121104    & 2013-07-12.4  & LDSS3  & 4080-10720  & 1  & VPH-All & none & 3000  & 1.77  \\
SN\,2006oz   & 2013-07-12.3 & LDSS3       & 4080-10720            & 1   & VPH-All    & none & 5400 & 1.23       \\
PS1-12zn     & 2013-01-10.3 & LDSS3 & 5850-9970   & 1 & VPH-Red  & OG590  & 5000 & 1.18 \\
PS1-12bqf    & 2013-10-05.2 & LDSS3 & 5310-9970   & 1 & VPH-Red  & none   & 3600 & 1.19 \\
PS1-11bdn    & 2013-01-13.1 & LDSS3 & 5850-9970   & 1 & VPH-Red  & OG590  & 5400 & 1.39 
\enddata
\tablenotetext{a}{Taken with the slit oriented through a nearby galaxy; see Section~\ref{sec:11xk} for details. Note that IMACS has an atmospheric dispersion corrector, so that observing away from parallactic angle does not affect relative line fluxes.}
\label{tab:spec}
\end{deluxetable*}

\begin{deluxetable*}{lcccccccc}
\tabletypesize{\scriptsize}
\tablecaption{Raw Measured Emission Line Fluxes ($10^{-15} {\rm~erg~s}^{-1}~{\rm cm}^{-2}$)\label{tab:lines} }
\tablehead{
\colhead{Object} & 
\colhead{[\ion{O}{2}]$\lambda$3727} &
\colhead{H$\gamma$} &
\colhead{[\ion{O}{3}]$\lambda$4363} &
\colhead{H$\beta$} &
\colhead{[\ion{O}{3}]$\lambda$4959} &
\colhead{[\ion{O}{3}]$\lambda$5007} &
\colhead{H$\alpha$} &
\colhead{[\ion{N}{2}]$\lambda$6584} 
}
\startdata
PTF10hgi     & \ldots  & \ldots  & \ldots  & 0.020 $\pm$ 0.006 & \ldots  & 0.024 $\pm$ 0.005 & 0.083 $\pm$ 0.008  & $< 0.01$  \\ 
SN\,2010kd   & 0.068 $\pm$ 0.013 & 0.033 $\pm$ 0.025  & \ldots   & 0.065 $\pm$ 0.011 & 0.106 $\pm$ 0.016 &  0.293 $\pm$ 0.013 & 0.120 $\pm$ 0.010  &  $< 0.012 $  \\ 
PTF12dam     & 12.12 $\pm$ 0.11 & 3.47 $\pm$ 0.07  & 0.67 $\pm$ 0.05 & 8.07 $\pm$ 0.08 & 15.82 $\pm$ 0.13  & 47.19 $\pm$ 0.24 & 24.20 $\pm$ 0.18 & 0.82 $\pm$ 0.06 \\ 
SN\,2011ke   & 0.88 $\pm$ 0.07 & 0.30 $\pm$ 0.02 & 0.046 $\pm$ 0.015  & 0.71 $\pm$ 0.02  &  0.95 $\pm$ 0.02 & 2.83 $\pm$ 0.03 & 1.77 $\pm$ 0.02 & $ <0.04 $ \\ 
SN\,2012il   & 0.32 $\pm$ 0.01  & 0.10 $\pm$ 0.01  & \ldots   & 0.24 $\pm$ 0.01  & 0.44 $\pm$ 0.02  &  1.38 $\pm 0.02$  & 0.70 $\pm$ 0.04 & $< 0.035$  \\ 
PTF11rks     & 0.65 $\pm$ 0.08   & \ldots  & \ldots  & 0.12 $\pm$ 0.02  & 0.07 $\pm$ 0.02 & 0.22 $\pm$ 0.03  & 0.40 $\pm$ 0.03 & 0.07 $\pm$ 0.03 \\
SN\,2011kf   & 0.047 $\pm$ 0.016 & \ldots   & \ldots  & 0.046 $\pm$ 0.01  & 0.032 $\pm$ 0.01 & 0.099 $\pm$ 0.01  & 0.094 $\pm$ 0.01  & $< 0.019$ \\ 
MLS121104    & 0.53 $\pm$ 0.02  & 0.095 $\pm$ 0.009 & \ldots  & 0.198 $\pm$ 0.007 & 0.152 $\pm$ 0.007 & 0.419 $\pm$ 0.007 &0.538 $\pm$ 0.009 & 0.059 $\pm$ 0.004 \\ 
SN\,2006oz   & 0.026 $\pm$ 0.032 &\ldots   & \ldots  & 0.018 $\pm$ 0.005  & 0.016 $\pm$ 0.003 & 0.047 $\pm$ 0.002 & 0.037 $\pm$ 0.01 & $< 0.006$ \\
PS1-12zn     & 0.236 $\pm$ 0.015  & $< 0.04$  & \ldots  & 0.069 $\pm$ 0.01 &  0.076 $\pm$ 0.02 & 0.29 $\pm$ 0.01 &\ldots  & \ldots   \\ 
PS1-12bqf    & 0.13 $\pm$ 0.01 & \ldots  & \ldots  & 0.050 $\pm$ 0.005 & 0.010 $\pm$ 0.004 & 0.046 $\pm$ 0.008 &\ldots  &  \ldots  \\ 
PS1-11bdn    & 0.043 $\pm$ 0.008  & \ldots  & \ldots  & $< 0.03$  &0.048 $\pm$ 0.01  & 0.13 $\pm$ 0.01  &\ldots  &   \ldots 
\enddata
\end{deluxetable*}

\begin{deluxetable}{lcccccccc}
\tabletypesize{\scriptsize}
\tablecaption{Derived Host Galaxy Properties \label{tab:results} }
\tablehead{
\colhead{Object} & 
\colhead{$M_B$} &
\colhead{$A_V$}  &
\colhead{$\log(M_*)$}  &
\colhead{$\log(M_*), A_V = 0$}  &
\colhead{$\log({\rm age}), A_V = 0$}  &
\colhead{SFR}    &
\colhead{$12 + \log({\rm O/H})$} &
\colhead{$12 + \log({\rm O/H})$} \\
\colhead{} &
\colhead{(mag)} &
\colhead{(mag)} &
\colhead{(${\rm M}_{\odot}$)} &
\colhead{(${\rm M}_{\odot}$)} &
\colhead{(yr)} &
\colhead{(${\rm M}_{\odot} {\rm ~yr}^{-1}$)} &
\colhead{T$_e$ method} &
\colhead{R$_{23}$ method}
}
\startdata
PTF10hgi  & $-15.78$  & 1.14$^{+0.87}_{-1.14}$ & $ 8.06^{+0.03}_{-0.57} $  & 7.87 $^{+0.58}_{-0.01}$ & $8.50^{+1.40}_{-0.02} $& $3.8 \times 10^{-2}$ &\ldots  & \ldots \\
SN\,2010kd & $-17.15$& 0.0  & 8.07$^{+0.15}_{-0.74}$  & $8.07^{+0.14}_{-0.15}$ & $8.10^{+0.66}_{-0.27}$ & $2.5 \times 10^{-2}$ &\ldots   & 8.16 / 8.71 \\
PTF12dam   &$-19.25$  & 0.16$^{+0.04}_{-0.04}$ &  8.77$^{+0.02}_{-0.04}$ &\ldots  & \ldots &  6.4   & 8.02\tablenotemark{a} & 8.40  \\
SN\,2007bi\tablenotemark{b} & $-16.40$  & 0.0$^{+1.09}_{-0.00}$  & 6.84$^{+0.22}_{-0.07}$ & $6.84^{+0.17}_{-0.08}$ &$6.50^{+0.30}_{-0.50}$  & $1.1 \times 10^{-2}$       & \ldots  & 8.33 \\
SN\,2011ke & $-16.65$ & 0.30$^{+0.36}_{-0.30}$  & 7.47$^{+0.14}_{-0.11}$ & $7.47^{+0.14}_{-0.07}$ & $7.70^{+0.71}_{-0.19}$ &  0.77    & 7.59 & 8.13  \\
SN\,2012il & $-17.58$ & 0.0$^{+0.23}_{-0.00}$ & 8.27$^{+0.06}_{-0.07}$ & $8.27^{+0.06}_{-0.07}$  & $8.10^{+0.16}_{-0.14}$  &  0.47     &  \ldots     & 8.27  \\
PTF11rks   & $-18.66$ & 0.30$^{+0.47}_{-0.30}$    & $9.00^{+0.09}_{-0.21} $ & $8.92^{+0.01}_{-0.08}$ & $8.40^{+0.17}_{-0.10}$  &   0.49      & \ldots     &  8.42 / 8.57  \\
SN\,2010gx\tablenotemark{c} & $-16.98$ & 0.96$^{+0.08}_{-0.08}$ & 8.19$^{+0.06}_{-0.07}$ &\ldots &\ldots &  1.30  & 7.46   & 8.27 \\
SN\,2011kf & $-16.87$ & 0.0 & 7.83$^{+0.23}_{-0.11}$ &$7.83^{+0.28}_{-0.15}$ &  $7.90^{+1.06}_{-0.14}$ &  0.14   & \ldots  & 8.05 / 8.84 \\
PTF09cnd\tablenotemark{d}   & $-16.75$ & 0.00$^{+0.54}_{-0.00}$    & 6.99$^{+0.51}_{-0.12}$ &  $6.99^{+0.38}_{-0.12}$ & $7.10^{+0.71}_{-0.22}$  & 0.22  & \ldots  & 8.33 / 8.64 \\
SN\,2005ap & $-16.73$  & 0.70$^{+2.45}_{-0.70}$    &  8.08$^{+0.56}_{-0.57}$ &  $8.17^{+0.22}_{-0.28}$ &$9.00^{+0.41}_{-1.03}$  & \ldots & \ldots & \ldots \\
MLS121104  &  $-19.39$ & 0.0 & 9.48$^{+0.04}_{-0.06}$ & $9.44^{+0.14}_{-0.07}$ & $9.10^{+0.38}_{-0.28}$  & 1.27  & \ldots  & 8.80   \\
PTF09cwl   & $> -16.5$  &  \ldots     & $< 7.9 $ &\ldots &\ldots & \ldots&  \ldots& \ldots  \\
SN\,2006oz & $-16.96$ & 0.0  & 8.42$^{+0.64}_{-0.16}$ & $8.43^{+0.55}_{-0.14}$ & $8.70^{+1.17}_{-0.24}$  &  0.23  & \ldots  & 8.00 / 8.88 \\
PTF09atu   & $> -16.7$  & \ldots      & $< 7.9$ & \ldots & \ldots & \ldots  &  \ldots  & \ldots \\
PS1-12bqf  & $-20.28$  & 0.90$^{+0.39}_{-0.90}$     & 9.68$^{+0.13}_{-0.10}$ & $9.60^{+0.15}_{-0.04}$  & $8.40^{+0.59}_{-0.02}$ &  1.73  &\ldots  & 7.84 / 9.04 \\
PS1-11ap\tablenotemark{e}   & $-18.83$ &  0.20$^{+0.32}_{-0.20}$    & 8.71$^{+0.15}_{-0.12}$ &  $8.69^{+0.06}_{-0.10}$ & $8.00^{+0.28}_{-0.10}$ & 0.49 &\ldots  & 8.34 / 8.64 \\
PS1-10bzj\tablenotemark{f}  & $-17.90$ &  0.0$^{+0.21}_{-0.00}$    & 7.44$^{+0.11}_{-0.05}$ &  $7.47^{+0.10}_{-0.11}$ & $7.10^{+0.22}_{-0.30}$ &  4.2   & 7.80 &  8.30 \\
PS1-12zn   & $-18.75$ &  0.0$^{+2.60}_{-0.00}$    & 8.57$^{+0.24}_{-0.32}$ &$8.85^{+0.03}_{-0.35}$ & $8.90^{+0.07}_{-1.10}$ & 6.7  &\ldots   &  8.45      \\
PS1-11bdn & $-17.35$ & 1.0$^{+1.88}_{-1.00}$    & 8.11$^{+0.77}_{-0.59}$  & $8.23^{+0.45}_{-0.57}$  & $8.80^{+0.56}_{-1.17}$ & 1.54  &\ldots   &   \ldots  \\
PS1-13gt  &$> -18.9$  & \ldots    & $< 8.7$  & \ldots &\ldots  &  $< 1.6$      &\ldots   &  \ldots  \\
PS1-10awh &  $-16.47$ & \ldots    & $7.76^{+0.33}_{-0.65}$  & \ldots & \ldots &   0.18     & \ldots  &  \ldots  \\
PS1-10ky  & $> -16.3$ & \ldots    & $< 7.6$  &\ldots  &\ldots  & $0.4$      &\ldots   &  \ldots  \\
PS1-11aib & $> -19.3$ & \ldots    & $< 8.9 $  &\ldots  &\ldots  &  $< 2.8$      &\ldots   & \ldots   \\
SCP06F6\tablenotemark{g}   & $> -16.9$ & \ldots      & $< 8.0$  &\ldots   &\ldots  &  \ldots     &\ldots   &  \ldots   \\
PS1-10pm  &  $-19.29$ & 0.90$^{+0.10}_{-0.90}$     & 8.97$^{+0.32}_{-0.34}$ & $8.95^{+0.34}_{-0.13}$  & $8.30^{+0.95}_{-0.43}$  &  1.5 &\ldots  & \ldots \\
PS1-11tt  &$-18.15$  &  0.30$^{+1.01}_{-0.30}$    & 7.62$^{+0.73}_{-0.16}$ & $7.74^{+0.18}_{-0.28}$ & $7.50^{+0.34}_{-0.52}$  &  1.1 &\ldots  & \ldots \\
PS1-10afx\tablenotemark{h} &$-22.03$  &  0.40$^{+2.00}_{-0.40}$   & 10.23$^{+0.16}_{-0.12}$ &  $10.26^{+0.03}_{-0.07}$ & $9.01^{+0.16}_{-0.31}$ &  13  & \ldots & \ldots \\
PS1-11afv & $-19.50$ &  0.20$^{+1.51}_{-0.20}$    &  8.97$^{+0.22}_{-0.56}$ &  $9.01^{+0.16}_{-0.31}$ &  $8.70^{+0.23}_{-0.98}$  &  2.2 & \ldots & \ldots \\
PS1-11bam\tablenotemark{i} & $-20.89$  & 0.50$^{+0.61}_{-0.50}$     &  8.89$^{+1.07}_{-0.25}$ & $9.59^{+0.53}_{-0.73}$  &  $8.60^{+0.61}_{-1.20}$ & 6.0 &\ldots  & \ldots \\
PS1-12bmy & $-20.36$ &  0.0$^{+0.76}_{-0.00}$     &  9.36$^{+0.60}_{-0.11}$ & 9.36$^{+0.30}_{-0.11}$ & 8.10$^{+0.60}_{-0.11}$ & 2.6 & \ldots & \ldots 
\enddata
\tablenotetext{a}{From Chen et al., in prep.}
\tablenotetext{b}{Based on data in \citet{ysv+10}.}
\tablenotetext{c}{Based on data in \citet{csb+13}.} 
\tablenotetext{d}{Metallicity and SFR derived from a spectrum published in \citet{qkk+11}; see Section~\ref{sec:cnd} for details.}  
\tablenotetext{e}{Based on data in \citet{msk+14}.}   
\tablenotetext{f}{Based on data in \citet{lcb+13}.}     
\tablenotetext{g}{Based on data in \citet{bdt+09}.}     
\tablenotetext{h}{Based on data in \citet{cbr+13}.}       
\tablenotetext{i}{Based on data in \citet{bcl+12}.}    
\end{deluxetable}

\begin{deluxetable}{lccccc}
\tablecaption{Results ($p$-values) from Statistical Tests  \label{tab:stats} }
\tablehead{
\colhead{Property} & \colhead{SLSN-LGRB} &  \colhead{SLSN-CCSN} & \colhead{SLSN-LGRB} & \colhead{SLSN-CCSN} & \colhead{Test used}\\
\colhead{}  & \colhead{} & \colhead{}   & \colhead{(PS1/MDS only)} & \colhead{(PS1/MDS only)} &
}
\startdata
$M_B$   &  0.0013  & $7 \times 10^{-5}$  & 0.26 & 0.15 & Rank-Sum \\
Mass    &  0.007  &  $1.6 \times 10^{-7}$ & 0.17 & 0.0009 & Rank-Sum \\
SFR     & 0.067  & $4 \times 10^{-4}$  & 0.79  & 0.076 &  Rank-Sum \\
sSFR    & 0.55  & 0.004  & 0.53 & 0.009  &  Kolmogorov-Smirnov  \\
H$\beta$ EW & 0.75 & \ldots & \ldots & \ldots &  Kolmogorov-Smirnov
\enddata
\end{deluxetable}

\appendix

\section{Notes on Individual Objects}

\subsection{SN\,2011ke}
\label{sec:11xk}
Inspection of archival CFHT images shows that SN\,2011ke exploded in a compact dwarf galaxy, with a redder and more extended companion. We obtained a spectrum with the slit going through both the SN site and the companion galaxy; while the spectrum was not taken at parallactic angle IMACS has an atmospheric dispersion corrector, so relative line fluxes should not be affected. We find that the two galaxies are at a similar redshift, though with a velocity offset of $\sim 100$~km~s$^{-1}$. A color image combining $g$- and $r$-band from CFHT with our own $z$-band images from IMACS is shown in Figure~\ref{fig:babypea}. The blue-green color of the dwarf galaxy is due to strong [\ion{O}{3}] emission in $r$-band, similar to the ``Green Pea'' galaxies found in SDSS \citep{css+09}.

The SDSS images of this system do not separate the two galaxies, and the SDSS catalog photometry includes light from both sources. To get host galaxy photometry, we perform photometry in a 1\arcsec aperture centered on the compact dwarf on the CFHT ($g$ and $r$), IMACS ($z$) and FourStar ($J$) images and apply an aperture correction in each band calculated from stars in the field. The photometry listed for SN\,2011ke in Table~\ref{tab:morephot} is for the dwarf galaxy only. Similarly, derived quantities listed are based on the spectroscopy and photometry of the dwarf.

\begin{figure}
\begin{center}
\includegraphics[width=5cm]{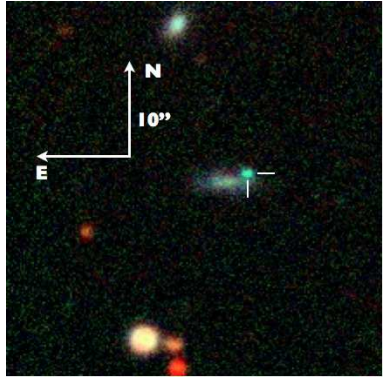}
\caption{Combined $grz$ image of the host galaxy of SN\,2011ke. The location of the SN is marked by the cross-hairs, and show that the SN went off in a compact dwarf galaxy. The redder, more extended galaxy next to it is at the same redshift, with a velocity offset of $\sim 100$~km~s$^{-1}$. Note the unusual color of the dwarf galaxy, due to the strong [\ion{O}{3}] emission that falls in $r$-band.
\label{fig:babypea} }
\end{center}
\end{figure}

\subsection{PTF09cnd}
\label{sec:cnd}
We obtained deep imaging of the field of PTF09cnd with MMTCam. As can be seen in Figure~\ref{fig:galpix} there are several sources near the reported location of the transient (marked by the green circle). We assume the closest source is the correct host, and use this photometry to construct a model SED. A spectrum confirming the redshift would be necessary, however, to make a definitive association.

To determine a metallicity for PTF09cnd, we download the archival spectra of the transient from \citet{qkk+11} from the WISEREP database \citep{yg12}. The late-time spectrum exhibits a number of galaxy emission lines, which we use to determine the host properties.


\begin{thebibliography}{}
\expandafter\ifx\csname natexlab\endcsname\relax\def\natexlab#1{#1}\fi

\bibitem[{{Ahn} {et~al.}(2012){Ahn}, {Alexandroff}, {Allende Prieto},
  {Anderson}, {Anderton}, {Andrews}, {Aubourg}, {Bailey}, {Balbinot}, {Barnes},
  \& et~al.}]{sdssdr9}
{Ahn}, C.~P., {Alexandroff}, R., {Allende Prieto}, C., {et~al.} 2012, \apjs,
  203, 21

\bibitem[{{Akerlof} {et~al.}(2003){Akerlof}, {Kehoe}, {McKay}, {Rykoff},
  {Smith}, {Casperson}, {McGowan}, {Vestrand}, {Wozniak}, {Wren}, {Ashley},
  {Phillips}, {Marshall}, {Epps}, \& {Schier}}]{akm+03}
{Akerlof}, C.~W., {Kehoe}, R.~L., {McKay}, T.~A., {et~al.} 2003, \pasp, 115,
  132

\bibitem[{{Ashby} {et~al.}(2013){Ashby}, {Willner}, {Fazio}, {Huang}, {Arendt},
  {Barmby}, {Barro}, {Bell}, {Bouwens}, {Cattaneo}, {Croton}, {Dav{\'e}},
  {Dunlop}, {Egami}, {Faber}, {Finlator}, {Grogin}, {Guhathakurta},
  {Hernquist}, {Hora}, {Illingworth}, {Kashlinsky}, {Koekemoer}, {Koo},
  {Labb{\'e}}, {Li}, {Lin}, {Moseley}, {Nandra}, {Newman}, {Noeske}, {Ouchi},
  {Peth}, {Rigopoulou}, {Robertson}, {Sarajedini}, {Simard}, {Smith}, {Wang},
  {Wechsler}, {Weiner}, {Wilson}, {Wuyts}, {Yamada}, \& {Yan}}]{awf+13}
{Ashby}, M.~L.~N., {Willner}, S.~P., {Fazio}, G.~G., {et~al.} 2013, \apj, 769,
  80

\bibitem[{{Barbary} {et~al.}(2009){Barbary}, {Dawson}, {Tokita}, {Aldering},
  {Amanullah}, {Connolly}, {Doi}, {Faccioli}, {Fadeyev}, {Fruchter},
  {Goldhaber}, {Goobar}, {Gude}, {Huang}, {Ihara}, {Konishi}, {Kowalski},
  {Lidman}, {Meyers}, {Morokuma}, {Nugent}, {Perlmutter}, {Rubin}, {Schlegel},
  {Spadafora}, {Suzuki}, {Swift}, {Takanashi}, {Thomas}, \& {Yasuda}}]{bdt+09}
{Barbary}, K., {Dawson}, K.~S., {Tokita}, K., {et~al.} 2009, \apj, 690, 1358

\bibitem[{{Bastian} {et~al.}(2010){Bastian}, {Covey}, \& {Meyer}}]{bcm10}
{Bastian}, N., {Covey}, K.~R., \& {Meyer}, M.~R. 2010, \araa, 48, 339

\bibitem[{{Berger} {et~al.}(2012){Berger}, {Chornock}, {Lunnan}, {Foley},
  {Czekala}, {Rest}, {Leibler}, {Soderberg}, {Roth}, {Narayan}, {Huber},
  {Milisavljevic}, {Sanders}, {Drout}, {Margutti}, {Kirshner}, {Marion},
  {Challis}, {Riess}, {Smartt}, {Burgett}, {Hodapp}, {Heasley}, {Kaiser},
  {Kudritzki}, {Magnier}, {McCrum}, {Price}, {Smith}, {Tonry}, \&
  {Wainscoat}}]{bcl+12}
{Berger}, E., {Chornock}, R., {Lunnan}, R., {et~al.} 2012, \apjl, 755, L29

\bibitem[{{Brammer} {et~al.}(2008){Brammer}, {van Dokkum}, \& {Coppi}}]{bvc08}
{Brammer}, G.~B., {van Dokkum}, P.~G., \& {Coppi}, P. 2008, \apj, 686, 1503

\bibitem[{{Cardamone} {et~al.}(2009){Cardamone}, {Schawinski}, {Sarzi},
  {Bamford}, {Bennert}, {Urry}, {Lintott}, {Keel}, {Parejko}, {Nichol},
  {Thomas}, {Andreescu}, {Murray}, {Raddick}, {Slosar}, {Szalay}, \&
  {Vandenberg}}]{css+09}
{Cardamone}, C., {Schawinski}, K., {Sarzi}, M., {et~al.} 2009, \mnras, 399,
  1191

\bibitem[{{Cardelli} {et~al.}(1989){Cardelli}, {Clayton}, \& {Mathis}}]{ccm89}
{Cardelli}, J.~A., {Clayton}, G.~C., \& {Mathis}, J.~S. 1989, \apj, 345, 245

\bibitem[{{Chatzopoulos} \& {Wheeler}(2012)}]{cw12}
{Chatzopoulos}, E., \& {Wheeler}, J.~C. 2012, \apj, 760, 154

\bibitem[{{Chen} {et~al.}(2013){Chen}, {Smartt}, {Bresolin}, {Pastorello},
  {Kudritzki}, {Kotak}, {McCrum}, {Fraser}, \& {Valenti}}]{csb+13}
{Chen}, T.-W., {Smartt}, S.~J., {Bresolin}, F., {et~al.} 2013, \apjl, 763, L28

\bibitem[{{Chevalier}(2012)}]{che12}
{Chevalier}, R.~A. 2012, \apjl, 752, L2

\bibitem[{{Chevalier} \& {Irwin}(2011)}]{ci11}
{Chevalier}, R.~A., \& {Irwin}, C.~M. 2011, \apjl, 729, L6

\bibitem[{{Chomiuk} {et~al.}(2011){Chomiuk}, {Chornock}, {Soderberg}, {Berger},
  {Chevalier}, {Foley}, {Huber}, {Narayan}, {Rest}, {Gezari}, {Kirshner},
  {Riess}, {Rodney}, {Smartt}, {Stubbs}, {Tonry}, {Wood-Vasey}, {Burgett},
  {Chambers}, {Czekala}, {Flewelling}, {Forster}, {Kaiser}, {Kudritzki},
  {Magnier}, {Martin}, {Morgan}, {Neill}, {Price}, {Roth}, {Sanders}, \&
  {Wainscoat}}]{ccs+11}
{Chomiuk}, L., {Chornock}, R., {Soderberg}, A.~M., {et~al.} 2011, \apj, 743,
  114

\bibitem[{{Chornock} {et~al.}(2013){Chornock}, {Berger}, {Rest},
  {Milisavljevic}, {Lunnan}, {Foley}, {Soderberg}, {Smartt}, {Burgasser},
  {Challis}, {Chomiuk}, {Czekala}, {Drout}, {Fong}, {Huber}, {Kirshner},
  {Leibler}, {McLeod}, {Marion}, {Narayan}, {Riess}, {Roth}, {Sanders},
  {Scolnic}, {Smith}, {Stubbs}, {Tonry}, {Valenti}, {Burgett}, {Chambers},
  {Hodapp}, {Kaiser}, {Kudritzki}, {Magnier}, \& {Price}}]{cbr+13}
{Chornock}, R., {Berger}, E., {Rest}, A., {et~al.} 2013, \apj, 767, 162

\bibitem[{{Copetti} {et~al.}(1986){Copetti}, {Pastoriza}, \& {Dottori}}]{cpd86}
{Copetti}, M.~V.~F., {Pastoriza}, M.~G., \& {Dottori}, H.~A. 1986, \aap, 156,
  111

\bibitem[{{Dessart} {et~al.}(2012){Dessart}, {Hillier}, {Waldman}, {Livne}, \&
  {Blondin}}]{dhw+12}
{Dessart}, L., {Hillier}, D.~J., {Waldman}, R., {Livne}, E., \& {Blondin}, S.
  2012, \mnras, 426, L76

\bibitem[{{Dexter} \& {Kasen}(2013)}]{dk13}
{Dexter}, J., \& {Kasen}, D. 2013, \apj, 772, 30

\bibitem[{{Drake} {et~al.}(2009){Drake}, {Djorgovski}, {Mahabal}, {Beshore},
  {Larson}, {Graham}, {Williams}, {Christensen}, {Catelan}, {Boattini},
  {Gibbs}, {Hill}, \& {Kowalski}}]{ddm+09}
{Drake}, A.~J., {Djorgovski}, S.~G., {Mahabal}, A., {et~al.} 2009, \apj, 696,
  870

\bibitem[{{Drake} {et~al.}(2011){Drake}, {Djorgovski}, {Mahabal}, {Graham},
  {Williams}, {Donalek}, {Prieto}, {Catelan}, {Christensen}, {Beshore},
  {Larson}, \& {McNaught}}]{ddm+11a}
{Drake}, A.~J., {Djorgovski}, S.~G., {Mahabal}, A.~A., {et~al.} 2011, The
  Astronomer's Telegram, 3343, 1

\bibitem[{{Drake} {et~al.}(2012{\natexlab{a}}){Drake}, {Djorgovski}, {Mahabal},
  {Graham}, {Williams}, {Prieto}, {Catelan}, {Christensen}, {Beshore}, \&
  {Larson}}]{ddm+12}
---. 2012{\natexlab{a}}, The Astronomer's Telegram, 3873, 1

\bibitem[{{Drake} {et~al.}(2012{\natexlab{b}}){Drake}, {Djorgovski}, {Mahabal},
  {Graham}, {Williams}, {Prieto}, {Catelan}, {Christensen}, {Larson}, \&
  {Beshore}}]{ddm+12b}
---. 2012{\natexlab{b}}, The Astronomer's Telegram, 4595, 1

\bibitem[{{Dressler} {et~al.}(2006){Dressler}, {Hare}, {Bigelow}, \&
  {Osip}}]{dhb+06}
{Dressler}, A., {Hare}, T., {Bigelow}, B.~C., \& {Osip}, D.~J. 2006, in Society
  of Photo-Optical Instrumentation Engineers (SPIE) Conference Series, Vol.
  6269, Society of Photo-Optical Instrumentation Engineers (SPIE) Conference
  Series

\bibitem[{{Fatkhullin} \& {Gabdeev}(2012)}]{fg12}
{Fatkhullin}, T., \& {Gabdeev}, M. 2012, The Astronomer's Telegram, 4599, 1

\bibitem[{{Fruchter} \& {Hook}(2002)}]{fh02}
{Fruchter}, A.~S., \& {Hook}, R.~N. 2002, \pasp, 114, 144

\bibitem[{{Fruchter} {et~al.}(2006){Fruchter}, {Levan}, {Strolger},
  {Vreeswijk}, {Thorsett}, {Bersier}, {Burud}, {Castro Cer{\'o}n},
  {Castro-Tirado}, {Conselice}, {Dahlen}, {Ferguson}, {Fynbo}, {Garnavich},
  {Gibbons}, {Gorosabel}, {Gull}, {Hjorth}, {Holland}, {Kouveliotou}, {Levay},
  {Livio}, {Metzger}, {Nugent}, {Petro}, {Pian}, {Rhoads}, {Riess}, {Sahu},
  {Smette}, {Tanvir}, {Wijers}, \& {Woosley}}]{fls+06}
{Fruchter}, A.~S., {Levan}, A.~J., {Strolger}, L., {et~al.} 2006, \nat, 441,
  463

\bibitem[{{Gal-Yam}(2012)}]{gal12}
{Gal-Yam}, A. 2012, Science, 337, 927

\bibitem[{{Gal-Yam} {et~al.}(2009){Gal-Yam}, {Mazzali}, {Ofek}, {Nugent},
  {Kulkarni}, {Kasliwal}, {Quimby}, {Filippenko}, {Cenko}, {Chornock},
  {Waldman}, {Kasen}, {Sullivan}, {Beshore}, {Drake}, {Thomas}, {Bloom},
  {Poznanski}, {Miller}, {Foley}, {Silverman}, {Arcavi}, {Ellis}, \&
  {Deng}}]{gmo+09}
{Gal-Yam}, A., {Mazzali}, P., {Ofek}, E.~O., {et~al.} 2009, \nat, 462, 624

\bibitem[{{Garg} {et~al.}(2007){Garg}, {Stubbs}, {Challis}, {Wood-Vasey},
  {Blondin}, {Huber}, {Cook}, {Nikolaev}, {Rest}, {Smith}, {Olsen}, {Suntzeff},
  {Aguilera}, {Prieto}, {Becker}, {Miceli}, {Miknaitis}, {Clocchiatti},
  {Minniti}, {Morelli}, \& {Welch}}]{gsc+07}
{Garg}, A., {Stubbs}, C.~W., {Challis}, P., {et~al.} 2007, \aj, 133, 403

\bibitem[{{Ginzburg} \& {Balberg}(2012)}]{gb12}
{Ginzburg}, S., \& {Balberg}, S. 2012, \apj, 757, 178

\bibitem[{{Gonzaga} {et~al.}(2012){Gonzaga}, {Hack}, {Fruchter}, \&
  {Mack}}]{ghf+12}
{Gonzaga}, S., {Hack}, W., {Fruchter}, A., \& {Mack}, J. 2012, {The DrizzlePac
  Handbook} (Baltimore, STScI)

\bibitem[{{Graham} \& {Fruchter}(2013)}]{gf13}
{Graham}, J.~F., \& {Fruchter}, A.~S. 2013, \apj, 774, 119

\bibitem[{{Henry} {et~al.}(2013){Henry}, {Martin}, {Finlator}, \&
  {Dressler}}]{hmf+13}
{Henry}, A., {Martin}, C.~L., {Finlator}, K., \& {Dressler}, A. 2013, \apj,
  769, 148

\bibitem[{{Hook} {et~al.}(2004){Hook}, {J{\o}rgensen}, {Allington-Smith},
  {Davies}, {Metcalfe}, {Murowinski}, \& {Crampton}}]{hja+04}
{Hook}, I.~M., {J{\o}rgensen}, I., {Allington-Smith}, J.~R., {et~al.} 2004,
  \pasp, 116, 425

\bibitem[{{Inserra} {et~al.}(2013){Inserra}, {Smartt}, {Jerkstrand}, {Valenti},
  {Fraser}, {Wright}, {Smith}, {Chen}, {Kotak}, {Pastorello}, {Nicholl},
  {Bresolin}, {Kudritzki}, {Benetti}, {Botticella}, {Burgett}, {Chambers},
  {Ergon}, {Flewelling}, {Fynbo}, {Geier}, {Hodapp}, {Howell}, {Huber},
  {Kaiser}, {Leloudas}, {Magill}, {Magnier}, {McCrum}, {Metcalfe}, {Price},
  {Rest}, {Sollerman}, {Sweeney}, {Taddia}, {Taubenberger}, {Tonry},
  {Wainscoat}, {Waters}, \& {Young}}]{isj+13}
{Inserra}, C., {Smartt}, S.~J., {Jerkstrand}, A., {et~al.} 2013, \apj, 770, 128

\bibitem[{{Kaiser} {et~al.}(2010){Kaiser}, {Burgett}, {Chambers}, {Denneau},
  {Heasley}, {Jedicke}, {Magnier}, {Morgan}, {Onaka}, \& {Tonry}}]{PS1_system}
{Kaiser}, N., {Burgett}, W., {Chambers}, K., {et~al.} 2010, in Society of
  Photo-Optical Instrumentation Engineers (SPIE) Conference Series, Vol. 7733,
  Society of Photo-Optical Instrumentation Engineers (SPIE) Conference Series

\bibitem[{{Kasen} \& {Bildsten}(2010)}]{kb10}
{Kasen}, D., \& {Bildsten}, L. 2010, \apj, 717, 245

\bibitem[{{Kelly} \& {Kirshner}(2012)}]{kk12}
{Kelly}, P.~L., \& {Kirshner}, R.~P. 2012, \apj, 759, 107

\bibitem[{{Kennicutt}(1998)}]{ken98}
{Kennicutt}, Jr., R.~C. 1998, \araa, 36, 189

\bibitem[{{Kewley} \& {Ellison}(2008)}]{ke08}
{Kewley}, L.~J., \& {Ellison}, S.~L. 2008, \apj, 681, 1183

\bibitem[{{Kobulnicky} \& {Kewley}(2004)}]{kk04}
{Kobulnicky}, H.~A., \& {Kewley}, L.~J. 2004, \apj, 617, 240

\bibitem[{{Kocevski} \& {West}(2011)}]{kw11}
{Kocevski}, D., \& {West}, A.~A. 2011, \apjl, 735, L8

\bibitem[{{Komatsu} {et~al.}(2011){Komatsu}, {Smith}, {Dunkley}, {Bennett},
  {Gold}, {Hinshaw}, {Jarosik}, {Larson}, {Nolta}, {Page}, {Spergel},
  {Halpern}, {Hill}, {Kogut}, {Limon}, {Meyer}, {Odegard}, {Tucker}, {Weiland},
  {Wollack}, \& {Wright}}]{ksd+11}
{Komatsu}, E., {Smith}, K.~M., {Dunkley}, J., {et~al.} 2011, \apjs, 192, 18

\bibitem[{{Koz{\l}owski} {et~al.}(2010){Koz{\l}owski}, {Kochanek}, {Stern},
  {Prieto}, {Stanek}, {Thompson}, {Assef}, {Drake}, {Szczygie{\l}},
  {Wo{\'z}niak}, {Nugent}, {Ashby}, {Beshore}, {Brown}, {Dey}, {Griffith},
  {Harrison}, {Jannuzi}, {Larson}, {Madsen}, {Pilecki}, {Pojma{\'n}ski},
  {Skowron}, {Vestrand}, \& {Wren}}]{kks+10}
{Koz{\l}owski}, S., {Kochanek}, C.~S., {Stern}, D., {et~al.} 2010, \apj, 722,
  1624

\bibitem[{{Kriek} {et~al.}(2009){Kriek}, {van Dokkum}, {Labb{\'e}}, {Franx},
  {Illingworth}, {Marchesini}, \& {Quadri}}]{kvl+09}
{Kriek}, M., {van Dokkum}, P.~G., {Labb{\'e}}, I., {et~al.} 2009, \apj, 700,
  221

\bibitem[{{Kroupa} {et~al.}(2013){Kroupa}, {Weidner}, {Pflamm-Altenburg},
  {Thies}, {Dabringhausen}, {Marks}, \& {Maschberger}}]{kwp+12}
{Kroupa}, P., {Weidner}, C., {Pflamm-Altenburg}, J., {et~al.} 2013, {The
  Stellar and Sub-Stellar Initial Mass Function of Simple and Composite
  Populations}, ed. T.~D. {Oswalt} \& G.~{Gilmore}, 115

\bibitem[{{Langer} \& {Norman}(2006)}]{ln06}
{Langer}, N., \& {Norman}, C.~A. 2006, \apjl, 638, L63

\bibitem[{{Laskar} {et~al.}(2011){Laskar}, {Berger}, \& {Chary}}]{lbc11}
{Laskar}, T., {Berger}, E., \& {Chary}, R.-R. 2011, \apj, 739, 1

\bibitem[{{Lavalley} {et~al.}(1992){Lavalley}, {Isobe}, \& {Feigelson}}]{asurv}
{Lavalley}, M.~P., {Isobe}, T., \& {Feigelson}, E.~D. 1992, in Bulletin of the
  American Astronomical Society, Vol.~24, Bulletin of the American Astronomical
  Society, 839--840

\bibitem[{{Law} {et~al.}(2009){Law}, {Kulkarni}, {Dekany}, {Ofek}, {Quimby},
  {Nugent}, {Surace}, {Grillmair}, {Bloom}, {Kasliwal}, {Bildsten}, {Brown},
  {Cenko}, {Ciardi}, {Croner}, {Djorgovski}, {van Eyken}, {Filippenko}, {Fox},
  {Gal-Yam}, {Hale}, {Hamam}, {Helou}, {Henning}, {Howell}, {Jacobsen},
  {Laher}, {Mattingly}, {McKenna}, {Pickles}, {Poznanski}, {Rahmer}, {Rau},
  {Rosing}, {Shara}, {Smith}, {Starr}, {Sullivan}, {Velur}, {Walters}, \&
  {Zolkower}}]{lkd+09}
{Law}, N.~M., {Kulkarni}, S.~R., {Dekany}, R.~G., {et~al.} 2009, \pasp, 121,
  1395

\bibitem[{{Leloudas} {et~al.}(2012){Leloudas}, {Chatzopoulos}, {Dilday},
  {Gorosabel}, {Vinko}, {Gallazzi}, {Wheeler}, {Bassett}, {Fischer}, {Frieman},
  {Fynbo}, {Goobar}, {Jel{\'{\i}}nek}, {Malesani}, {Nichol}, {Nordin},
  {{\"O}stman}, {Sako}, {Schneider}, {Smith}, {Sollerman}, {Stritzinger},
  {Th{\"o}ne}, \& {de Ugarte Postigo}}]{lcd+12}
{Leloudas}, G., {Chatzopoulos}, E., {Dilday}, B., {et~al.} 2012, \aap, 541,
  A129

\bibitem[{{Levesque} {et~al.}(2010{\natexlab{a}}){Levesque}, {Berger},
  {Kewley}, \& {Bagley}}]{lbk+10}
{Levesque}, E.~M., {Berger}, E., {Kewley}, L.~J., \& {Bagley}, M.~M.
  2010{\natexlab{a}}, \aj, 139, 694

\bibitem[{{Levesque} {et~al.}(2010{\natexlab{b}}){Levesque}, {Kewley},
  {Berger}, \& {Zahid}}]{lkb+10}
{Levesque}, E.~M., {Kewley}, L.~J., {Berger}, E., \& {Zahid}, H.~J.
  2010{\natexlab{b}}, \aj, 140, 1557

\bibitem[{{Levesque} {et~al.}(2010{\natexlab{c}}){Levesque}, {Kewley},
  {Graham}, \& {Fruchter}}]{lkg+10}
{Levesque}, E.~M., {Kewley}, L.~J., {Graham}, J.~F., \& {Fruchter}, A.~S.
  2010{\natexlab{c}}, \apjl, 712, L26

\bibitem[{{Lunnan} {et~al.}(2013){Lunnan}, {Chornock}, {Berger},
  {Milisavljevic}, {Drout}, {Sanders}, {Challis}, {Czekala}, {Foley}, {Fong},
  {Huber}, {Kirshner}, {Leibler}, {Marion}, {McCrum}, {Narayan}, {Rest},
  {Roth}, {Scolnic}, {Smartt}, {Smith}, {Soderberg}, {Stubbs}, {Tonry},
  {Burgett}, {Chambers}, {Kudritzki}, {Magnier}, \& {Price}}]{lcb+13}
{Lunnan}, R., {Chornock}, R., {Berger}, E., {et~al.} 2013, \apj, 771, 97

\bibitem[{{MacFadyen} \& {Woosley}(1999)}]{mw99}
{MacFadyen}, A.~I., \& {Woosley}, S.~E. 1999, \apj, 524, 262

\bibitem[{{Magnier}(2006)}]{PS1_IPP}
{Magnier}, E. 2006, in The Advanced Maui Optical and Space Surveillance
  Technologies Conference

\bibitem[{{Magnier} {et~al.}(2008){Magnier}, {Liu}, {Monet}, \&
  {Chambers}}]{PS1_astrometry}
{Magnier}, E.~A., {Liu}, M., {Monet}, D.~G., \& {Chambers}, K.~C. 2008, in IAU
  Symposium, Vol. 248, IAU Symposium, ed. W.~J. {Jin}, I.~{Platais}, \&
  M.~A.~C. {Perryman}, 553--559

\bibitem[{{Mannucci} {et~al.}(2010){Mannucci}, {Cresci}, {Maiolino}, {Marconi},
  \& {Gnerucci}}]{mcm+10}
{Mannucci}, F., {Cresci}, G., {Maiolino}, R., {Marconi}, A., \& {Gnerucci}, A.
  2010, \mnras, 408, 2115

\bibitem[{{Mannucci} {et~al.}(2011){Mannucci}, {Salvaterra}, \&
  {Campisi}}]{msc11}
{Mannucci}, F., {Salvaterra}, R., \& {Campisi}, M.~A. 2011, \mnras, 414, 1263

\bibitem[{{Maraston}(2005)}]{mar05}
{Maraston}, C. 2005, \mnras, 362, 799

\bibitem[{{Mauduit} {et~al.}(2012){Mauduit}, {Lacy}, {Farrah}, {Surace},
  {Jarvis}, {Oliver}, {Maraston}, {Vaccari}, {Marchetti}, {Zeimann},
  {Gonz{\'a}les-Solares}, {Pforr}, {Petric}, {Henriques}, {Thomas}, {Afonso},
  {Rettura}, {Wilson}, {Falder}, {Geach}, {Huynh}, {Norris}, {Seymour},
  {Richards}, {Stanford}, {Alexander}, {Becker}, {Best}, {Bizzocchi},
  {Bonfield}, {Castro}, {Cava}, {Chapman}, {Christopher}, {Clements}, {Covone},
  {Dubois}, {Dunlop}, {Dyke}, {Edge}, {Ferguson}, {Foucaud}, {Franceschini},
  {Gal}, {Grant}, {Grossi}, {Hatziminaoglou}, {Hickey}, {Hodge}, {Huang},
  {Ivison}, {Kim}, {LeFevre}, {Lehnert}, {Lonsdale}, {Lubin}, {McLure},
  {Messias}, {Mart{\'{\i}}nez-Sansigre}, {Mortier}, {Nielsen}, {Ouchi},
  {Parish}, {Perez-Fournon}, {Pierre}, {Rawlings}, {Readhead}, {Ridgway},
  {Rigopoulou}, {Romer}, {Rosebloom}, {Rottgering}, {Rowan-Robinson}, {Sajina},
  {Simpson}, {Smail}, {Squires}, {Stevens}, {Taylor}, {Trichas}, {Urrutia},
  {van Kampen}, {Verma}, \& {Xu}}]{servs}
{Mauduit}, J.-C., {Lacy}, M., {Farrah}, D., {et~al.} 2012, \pasp, 124, 714

\bibitem[{{McCrum} {et~al.}(2014{\natexlab{a}}){McCrum}, {Smartt}, {Rest},
  {Smith}, {Kotak}, {Rodney}, {Young}, {Chornock}, {Berger}, {Foley}, {Fraser},
  {Wright}, {Scolnic}, {Tonry}, {Urata}, {Huang}, {Pastorello}, {Botticella},
  {Valenti}, {Mattila}, {Kankare}, {Farrow}, {Huber}, {Stubbs}, {Kirshner},
  {Bresolin}, {Burgett}, {Chambers}, {Draper}, {Flewelling}, {Jedicke},
  {Kaiser}, {Magnier}, {Metcalfe}, {Morgan}, {Price}, {Sweeney}, {Wainscoat},
  \& {Waters}}]{msr+14}
{McCrum}, M., {Smartt}, S.~J., {Rest}, A., {et~al.} 2014{\natexlab{a}}, ArXiv
  e-prints, arXiv:1402.1631

\bibitem[{{McCrum} {et~al.}(2014{\natexlab{b}}){McCrum}, {Smartt}, {Kotak},
  {Rest}, {Jerkstrand}, {Inserra}, {Rodney}, {Chen}, {Howell}, {Huber},
  {Pastorello}, {Tonry}, {Bresolin}, {Kudritzki}, {Chornock}, {Berger},
  {Smith}, {Botticella}, {Foley}, {Fraser}, {Milisavljevic}, {Nicholl},
  {Riess}, {Stubbs}, {Valenti}, {Wood-Vasey}, {Wright}, {Young}, {Drout},
  {Czekala}, {Burgett}, {Chambers}, {Draper}, {Flewelling}, {Hodapp}, {Kaiser},
  {Magnier}, {Metcalfe}, {Price}, {Sweeney}, \& {Wainscoat}}]{msk+14}
{McCrum}, M., {Smartt}, S.~J., {Kotak}, R., {et~al.} 2014{\natexlab{b}},
  \mnras, 437, 656

\bibitem[{{Miknaitis} {et~al.}(2007){Miknaitis}, {Pignata}, {Rest},
  {Wood-Vasey}, {Blondin}, {Challis}, {Smith}, {Stubbs}, {Suntzeff}, {Foley},
  {Matheson}, {Tonry}, {Aguilera}, {Blackman}, {Becker}, {Clocchiatti},
  {Covarrubias}, {Davis}, {Filippenko}, {Garg}, {Garnavich}, {Hicken}, {Jha},
  {Krisciunas}, {Kirshner}, {Leibundgut}, {Li}, {Miceli}, {Narayan}, {Prieto},
  {Riess}, {Salvo}, {Schmidt}, {Sollerman}, {Spyromilio}, \&
  {Zenteno}}]{mpr+07}
{Miknaitis}, G., {Pignata}, G., {Rest}, A., {et~al.} 2007, \apj, 666, 674

\bibitem[{{Milisavljevic} {et~al.}(2013){Milisavljevic}, {Soderberg},
  {Margutti}, {Drout}, {Howie Marion}, {Sanders}, {Hsiao}, {Lunnan},
  {Chornock}, {Fesen}, {Parrent}, {Levesque}, {Berger}, {Foley}, {Challis},
  {Kirshner}, {Dittmann}, {Bieryla}, {Kamble}, {Chakraborti}, {De Rosa},
  {Fausnaugh}, {Hainline}, {Chen}, {Hickox}, {Morrell}, {Phillips}, \&
  {Stritzinger}}]{msm+13}
{Milisavljevic}, D., {Soderberg}, A.~M., {Margutti}, R., {et~al.} 2013, \apjl,
  770, L38

\bibitem[{{Modjaz} {et~al.}(2011){Modjaz}, {Kewley}, {Bloom}, {Filippenko},
  {Perley}, \& {Silverman}}]{mkb+11}
{Modjaz}, M., {Kewley}, L., {Bloom}, J.~S., {et~al.} 2011, \apjl, 731, L4

\bibitem[{{Modjaz} {et~al.}(2008){Modjaz}, {Kewley}, {Kirshner}, {Stanek},
  {Challis}, {Garnavich}, {Greene}, {Kelly}, \& {Prieto}}]{mkk+08}
{Modjaz}, M., {Kewley}, L., {Kirshner}, R.~P., {et~al.} 2008, \aj, 135, 1136

\bibitem[{{Moriya} {et~al.}(2013){Moriya}, {Blinnikov}, {Tominaga}, {Yoshida},
  {Tanaka}, {Maeda}, \& {Nomoto}}]{mbt+13}
{Moriya}, T.~J., {Blinnikov}, S.~I., {Tominaga}, N., {et~al.} 2013, \mnras,
  428, 1020

\bibitem[{{Moriya} \& {Maeda}(2012)}]{mm12}
{Moriya}, T.~J., \& {Maeda}, K. 2012, \apjl, 756, L22

\bibitem[{{Neill} {et~al.}(2011){Neill}, {Sullivan}, {Gal-Yam}, {Quimby},
  {Ofek}, {Wyder}, {Howell}, {Nugent}, {Seibert}, {Martin}, {Overzier},
  {Barlow}, {Foster}, {Friedman}, {Morrissey}, {Neff}, {Schiminovich},
  {Bianchi}, {Donas}, {Heckman}, {Lee}, {Madore}, {Milliard}, {Rich}, \&
  {Szalay}}]{nsg+11}
{Neill}, J.~D., {Sullivan}, M., {Gal-Yam}, A., {et~al.} 2011, \apj, 727, 15

\bibitem[{{Nicholl} {et~al.}(2013){Nicholl}, {Smartt}, {Jerkstrand}, {Inserra},
  {McCrum}, {Kotak}, {Fraser}, {Wright}, {Chen}, {Smith}, {Young}, {Sim},
  {Valenti}, {Howell}, {Bresolin}, {Kudritzki}, {Tonry}, {Huber}, {Rest},
  {Pastorello}, {Tomasella}, {Cappellaro}, {Benetti}, {Mattila}, {Kankare},
  {Kangas}, {Leloudas}, {Sollerman}, {Taddia}, {Berger}, {Chornock}, {Narayan},
  {Stubbs}, {Foley}, {Lunnan}, {Soderberg}, {Sanders}, {Milisavljevic},
  {Margutti}, {Kirshner}, {Elias-Rosa}, {Morales-Garoffolo}, {Taubenberger},
  {Botticella}, {Gezari}, {Urata}, {Rodney}, {Riess}, {Scolnic}, {Wood-Vasey},
  {Burgett}, {Chambers}, {Flewelling}, {Magnier}, {Kaiser}, {Metcalfe},
  {Morgan}, {Price}, {Sweeney}, \& {Waters}}]{nsj+13}
{Nicholl}, M., {Smartt}, S.~J., {Jerkstrand}, A., {et~al.} 2013, \nat, 502, 346

\bibitem[{{Ofek} {et~al.}(2007){Ofek}, {Cameron}, {Kasliwal}, {Gal-Yam}, {Rau},
  {Kulkarni}, {Frail}, {Chandra}, {Cenko}, {Soderberg}, \& {Immler}}]{ock+07}
{Ofek}, E.~O., {Cameron}, P.~B., {Kasliwal}, M.~M., {et~al.} 2007, \apjl, 659,
  L13

\bibitem[{Osterbrock(1989)}]{ost89}
Osterbrock, D. 1989, Astrophysics of gaseous nebulae and active galactic
  nuclei, A series of books in astronomy (University Science Books)

\bibitem[{{Oyaizu} {et~al.}(2008){Oyaizu}, {Lima}, {Cunha}, {Lin}, {Frieman},
  \& {Sheldon}}]{olc+08}
{Oyaizu}, H., {Lima}, M., {Cunha}, C.~E., {et~al.} 2008, \apj, 674, 768

\bibitem[{{Pastorello} {et~al.}(2010){Pastorello}, {Smartt}, {Botticella},
  {Maguire}, {Fraser}, {Smith}, {Kotak}, {Magill}, {Valenti}, {Young},
  {Gezari}, {Bresolin}, {Kudritzki}, {Howell}, {Rest}, {Metcalfe}, {Mattila},
  {Kankare}, {Huang}, {Urata}, {Burgett}, {Chambers}, {Dombeck}, {Flewelling},
  {Grav}, {Heasley}, {Hodapp}, {Kaiser}, {Luppino}, {Lupton}, {Magnier},
  {Monet}, {Morgan}, {Onaka}, {Price}, {Rhoads}, {Siegmund}, {Stubbs},
  {Sweeney}, {Tonry}, {Wainscoat}, {Waterson}, {Waters}, \&
  {Wynn-Williams}}]{psb+10}
{Pastorello}, A., {Smartt}, S.~J., {Botticella}, M.~T., {et~al.} 2010, \apjl,
  724, L16

\bibitem[{{Peng} {et~al.}(2002){Peng}, {Ho}, {Impey}, \& {Rix}}]{phi+02}
{Peng}, C.~Y., {Ho}, L.~C., {Impey}, C.~D., \& {Rix}, H.-W. 2002, \aj, 124, 266

\bibitem[{{Perley} {et~al.}(2013){Perley}, {Levan}, {Tanvir}, {Cenko}, {Bloom},
  {Hjorth}, {Kr{\"u}hler}, {Filippenko}, {Fruchter}, {Fynbo}, {Jakobsson},
  {Kalirai}, {Milvang-Jensen}, {Morgan}, {Prochaska}, \& {Silverman}}]{plt+13}
{Perley}, D.~A., {Levan}, A.~J., {Tanvir}, N.~R., {et~al.} 2013, \apj, 778, 128

\bibitem[{{Persson} {et~al.}(2013){Persson}, {Murphy}, {Smee}, {Birk},
  {Monson}, {Uomoto}, {Koch}, {Shectman}, {Barkhouser}, {Orndorff}, {Hammond},
  {Harding}, {Scharfstein}, {Kelson}, {Marshall}, \& {McCarthy}}]{pms+13}
{Persson}, S.~E., {Murphy}, D.~C., {Smee}, S., {et~al.} 2013, \pasp, 125, 654

\bibitem[{{Prieto} {et~al.}(2012){Prieto}, {Drake}, {Mahabal}, {Djorgovski},
  {Graham}, {Williams}, {Catelan}, {Christensen}, {Beshore}, \&
  {Larson}}]{pdm+12}
{Prieto}, J.~L., {Drake}, A.~J., {Mahabal}, A.~A., {et~al.} 2012, The
  Astronomer's Telegram, 3883, 1

\bibitem[{{Quimby} {et~al.}(2007){Quimby}, {Aldering}, {Wheeler},
  {H{\"o}flich}, {Akerlof}, \& {Rykoff}}]{qaw+07}
{Quimby}, R.~M., {Aldering}, G., {Wheeler}, J.~C., {et~al.} 2007, \apjl, 668,
  L99

\bibitem[{{Quimby} {et~al.}(2011{\natexlab{a}}){Quimby}, {Gal-Yam}, {Arcavi},
  {Yaron}, {Horesh}, \& {Mooley}}]{qga+11}
{Quimby}, R.~M., {Gal-Yam}, A., {Arcavi}, I., {et~al.} 2011{\natexlab{a}}, The
  Astronomer's Telegram, 3841, 1

\bibitem[{{Quimby} {et~al.}(2013{\natexlab{a}}){Quimby}, {Yuan}, {Akerlof}, \&
  {Wheeler}}]{qya+13}
{Quimby}, R.~M., {Yuan}, F., {Akerlof}, C., \& {Wheeler}, J.~C.
  2013{\natexlab{a}}, \mnras, 431, 912

\bibitem[{{Quimby} {et~al.}(2010){Quimby}, {Kulkarni}, {Ofek}, {Kasliwal},
  {Gal-Yam}, {Arcavi}, {Ben-Ami}, {Xu}, {Sternberg}, {Silverman}, {Cenko},
  {Kleiser}, {Nugent}, \& {Howell}}]{qko+10}
{Quimby}, R.~M., {Kulkarni}, S., {Ofek}, E., {et~al.} 2010, The Astronomer's
  Telegram, 2740, 1

\bibitem[{{Quimby} {et~al.}(2011{\natexlab{b}}){Quimby}, {Kulkarni},
  {Kasliwal}, {Gal-Yam}, {Arcavi}, {Sullivan}, {Nugent}, {Thomas}, {Howell},
  {Nakar}, {Bildsten}, {Theissen}, {Law}, {Dekany}, {Rahmer}, {Hale}, {Smith},
  {Ofek}, {Zolkower}, {Velur}, {Walters}, {Henning}, {Bui}, {McKenna},
  {Poznanski}, {Cenko}, \& {Levitan}}]{qkk+11}
{Quimby}, R.~M., {Kulkarni}, S.~R., {Kasliwal}, M.~M., {et~al.}
  2011{\natexlab{b}}, \nat, 474, 487

\bibitem[{{Quimby} {et~al.}(2012){Quimby}, {Arcavi}, {Sternberg}, {Ben-Ami},
  {Yaron}, {Gal-Yam}, {Graham}, {Cenko}, {Filippenko}, {Perley}, {Cao}, \&
  {Kulkarni}}]{qas+12}
{Quimby}, R.~M., {Arcavi}, I., {Sternberg}, A., {et~al.} 2012, The Astronomer's
  Telegram, 4121, 1

\bibitem[{{Quimby} {et~al.}(2013{\natexlab{b}}){Quimby}, {Werner}, {Oguri},
  {More}, {More}, {Tanaka}, {Nomoto}, {Moriya}, {Folatelli}, {Maeda}, \&
  {Bersten}}]{qwo+13}
{Quimby}, R.~M., {Werner}, M.~C., {Oguri}, M., {et~al.} 2013{\natexlab{b}},
  \apjl, 768, L20

\bibitem[{{Rest} {et~al.}(2005){Rest}, {Stubbs}, {Becker}, {Miknaitis},
  {Miceli}, {Covarrubias}, {Hawley}, {Smith}, {Suntzeff}, {Olsen}, {Prieto},
  {Hiriart}, {Welch}, {Cook}, {Nikolaev}, {Huber}, {Prochtor}, {Clocchiatti},
  {Minniti}, {Garg}, {Challis}, {Keller}, \& {Schmidt}}]{rsb+05}
{Rest}, A., {Stubbs}, C., {Becker}, A.~C., {et~al.} 2005, \apj, 634, 1103

\bibitem[{{Rest} {et~al.}(2011){Rest}, {Foley}, {Gezari}, {Narayan}, {Draine},
  {Olsen}, {Huber}, {Matheson}, {Garg}, {Welch}, {Becker}, {Challis},
  {Clocchiatti}, {Cook}, {Damke}, {Meixner}, {Miknaitis}, {Minniti}, {Morelli},
  {Nikolaev}, {Pignata}, {Prieto}, {Smith}, {Stubbs}, {Suntzeff}, {Walker},
  {Wood-Vasey}, {Zenteno}, {Wyrzykowski}, {Udalski}, {Szyma{\'n}ski}, {Kubiak},
  {Pietrzy{\'n}ski}, {Soszy{\'n}ski}, {Szewczyk}, {Ulaczyk}, \&
  {Poleski}}]{rfg+11}
{Rest}, A., {Foley}, R.~J., {Gezari}, S., {et~al.} 2011, \apj, 729, 88

\bibitem[{{Rest} {et~al.}(2013){Rest}, {Scolnic}, {Foley}, {Huber}, {Chornock},
  {Narayan}, {Tonry}, {Berger}, {Soderberg}, {Stubbs}, {Riess}, {Kirshner},
  {Smartt}, {Schlafly}, {Rodney}, {Botticella}, {Brout}, {Challis}, {Czekala},
  {Drout}, {Hudson}, {Kotak}, {Leibler}, {Lunnan}, {Marion}, {McCrum},
  {Milisavljevic}, {Pastorello}, {Sanders}, {Smith}, {Stafford}, {Thilker},
  {Valenti}, {Wood-Vasey}, {Zheng}, {Burgett}, {Chambers}, {Denneau}, {Draper},
  {Flewelling}, {Hodapp}, {Kaiser}, {Kudritzki}, {Magnier}, {Metcalfe},
  {Price}, {Sweeney}, {Wainscoat}, \& {Waters}}]{rsf+13}
{Rest}, A., {Scolnic}, D., {Foley}, R.~J., {et~al.} 2013, ArXiv e-prints,
  arXiv:1310.3828

\bibitem[{{Rix} {et~al.}(2004){Rix}, {Barden}, {Beckwith}, {Bell}, {Borch},
  {Caldwell}, {H{\"a}ussler}, {Jahnke}, {Jogee}, {McIntosh}, {Meisenheimer},
  {Peng}, {Sanchez}, {Somerville}, {Wisotzki}, \& {Wolf}}]{rbb+04}
{Rix}, H.-W., {Barden}, M., {Beckwith}, S.~V.~W., {et~al.} 2004, \apjs, 152,
  163

\bibitem[{{Sako} {et~al.}(2011){Sako}, {Bassett}, {Connolly}, {Dilday},
  {Cambell}, {Frieman}, {Gladney}, {Kessler}, {Lampeitl}, {Marriner}, {Miquel},
  {Nichol}, {Schneider}, {Smith}, \& {Sollerman}}]{sbc+11}
{Sako}, M., {Bassett}, B., {Connolly}, B., {et~al.} 2011, \apj, 738, 162

\bibitem[{{Sanders} {et~al.}(2007){Sanders}, {Salvato}, {Aussel}, {Ilbert},
  {Scoville}, {Surace}, {Frayer}, {Sheth}, {Helou}, {Brooke}, {Bhattacharya},
  {Yan}, {Kartaltepe}, {Barnes}, {Blain}, {Calzetti}, {Capak}, {Carilli},
  {Carollo}, {Comastri}, {Daddi}, {Ellis}, {Elvis}, {Fall}, {Franceschini},
  {Giavalisco}, {Hasinger}, {Impey}, {Koekemoer}, {Le F{\`e}vre}, {Lilly},
  {Liu}, {McCracken}, {Mobasher}, {Renzini}, {Rich}, {Schinnerer}, {Shopbell},
  {Taniguchi}, {Thompson}, {Urry}, \& {Williams}}]{scosmos}
{Sanders}, D.~B., {Salvato}, M., {Aussel}, H., {et~al.} 2007, \apjs, 172, 86

\bibitem[{{Sanders} {et~al.}(2012){Sanders}, {Soderberg}, {Levesque}, {Foley},
  {Chornock}, {Milisavljevic}, {Margutti}, {Berger}, {Drout}, {Czekala}, \&
  {Dittmann}}]{ssl+12}
{Sanders}, N.~E., {Soderberg}, A.~M., {Levesque}, E.~M., {et~al.} 2012, \apj,
  758, 132

\bibitem[{{Savaglio} {et~al.}(2009){Savaglio}, {Glazebrook}, \& {Le
  Borgne}}]{sgl09}
{Savaglio}, S., {Glazebrook}, K., \& {Le Borgne}, D. 2009, \apj, 691, 182

\bibitem[{{Schaerer} \& {Vacca}(1998)}]{sv98}
{Schaerer}, D., \& {Vacca}, W.~D. 1998, \apj, 497, 618

\bibitem[{{Schlafly} \& {Finkbeiner}(2011)}]{sf11}
{Schlafly}, E.~F., \& {Finkbeiner}, D.~P. 2011, \apj, 737, 103

\bibitem[{{Schlegel} {et~al.}(1998){Schlegel}, {Finkbeiner}, \&
  {Davis}}]{sfd98}
{Schlegel}, D.~J., {Finkbeiner}, D.~P., \& {Davis}, M. 1998, \apj, 500, 525

\bibitem[{{Schmidt} {et~al.}(1989){Schmidt}, {Weymann}, \& {Foltz}}]{swf89}
{Schmidt}, G.~D., {Weymann}, R.~J., \& {Foltz}, C.~B. 1989, \pasp, 101, 713

\bibitem[{{Shaw} \& {Dufour}(1994)}]{sd94}
{Shaw}, R.~A., \& {Dufour}, R.~J. 1994, in Astronomical Society of the Pacific
  Conference Series, Vol.~61, Astronomical Data Analysis Software and Systems
  III, ed. D.~R. {Crabtree}, R.~J. {Hanisch}, \& J.~{Barnes}, 327

\bibitem[{{Shi} {et~al.}(2006){Shi}, {Kong}, \& {Cheng}}]{skc06}
{Shi}, F., {Kong}, X., \& {Cheng}, F.~Z. 2006, \aap, 453, 487

\bibitem[{{Smith} {et~al.}(2010){Smith}, {Chornock}, {Silverman}, {Filippenko},
  \& {Foley}}]{scs+10}
{Smith}, N., {Chornock}, R., {Silverman}, J.~M., {Filippenko}, A.~V., \&
  {Foley}, R.~J. 2010, \apj, 709, 856

\bibitem[{{Smith} {et~al.}(2007){Smith}, {Li}, {Foley}, {Wheeler}, {Pooley},
  {Chornock}, {Filippenko}, {Silverman}, {Quimby}, {Bloom}, \&
  {Hansen}}]{slf+07}
{Smith}, N., {Li}, W., {Foley}, R.~J., {et~al.} 2007, \apj, 666, 1116

\bibitem[{{Stanek} {et~al.}(2006){Stanek}, {Gnedin}, {Beacom}, {Gould},
  {Johnson}, {Kollmeier}, {Modjaz}, {Pinsonneault}, {Pogge}, \&
  {Weinberg}}]{sgb+06}
{Stanek}, K.~Z., {Gnedin}, O.~Y., {Beacom}, J.~F., {et~al.} 2006, Acta Astron.,
  56, 333

\bibitem[{{Stasi{\'n}ska}(1982)}]{sta82}
{Stasi{\'n}ska}, G. 1982, \aaps, 48, 299

\bibitem[{{Stoll} {et~al.}(2011){Stoll}, {Prieto}, {Stanek}, {Pogge},
  {Szczygie{\l}}, {Pojma{\'n}ski}, {Antognini}, \& {Yan}}]{sps+11}
{Stoll}, R., {Prieto}, J.~L., {Stanek}, K.~Z., {et~al.} 2011, \apj, 730, 34

\bibitem[{{Strolger} {et~al.}(2004){Strolger}, {Riess}, {Dahlen}, {Livio},
  {Panagia}, {Challis}, {Tonry}, {Filippenko}, {Chornock}, {Ferguson},
  {Koekemoer}, {Mobasher}, {Dickinson}, {Giavalisco}, {Casertano}, {Hook},
  {Blondin}, {Leibundgut}, {Nonino}, {Rosati}, {Spinrad}, {Steidel}, {Stern},
  {Garnavich}, {Matheson}, {Grogin}, {Hornschemeier}, {Kretchmer}, {Laidler},
  {Lee}, {Lucas}, {de Mello}, {Moustakas}, {Ravindranath}, {Richardson}, \&
  {Taylor}}]{srd+04}
{Strolger}, L.-G., {Riess}, A.~G., {Dahlen}, T., {et~al.} 2004, \apj, 613, 200

\bibitem[{{Svensson} {et~al.}(2010){Svensson}, {Levan}, {Tanvir}, {Fruchter},
  \& {Strolger}}]{slt+10}
{Svensson}, K.~M., {Levan}, A.~J., {Tanvir}, N.~R., {Fruchter}, A.~S., \&
  {Strolger}, L.-G. 2010, \mnras, 405, 57

\bibitem[{{Tonry} \& {Onaka}(2009)}]{PS1_GPCA}
{Tonry}, J., \& {Onaka}, P. 2009, in Advanced Maui Optical and Space
  Surveillance Technologies Conference,

\bibitem[{{Tonry} {et~al.}(2012){Tonry}, {Stubbs}, {Lykke}, {Doherty},
  {Shivvers}, {Burgett}, {Chambers}, {Hodapp}, {Kaiser}, {Kudritzki},
  {Magnier}, {Morgan}, {Price}, \& {Wainscoat}}]{tsl+12}
{Tonry}, J.~L., {Stubbs}, C.~W., {Lykke}, K.~R., {et~al.} 2012, \apj, 750, 99

\bibitem[{{Tremonti} {et~al.}(2004){Tremonti}, {Heckman}, {Kauffmann},
  {Brinchmann}, {Charlot}, {White}, {Seibert}, {Peng}, {Schlegel}, {Uomoto},
  {Fukugita}, \& {Brinkmann}}]{thk+04}
{Tremonti}, C.~A., {Heckman}, T.~M., {Kauffmann}, G., {et~al.} 2004, \apj, 613,
  898

\bibitem[{{van den Heuvel} \& {Portegies Zwart}(2013)}]{vp13}
{van den Heuvel}, E.~P.~J., \& {Portegies Zwart}, S.~F. 2013, \apj, 779, 114

\bibitem[{{Vink} \& {de Koter}(2005)}]{vd05}
{Vink}, J.~S., \& {de Koter}, A. 2005, \aap, 442, 587

\bibitem[{{Vinko} {et~al.}(2010){Vinko}, {Zheng}, {Romadan}, {Quimby},
  {Whallon}, {Pandey}, {Fang}, {Akerlof}, {Pasque}, {Verkinderen}, {Wheeler},
  {Chatzopoulos}, \& {Caldwell}}]{vzr+10}
{Vinko}, J., {Zheng}, W., {Romadan}, A., {et~al.} 2010, Central Bureau
  Electronic Telegrams, 2556, 1

\bibitem[{{Willmer} {et~al.}(2006){Willmer}, {Faber}, {Koo}, {Weiner},
  {Newman}, {Coil}, {Connolly}, {Conroy}, {Cooper}, {Davis}, {Finkbeiner},
  {Gerke}, {Guhathakurta}, {Harker}, {Kaiser}, {Kassin}, {Konidaris}, {Lin},
  {Luppino}, {Madgwick}, {Noeske}, {Phillips}, \& {Yan}}]{wfk+06}
{Willmer}, C.~N.~A., {Faber}, S.~M., {Koo}, D.~C., {et~al.} 2006, \apj, 647,
  853

\bibitem[{{Wirth} {et~al.}(2004){Wirth}, {Willmer}, {Amico}, {Chaffee},
  {Goodrich}, {Kwok}, {Lyke}, {Mader}, {Tran}, {Barger}, {Cowie}, {Capak},
  {Coil}, {Cooper}, {Conrad}, {Davis}, {Faber}, {Hu}, {Koo}, {Le Mignant},
  {Newman}, \& {Songaila}}]{wwa+04}
{Wirth}, G.~D., {Willmer}, C.~N.~A., {Amico}, P., {et~al.} 2004, \aj, 127, 3121

\bibitem[{{Woosley}(2010)}]{woo10}
{Woosley}, S.~E. 2010, \apjl, 719, L204

\bibitem[{{Woosley} {et~al.}(2007){Woosley}, {Blinnikov}, \& {Heger}}]{wbh07}
{Woosley}, S.~E., {Blinnikov}, S., \& {Heger}, A. 2007, \nat, 450, 390

\bibitem[{{Yaron} \& {Gal-Yam}(2012)}]{yg12}
{Yaron}, O., \& {Gal-Yam}, A. 2012, \pasp, 124, 668

\bibitem[{{Yoon} \& {Langer}(2005)}]{yl05}
{Yoon}, S.-C., \& {Langer}, N. 2005, \aap, 443, 643

\bibitem[{{Young} {et~al.}(2010){Young}, {Smartt}, {Valenti}, {Pastorello},
  {Benetti}, {Benn}, {Bersier}, {Botticella}, {Corradi}, {Harutyunyan},
  {Hrudkova}, {Hunter}, {Mattila}, {de Mooij}, {Navasardyan}, {Snellen},
  {Tanvir}, \& {Zampieri}}]{ysv+10}
{Young}, D.~R., {Smartt}, S.~J., {Valenti}, S., {et~al.} 2010, \aap, 512, A70

\bibitem[{{Zahid} {et~al.}(2013){Zahid}, {Geller}, {Kewley}, {Hwang},
  {Fabricant}, \& {Kurtz}}]{zgk+13}
{Zahid}, H.~J., {Geller}, M.~J., {Kewley}, L.~J., {et~al.} 2013, \apjl, 771,
  L19

\bibitem[{{Zaritsky} {et~al.}(1994){Zaritsky}, {Kennicutt}, \&
  {Huchra}}]{zkh94}
{Zaritsky}, D., {Kennicutt}, Jr., R.~C., \& {Huchra}, J.~P. 1994, \apj, 420, 87

\end{thebibliography}
\end{document}